\documentclass{jfm}
\usepackage{graphicx}
\usepackage{amsmath,amssymb}
\usepackage{array}
\usepackage{color}
\usepackage[normalem]{ulem}

\usepackage{graphicx}
\usepackage{epstopdf}
\usepackage{caption,subcaption}
\usepackage{float}

\def\ULS{{U_{ls}^2}}

\shorttitle{Condensates in thin-layer turbulence}
\shortauthor{A. van Kan, A. Alexakis}

\title{Condensates in thin-layer turbulence}

\author{Adrian van Kan\aff{1,2}
  \corresp{\email{avankan@lps.ens.fr}},
Alexandros Alexakis\aff{2}}

\affiliation{\aff{1} Fakult\"at f\"ur Physik und Astronomie, Universit\"at Heidelberg, Im Neuenheimer Feld 226, D-69120 Heidelberg, Germany
\aff{2} Laboratoire de Physique Statistique, D\'epartement de Physique de l’ Ecole Normale Sup\'erieure, ´
PSL Research University, Universit\'e Paris Diderot, Sorbonne Paris Cit\'e, Sorbonne Universit\'es,
UPMC Univ. Paris 06, CNRS, 75005 Paris, France}

\begin{document}

\maketitle
\begin{abstract}
We examine the steady state of turbulent flows in thin layers using direct numerical simulations. It is shown that when the layer thickness is smaller than a critical height, an inverse cascade arises which leads to the formation of a steady state condensate where most of the energy is concentrated in the largest scale of the system. 
For layers of thickness smaller than a second critical height, the flow at steady state becomes exactly two-dimensional. The amplitude of the condensate is studied as a function of layer thickness and Reynolds number.
Bi-stability and intermittent bursts are found close to the two critical points.
The results are interpreted based on a mean-field three-scale model that reproduces 
some of the basic features of the numerical results.
\end{abstract}

\section{Introduction}
\label{sec:Intro}
Turbulent flows in geophysical and astrophysical contexts are often subject to geometrical constraints, such as thinness in a particular direction, that can strongly affect the behaviour of the flow.  
This occurs, for instance, in planetary atmospheres and oceans \citep{pedlosky2013geophysical} whose behaviour can strongly deviate from the classical three-dimensional homogeneous and  isotropic turbulence. 
This is related to the well-known fact that the behaviour of flows at large Reynolds numbers $\Rey$ depends on the dimensionality of the system. In three dimensions (3D), vortex stretching transfers energy to small scales in a \textit{direct cascade} \citep{frisch1995turbulence}. By contrast, in two-dimensions (2D), the conservation of enstrophy in addition to energy gives rise to an inverse energy cascade, a transfer of energy to the large scales \citep{bofetta2012twodimensionalturbulence}. 
Flows in thin layers display properties of both systems, with the large scales behaving like a 2D flow and the small scales 
behaving like a 3D flow. As a result, such systems are known to cascade energy both to large and to small scales \citep{smith1996crossover}. In fact, it has been shown in \citep{celani2010morethantwo, musacchio2017split, benavides_alexakis_2017} that as the height of the layer $H$ is varied, the system transitions from a state where energy cascades only to the small scales for large $H$, to a state where energy cascades to both large and small scales when $H$ is smaller than approximately half the size of the forcing length scale $\ell$. In particular, \citep{benavides_alexakis_2017}, using a Galerkin truncated model of the full Navier-Stokes equations, were able to provide strong evidence of the criticality of the transition. 
In addition, they observed a second transition to exact two-dimensionalisation for layers of very small thickness $H\propto \ell \Rey^{-1/2}$. This transition had been predicted theoretically using bounding techniques by \citep{gallet_doering_2015}. Similar transitions from a strictly forward cascade to an inverse cascade have been observed in other systems like rotating turbulence \citep{deusebio2014dimensional}, stratified turbulence \citep{sozza2015dimensional}, rotating and stratified flows \citep{marino2015resolving}, magneto-hydrodynamic systems \citep{alexakis2011two,seshasayanan2014edge,seshasayanan2016critical} and helically constrained flows \citep{sahoo2015disentangling, sahoo2017discontinuous}, to mention a few 
(see \cite{alexakis2018cascades} for a review).

The thin layer, however, remains possibly the simplest model exhibiting such transitions and it thus deserves a detailed study at the different stages of inverse cascade evolution.
In the presence of an inverse cascade, for finite systems and in the absence of a large-scale dissipation term, there are two stages in the development of the flow. In the first stage (at early times), energy is transferred to larger and larger scales by the inverse cascade. This process stops, however, when scales comparable to the system size are reached, after which energy starts to pile up at these largest scales. In the long-time limit, the increase of the large-scale energy saturates and a condensate is formed, where nearly all energy is found in the first few Fourier modes. 
For 2D Navier-Stokes turbulence, the possibility of such a condensation phenomenon was first conjectured in the seminal paper of \citep{kraichnan1967inertial}, first seen in DNS by \citep{hossain1983long}, further explored quantitatively by \citep{smith1993bose, smithr1994finite},  and more recently by \citep{chertkov2007dynamics, bouchet2009random, chan2012dyn, frishman2017turbulence,frishman2017jets}. Spectral condensation has also been studied in other quasi-2D systems such as quasi-geostrophic flows \citep[see][]{kukharkin1995quasicrystallization, kurkharkin1996generation, vallis1993generation, venaille2011oceanic}.
In terms of the real space flow field, this spectral condensation corresponds to coherent system-size vortices or shear layers. In 2D, where the cascade of energy is strictly inverse, a steady state in the condensate regime is realised when the energy of the condensate is so large that the dissipation due to viscosity at large scales balances the energy injection due to the forcing. 
For split cascading systems, this is not necessarily true due to the presence of non-vanishing 3D flow variations associated with the direct cascade. Therefore, in this case other processes exist that can redirect the energy back to the small scales where viscous dissipation is more efficient.
Such mechanisms have been demonstrated for rotating turbulence, where a flux-loop mechanism has been identified \citep[cf.][]{bartello1994coherent, alexakis2015rotating,  seshasayanan2018condensates}.  Similar condensates have also been observed in 3D fast rotating convection \citep{favier2014inverse, rubio2014upscale, guervilly2014large}.

Condensates in thin layers have been observed experimentally: the first study by \citep{sommeria1986experimental} was followed by the important contributions of \citep{paret1997experimental, paret1998intermittency} and more recently by \citep{shats2005spectral, shats2007suppression, xia2008turbulence, xia2009spectrallycondensed, xia2011upscale, byrne2011robust}. An up-to-date review of relevant experiments is presented in \citep{xia2017two}. These experiments operate primarily in the long-time limit in which the condensate is fully developed. This wealth of experimental studies of thin-layer condensates is in striking contrast with the existing numerical results which have focused exclusively on the transient growth of total kinetic energy due to the inverse cascade. In these numerical simulations, the condensate state reached after long time in the thin-layer case has not yet been examined due to the long computation time needed.
In this study, we aim to fill this gap and investigate the behaviour of turbulent flow at the condensate stage for a thin layer forced at intermediate scales, using direct numerical simulations (DNS) and low-order modelling. The DNS provide a detailed picture of the behaviour of the full system, while the modelling shines light on the main physical processes involved in the problem. \\

The remainder of this article is structured as follows. In section \ref{sec:setup}, we present the set-up and define the quantities we will be measuring. In section \ref{sec:DNS}, 
we present the results of a large number of direct numerical simulations (DNS) of thin-layer turbulence. Next, in section \ref{sec:close_to_transitions}, we discuss the behaviour close to the two critical points and in section \ref{sec:spec_flux}, we present spectra and spectral fluxes of energy. In section \ref{sec:3mode}, we introduce a low-order model which captures many features of the DNS results. Finally, in section \ref{sec:discussion}, we discuss our results and summarise.

\section{Physical setup}
\label{sec:setup}
\begin{figure}                                                                          %
\centering                                                                              %
\includegraphics[width=0.7\textwidth]{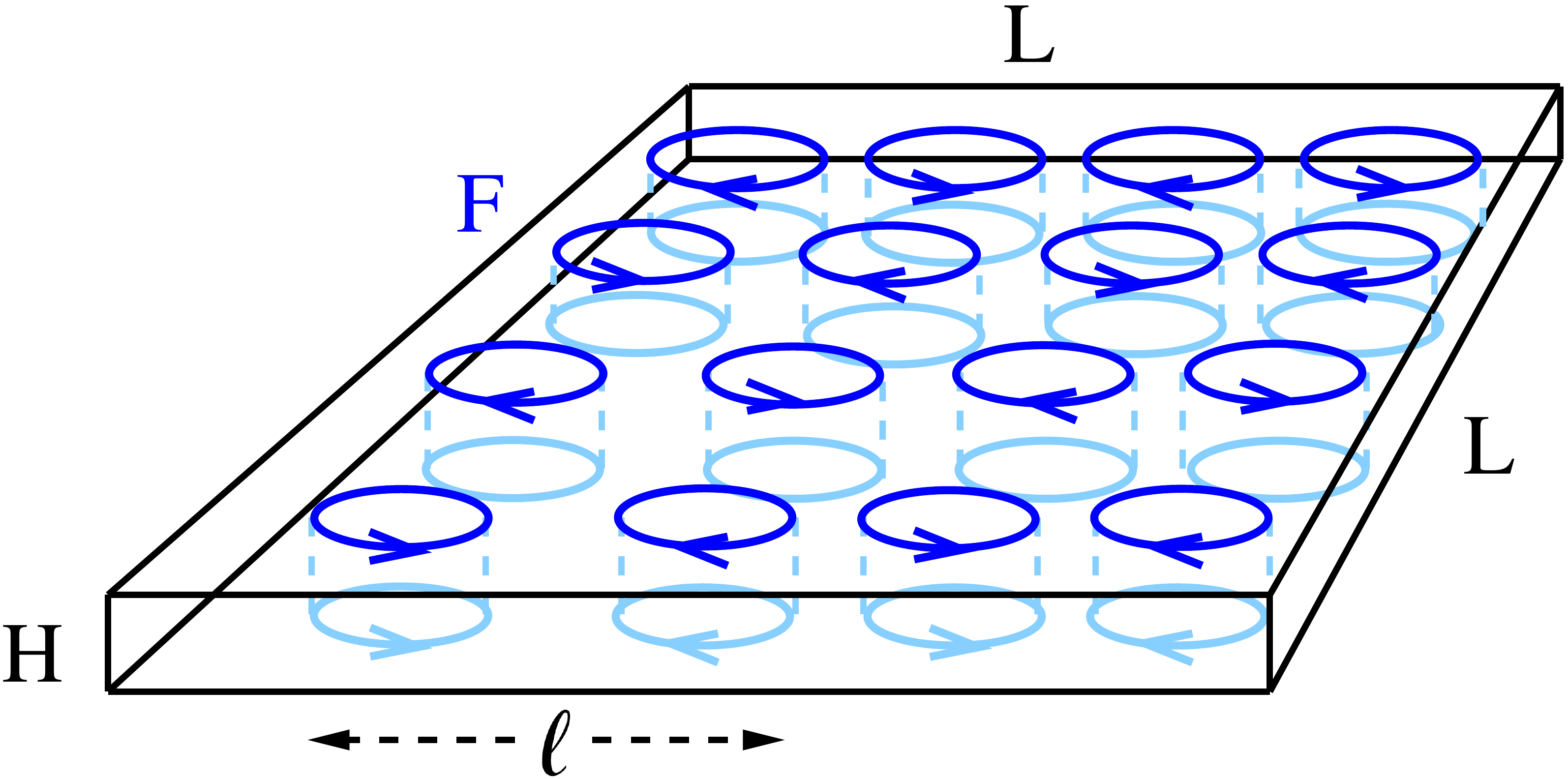}                                        %
\caption{The domain used is a box of height $H$ and square base of side length $L$.     %
The forcing is invariant along the thin direction and stochastic with fixed mean        %
rate of energy input, while involving only wavenumbers $\mathbf{k}$ with                   %
$|\mathbf{k}|=k_f=2\pi/\ell$. The thin direction will be referred to as the                %
\textit{vertical}, the others as the \textit{horizontal} directions.}                   %
\label{fig:DNS_illustr}                                                                 %
\end{figure}

In this section, we describe the set-up to be investigated.
We consider the idealised case of forced incompressible three-dimensional flow in a triply periodic box of dimensions $L\times  L \times H$. The thin direction $H$ will be referred to as the \textit{vertical} `$z$' direction and the remaining two as the \textit{horizontal} `$x$' and `$y$' directions. The geometry of the domain is illustrated in figure \ref{fig:DNS_illustr}. 
The flow obeys the incompressible Navier-Stokes equation:
\begin{subeqnarray}
\partial_t \mathbf{u} + \mathbf{u}\cdot \nabla \mathbf{u} =& -& \nabla P + \nu \nabla^2 \mathbf{u} + \mathbf{f},\\
\nabla \cdot \mathbf{u} =& 0& \label{eq:NS_inc}
\end{subeqnarray}
where $\mathbf{u}$ is the velocity field, $P$ is physical pressure divided by constant density and $\nu$ is (kinematic) viscosity. Energy is injected into the system by $\mathbf{f}$, a stochastic force, that depends only on $x$ and $y$ and has only $x$ and $y$ components, i.e. that is a two-dimensional-two-component (2D2C) field. We make this assumption firstly to specifically force the inversely cascading components of the flow and secondly because it is widely used in previous studies such as \citep{smith1996crossover, celani2010morethantwo,deusebio2014dimensional, gallet_doering_2015} and thus enables us to compare more easily with the literature. The force is divergence-free, hence it can be written as $\mathbf{f} = (-\partial_y \psi, \partial_x \psi,0)$. The spectrum of $\bf f$ is concentrated in a ring of wavenumbers of radius $k_f\equiv 2\pi/\ell$. It is delta-correlated in time, which leads to a fixed mean injection of energy $\langle  \mathbf{u}\cdot \mathbf{f}\rangle  = \epsilon$, where $\langle \cdot \rangle$ denotes an ensemble average over infinitely many realisations. We use random initial conditions whose small energy is spread out over a range of wave numbers.  In some cases, in order to compare with previous studies,  we used a hyper-viscosity, which amounts to replacing $\nu \nabla^2 \mathbf{u}$ by $-\nu_n (-\nabla^2)^n \mathbf{u}$.

 The system (\ref{eq:NS_inc}) is characterised by three non-dimensional parameters: 
the Reynolds number based on the energy injection rate $\Rey = (\epsilon \ell^4)^{1/3}/\nu$,
the ratio between forcing scale and domain height $Q=\ell/H$
and the  ratio between forcing scale and the horizontal domain size $K=\ell /L$.
The ratio between $K$ and $Q$ gives the aspect ratio $A=K/Q=H/L$ of the domain. 
The Kolmogorov dissipation length is denoted as $\eta =\nu^{3/4}/ \epsilon^{1/4}=\ell \Rey^{3/4}$.\\

The simulations performed for this work used an adapted version of the Geophysical High-Order Suite for Turbulence (GHOST) which uses pseudo-spectral methods including 2/3 aliasing to solve for the flow in the triply periodic domain, \citep[see][]{mininni2011hybrid}.
The resolution was varied from $256^2\times 16$ grid points to $2048^2\times 128$ grid points depending on the value of the parameters.
To explore the space spanned by these three parameters, we have performed systematic numerical experiments: for a fixed value of $\Rey$ and $K=1/8$, different simulations are performed with $Q$ varying from small to large values. The runs are continued until a steady state is reached where all quantities fluctuate around their mean value.
This is repeated for eight different values of $\Rey$ from $\Rey = 203$ (resolution $256^2\times 16$) to $4062$ (resolution $2048^2\times 128$) and for one value of hyperviscosity ($n=8$, $\nu_8 = 10^{-38}$ as in \citep{celani2010morethantwo}), as a consistency check, since many of the previous studies of thin-layer turbulence used hyper-viscosity. For $\Rey = 305$, we also perform a run with $K=1/16$ ($L\to 2L$). The number of runs performed for each $\Rey$ are summarised in table \ref{tab:runs}. 

\begin{table}                                                                         %
\begin{center}                                                                        %
   \begin{tabular}{ | c | c | c | c | c | c | c | c | c |  }                          %
\hline                                                                                %
  $\Rey$     & 203 &        305  & 406 & 609 & 870 & 2031 & 4062 & Hyper \\ \hline   %
  $1/K$      &   8 &   8 \&   16 &   8 &   8 &    8 &    8 &    8 &    8  \\ 
  $ n_x=n_y$ & 256 & 256 \& 512  & 256 & 512 &  512 & 1024 & 2048 & 1024  \\ 
  $ n_z   $  &  16 &         16  &  16 &  32 &   32 &   64 &  128 &   64  \\ 
    \# runs  & 40  &         40  &  40 &  30 &   30 &   10 &    2 &    4  \\  \hline  %
  \end{tabular}                                                                       %
   \caption{Summary of the different runs performed. For each $Re$ and $K$ several    %
   runs for different values of $Q$ have been performed. The horizontal resolution is %
   $n_x,n_y$, while $n_z$ stands for the vertical resolution at $Q=2$. The vertical   %
   resolution was changed with $Q$ to maintain an isotropic grid,                     %
   $ K n_x =K n_y = Q n_z$ wherever possible.}                                        %
  \label{tab:runs}                                                                    %
\end{center}                                                                          %
\end{table}                                                                           %

To quantify the energy distribution among different scales it is convenient to work in Fourier space. The Fourier series expansion of the velocity reads
\begin{equation}
\mathbf{u}(\mathbf{x},t) = \sum_{\mathbf{k}   } \hat{\mathbf{u}}_\mathbf{k} e^{i\mathbf{k}\cdot \mathbf{x}},\qquad 
\hat{\mathbf{u}}_\mathbf{k} = \frac{1}{L^2 H}\int \mathbf{u}(\mathbf{x},t) e^{-i\mathbf{k}\cdot \mathbf{x}} d\bf x
\end{equation}
where $\hat{\mathbf{u}}_\mathbf{k} = (\hat{u}_\mathbf{k}^{(x)},\hat{u}_\mathbf{k}^{(y)},\hat{u}_\mathbf{k}^{(z)})$ and the sum runs over all $\mathbf{k} \in \left(\frac{2 \pi}{L} \mathbb{Z} \right)^2 \times \frac{2\pi}{H} \mathbb{Z}$. In the pseudo-spectral calculations, this sum is truncated at a finite $k_{res}$. Since flow in a thin layer is a highly anisotropic system, it is important to consider quantities in the vertical and horizontal directions separately. For this purpose, we monitor various quantities in our simulations:
first of all, the total energy spectrum as a function of horizontal wavenumber
\begin{equation}
E_{tot}(k_h) = \frac{1}{2}\sum_{\mathbf{k} \atop {k_x^2+k_y^2=k_h^2} } \left|\hat{{\bf u}}_{\mathbf{k}} \right|^2.
\label{eq:spec_tot_ksphr}
\end{equation}
In addition, we monitor different components of domain-integrated energy, namely the total horizontal kinetic energy 
\begin{equation} 
\frac{1}{2} U_h^2 = \frac{1}{2}\sum_{\mathbf{k} \atop k_z=0 } \left(\left|\hat{{u}}_\mathbf{k}^{(x)}\right|^2 + \left|\hat{{u}}_\mathbf{k}^{(y)}\right|^2\right)
\label{eq:horiz_energy}
\end{equation}
(based on the (vertically averaged) 2D2C field only),
the large-scale horizontal kinetic energy 
\begin{equation}
\frac{1}{2} U_{ls}^2 = \frac{1}{2}\sum\limits_{\mathbf{k} \atop {k<k_{max} \atop  k_z=0}} \left(\left|\hat{{u}}_\mathbf{k}^{(x)}\right|^2 + \left|\hat{{u}}_\mathbf{k}^{(y)}\right|^2\right),
\label{eq:horiz_ls_energy}
\end{equation}
where $k_{max}= \sqrt{2} \frac{2\pi}{L}$, as well as 
the (vertically averaged) large-scale kinetic energy in the $z$ component 
\begin{equation}
\frac{1}{2} U_z^2 =\frac{1}{2} \sum_{\mathbf{k} \atop {k<k_{max} \atop k_z=0}} \left|\hat{{u}}_\mathbf{k}^{(z)}\right|^2
\label{eq:vert_energy}
\end{equation}
and the three-dimensional kinetic energy (\textit{3D energy}), defined as 
\begin{equation} 
 \frac{1}{2}U_{3D}^2 = \frac{1}{2} \sum _{\mathbf{k} \atop {k_z\neq 0\atop }} |\hat{\mathbf{u}}_\mathbf{k}|^2.
\label{eq:3D_energy}
\end{equation}

\section{Results from the direct numerical simulations}  
\label{sec:DNS}                                          
In this section, we present the results obtained from our simulations. 
For a given set of parameters $Re,Q,K$, two different behaviours are possible.
For thick layers $Q\ll1$, 3D turbulence is observed, i.e. there is no inverse cascade and the energy injected by the forcing is transferred to the small scales where it is dissipated. No system-size structures appear in this case.
For thin layers $Q\gg1$, a split cascade is present with part of the energy cascading inversely to the large scales and part of the energy cascading forward to the small scales. For these layers, at steady state, coherent system-size vortices appear with very large amplitudes.

\begin{figure}                                                                        
\centering                                                                            
\begin{subfigure}[b]{0.47\linewidth}                                                  
\includegraphics[width = 0.9\linewidth]{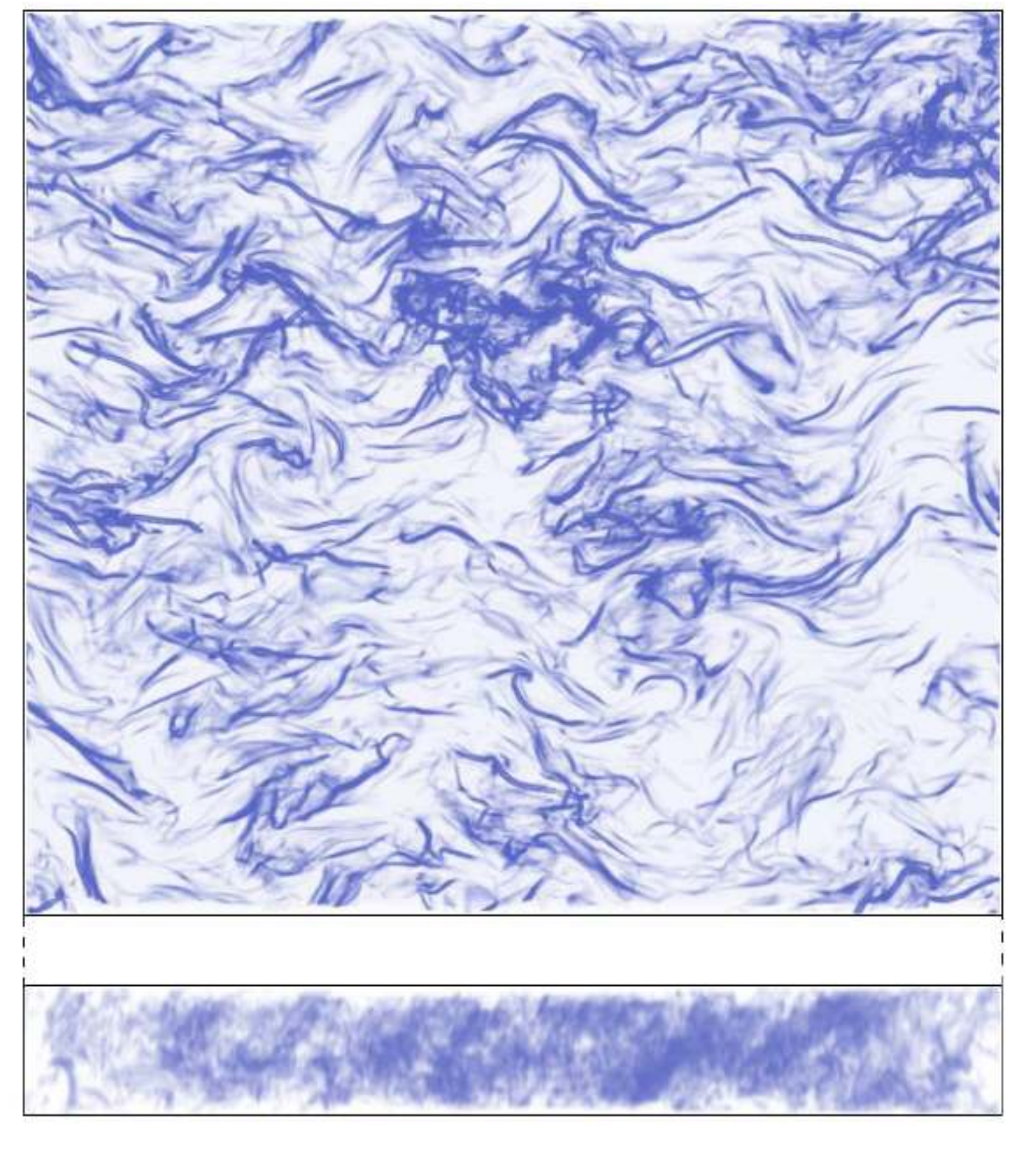} \caption{\label{fig:3Dturb}}  
\end{subfigure}                                                                       
\begin{subfigure}[b]{0.4765\linewidth}                                                  
\includegraphics[width = 0.9\linewidth]{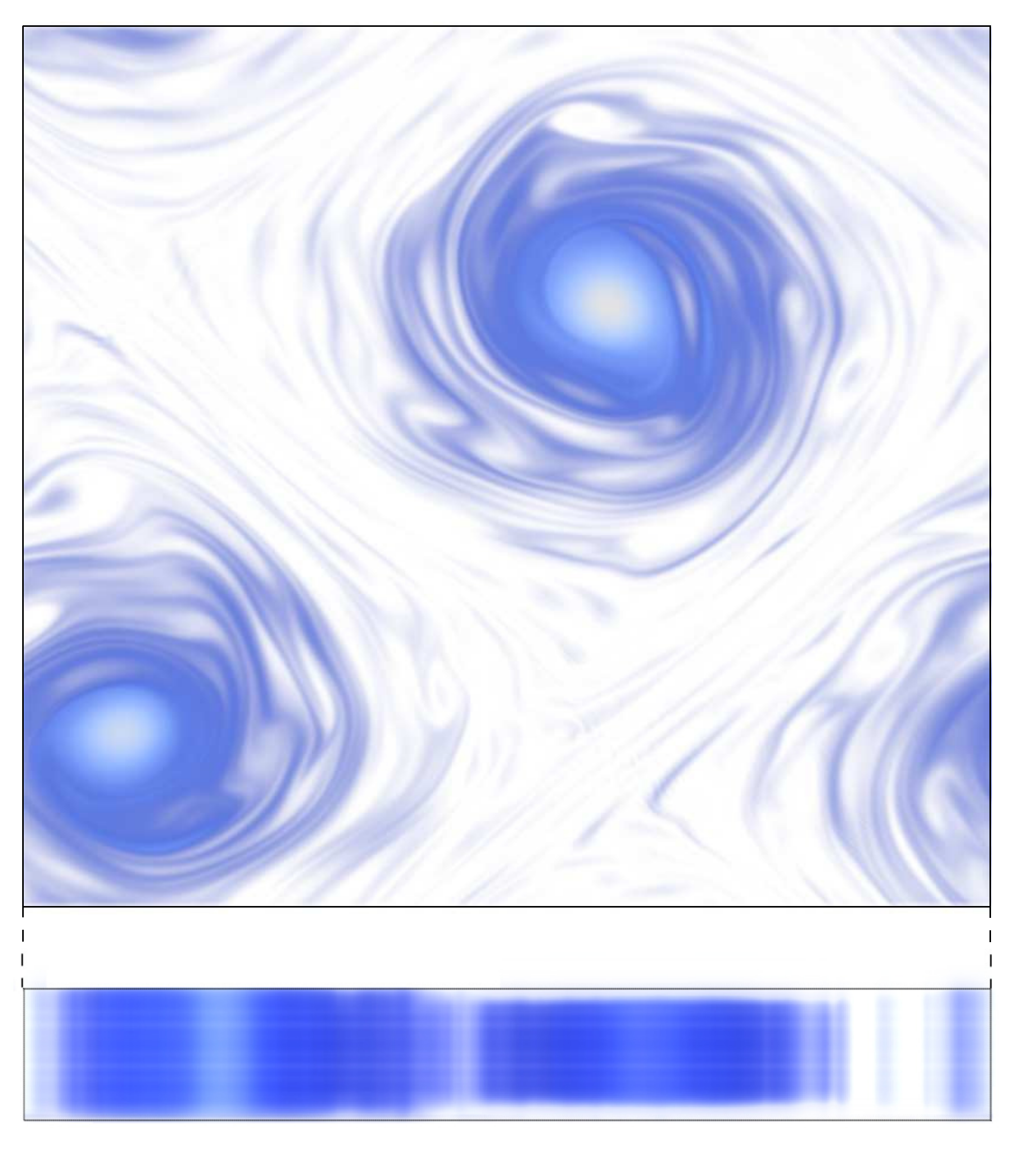} \caption{\label{fig:condturb}}    
\end{subfigure}                                                                       
\caption{Typical flow fields in the steady state of 3D turbulence (\ref{fig:3Dturb})  
and 2D turbulence (\ref{fig:condturb}) regimes, visualised using squared 
vorticity. The boxes below show the corresponding side views. Note the astonishing 
similarity between this figure and figures 1 a), b) of the experimental study by Xia et al 2011.} \label{fig:vis}  
\end{figure}
\begin{figure}                                                                     
\centering                                                                         
\begin{subfigure}[b]{0.485\linewidth}                                              
\includegraphics[width = 7.2cm]{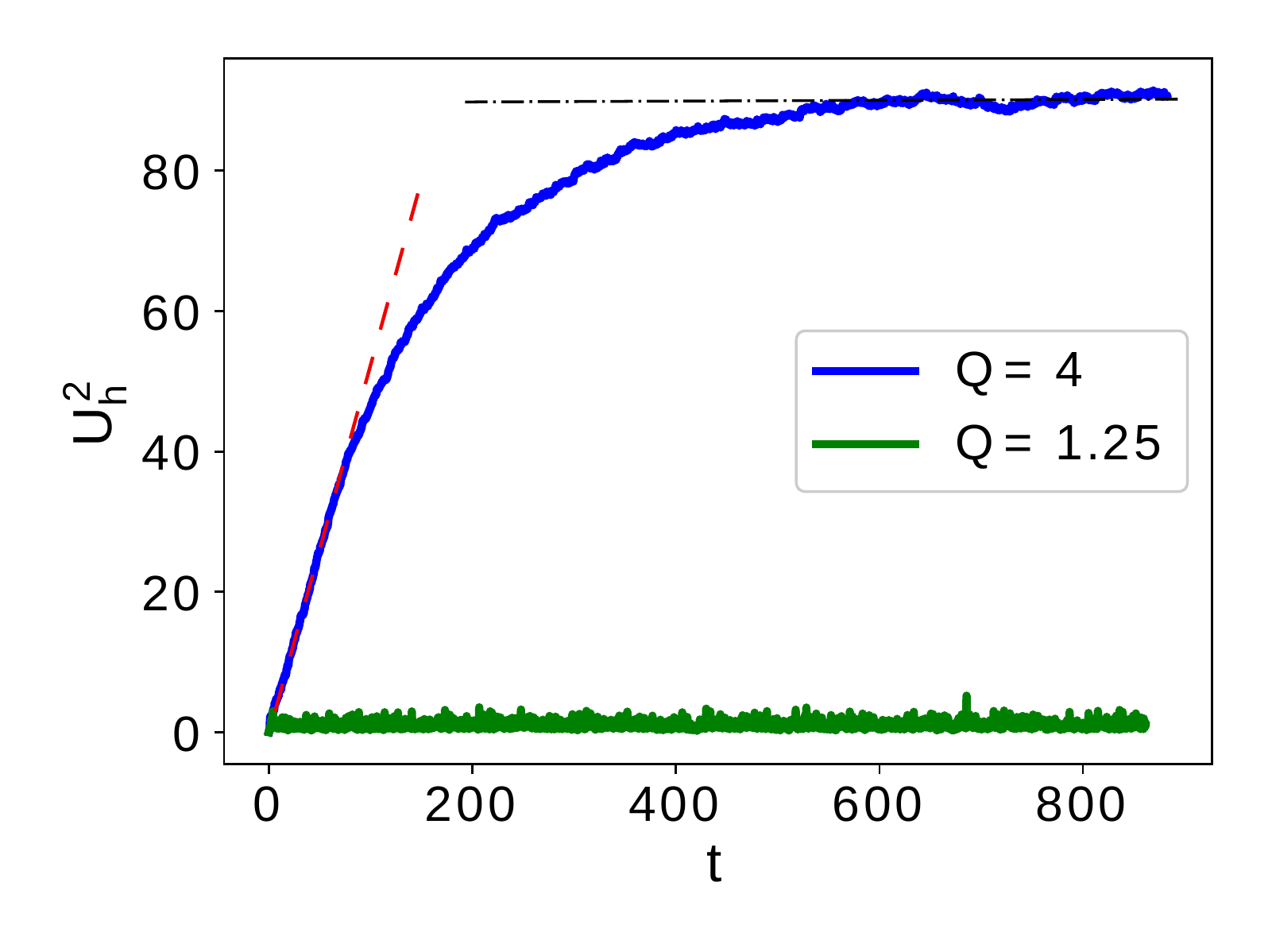}    \caption{\label{fig:typ1}}         
\end{subfigure} \hspace{0.2cm}                                                     
\begin{subfigure}[b]{0.485\linewidth}                
\includegraphics[width=7.2cm]{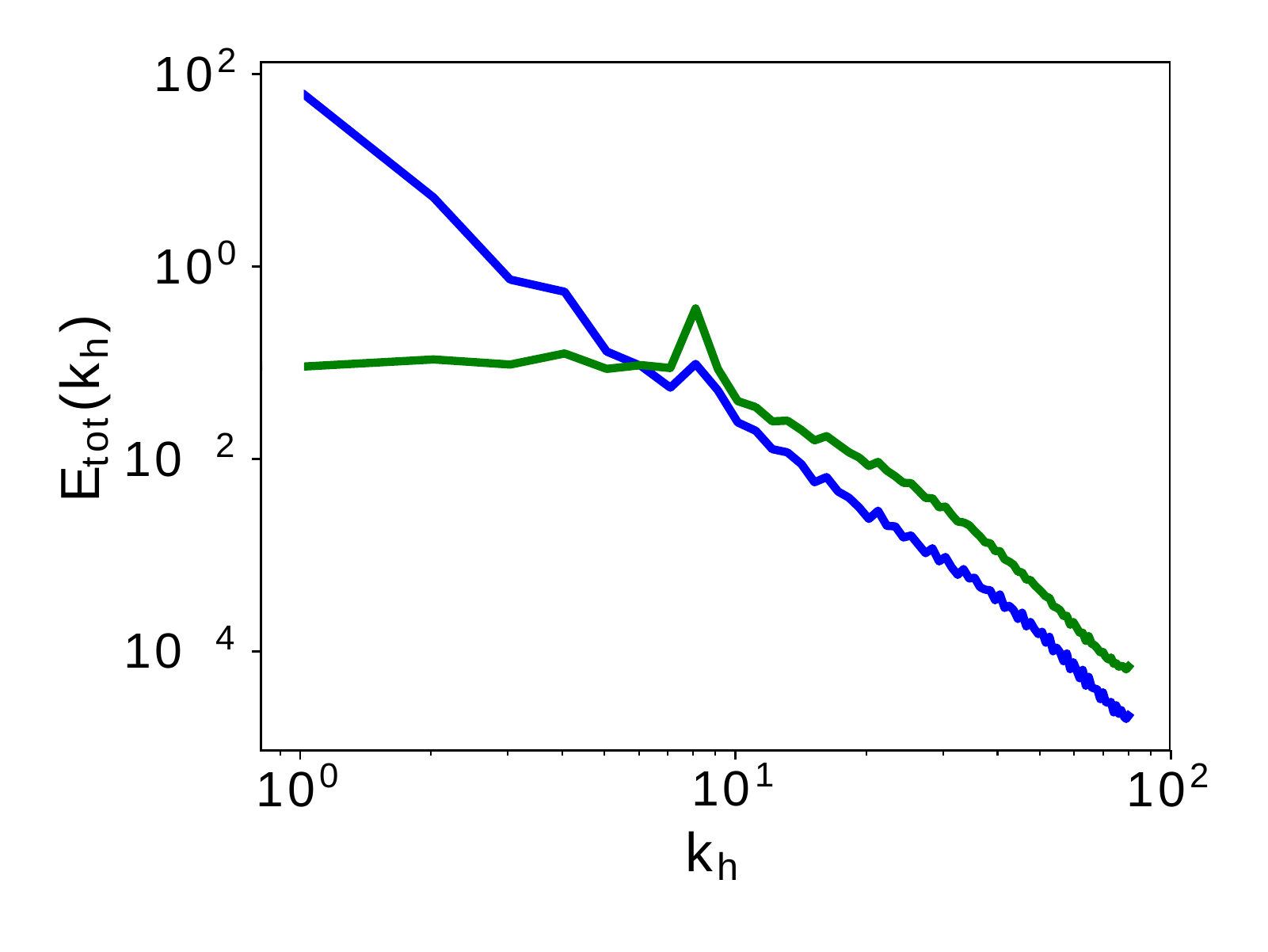}     \caption{\label{fig:typ2}}     
\end{subfigure}                                                                    
\caption{Panel (a) shows the typical time evolution of $U_h^2$ 
for $Q=1.25<Q_{3D}$ and $Q=4>Q_{3D}$. In the former case, $U_h^2$ remains small. In the latter, there is an initially linear increase whose slope measures the rate of inverse energy transfer. After long time, $U_h^2$ reaches its steady state value. Two quantities are measured:
the initial slope (\textbf{red-dashed} line) and the condensate value 
(horizontal black \textbf{dashed-dot} line).  A similar evolution observed in an experiment is shown in \citep{xia2009spectrallycondensed} figure 6. Panel (b) shows the corresponding 
spectra: in the presence of an inverse cascade there is a maximum at the largest 
scale, while in its absence the maximum is near the forcing scale. }   
\label{fig:time}                                                                   
\end{figure}

                                                                          
A visualisation of the flow field in these two different states is shown in figure \ref{fig:vis}
for the 3D turbulence and condensate states. 
Typical time-series of $U_h^2$ for a thick layer (forward cascade)
and a thin layer (inverse cascade) are shown in figure \ref{fig:typ1}. 
For the thick layer, the total energy fluctuates around a mean value of order $(\epsilon\ell)^{2/3}$,
while for the thin layer, the energy saturates to a much larger value. The energy spectra for the two runs of figure \ref{fig:typ1} at the steady state are shown in figure \ref{fig:typ2},
showing quantitatively that energy is concentrated in the large scales for the two different cases.
In more detail, $U_h^2$ for the thin layer shows two different stages: 
first, at early times, there is a linear increase with time and second, there is saturation at late times. Therefore, to fully describe the evolution of the system, we need to quantify the rate of the initial energy increase and the energy at which it saturates. The {\bf red-dashed} line indicates a fit to the initial linear increase. This slope provides a measurement of the rate $\epsilon_{inv}$ at which energy cascades inversely. For the steady state stage, the black {\bf dashed-dot} line indicates the mean value at late times. For all runs, we measure the slope of the $U_h^2$ curve and the steady state mean values of all corresponding energies defined in the previous section. For the runs of high resolution, to accelerate convergence, the large-scale velocity $\mathbf{u}_{k=1}$ (from a run at the early stage) was increased artificially and the run continued. Alternatively, an output of a converged run was used as initial condition. However, all cases were run sufficiently long to demonstrate that they have reached a steady state.

\begin{figure}                                                                       %
\centering                                                                           %
\begin{subfigure}[b]{0.485\linewidth}                                                
\includegraphics[width=\textwidth]{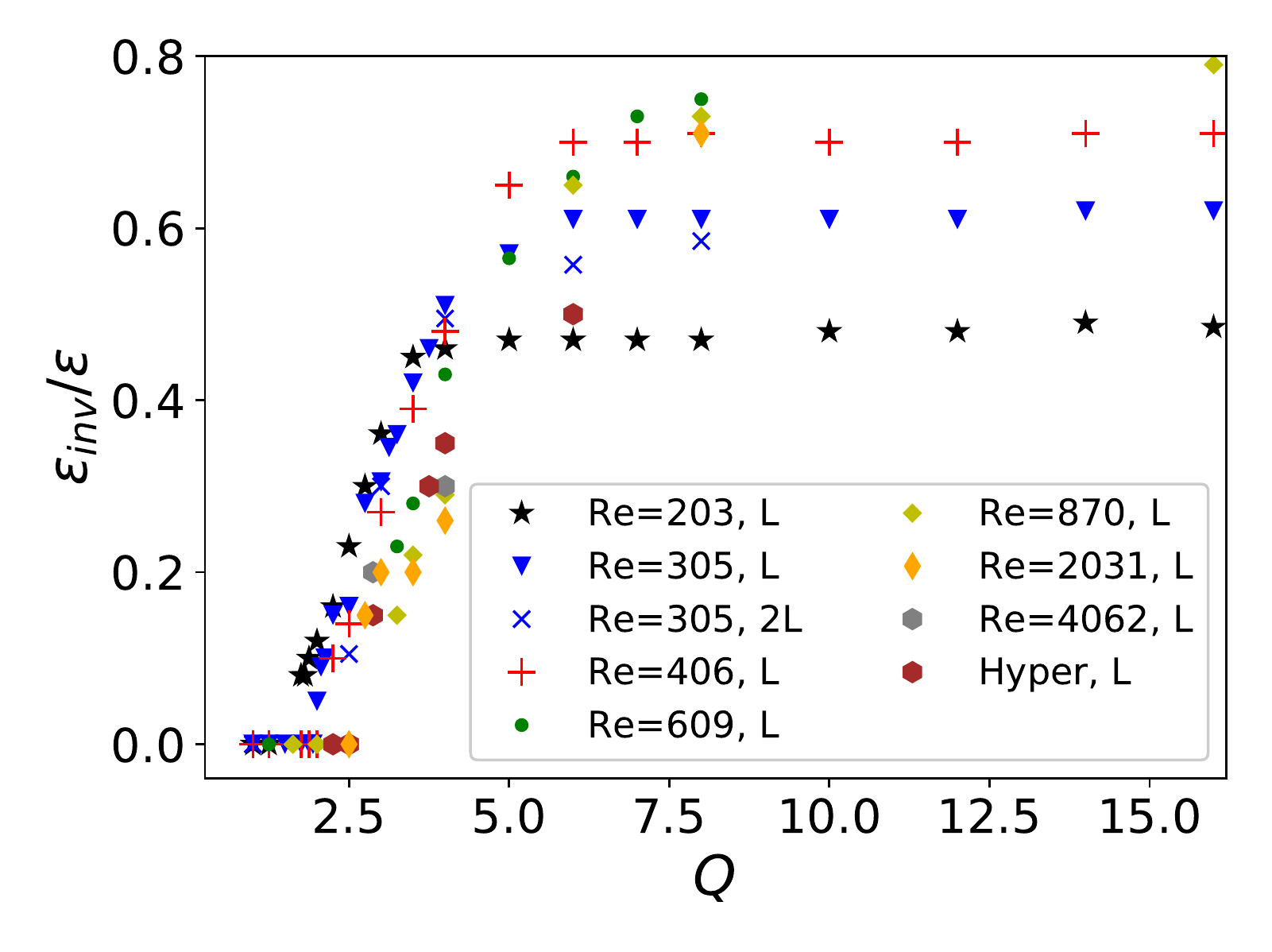} \caption{\label{fig:slopes1}}       
\end{subfigure}                                                                      
\begin{subfigure}[b]{0.485\linewidth}                                                
\includegraphics[width = \textwidth]{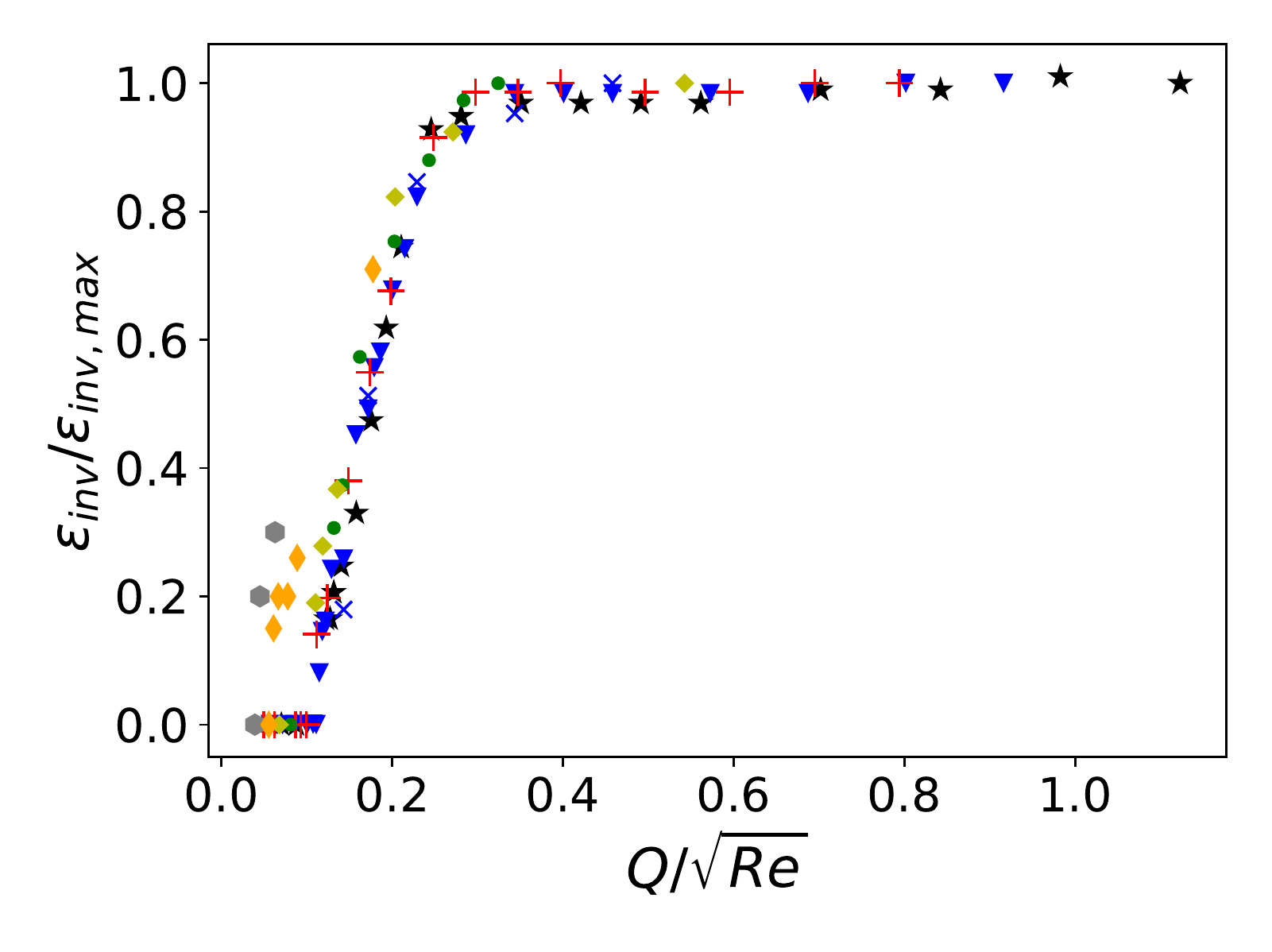} \caption{\label{fig:slopes2}}
\end{subfigure}                                                                      
 \caption{Panel (a): initial slopes, measured as indicated in    
 figure \ref{fig:typ1}, non-dimensionalised by the energy injection rate $\epsilon$, 
 as a function of $Q$ $(\propto 1/H)$ for all  $\Rey$ used. The same symbols are 
 used in all plots in this section. Thick layers are at small $Q$ (left) and thin layers at large $Q$ (right). Panel (b): the same data collapsed by a rescaling of the abscissa by $\sqrt{Re}$ and the coordinate by the maximum value obtained for that Reynolds number.  }
\label{fig:slopes}                                                                   %
\end{figure}                                                                         %
\begin{figure}                                                                          %
\centering                                                                              %
	\includegraphics[width=0.55\textwidth]{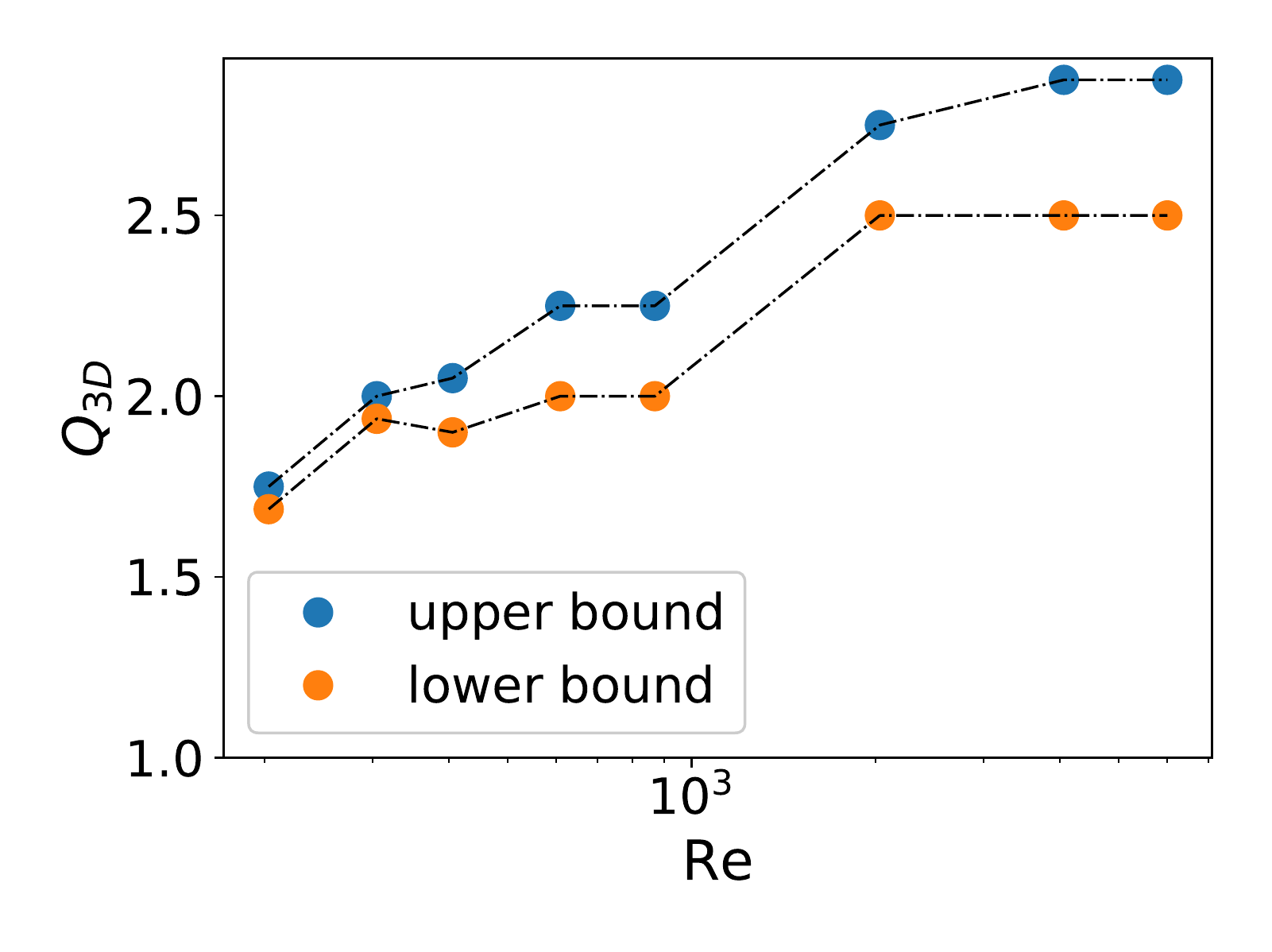} 
\caption{ Estimated value of $Q_{3D}$ as a function of $Re$. The top line shows the 
smallest value of $Q$ for which an inverse cascade was observed and the bottom line shows
the largest value of $Q$ for which no inverse cascade was observed. 
The rightmost point indicates the results from the hyper-viscous runs.}               
\label{fig:Q3D}                                                                         %
\end{figure}                                                                            %
Figure \ref{fig:slopes} shows the slopes of the initial total energy increase $\epsilon_{inv}$,
measured as illustrated in figure \ref{fig:typ1} for all our numerical simulations. The slopes are non-dimensionalised by the energy input rate $\epsilon$ and plotted versus $Q$ for all different values of $\Rey$ including the hyper-viscous runs. The slope at this early stage measures the strength of inverse energy transfer. At small $Q$ (deep layers), the slope vanishes for all runs, showing that no inverse cascade is present. Moving to larger $Q$, for every $\Rey$, there is a critical value $Q_{3D}(\Rey)$ of $Q$ above which the slope becomes non-zero. This is the birth of the inverse cascade. 
Figure \ref{fig:Q3D} shows estimates of $Q_{3D}$ as a function of $\Rey$: the upper curve shows the
smallest $Q$ for which an inverse cascade was observed for that given $\Rey$
while the lower curve shows the largest $Q$ for which no inverse cascade was observed.
The critical value $Q_{3D}$ lies between these two curves.
The point $Q_{3D}$ shifts to larger $Q$ as $\Rey$ is increased but eventually for the two largest $\Rey$ simulated, namely $\Rey=2031$ and $\Rey=4062$, as well as the hyper-viscous run, $Q_{3D}$ saturates at $Q_{3D}\approx 2.5$. (Previous findings \citep{celani2010morethantwo} estimated this value to $Q_{3D}\approx 2$, however in that work too limited a range of values of $Q$ was used to be able to precisely pinpoint $Q_{3D}$. Another possible reason for the different result is the different value of $1/K$ associated with the different forcing wavenumber $k_f=16$ used). The saturation of $Q_{3D}\approx 2.5$ indicates that $Q_{3D}$ converges to this value at large $\Rey$. For $Q>Q_{3D}$, the slope begins increasing linearly $\epsilon_{inv} \propto Q-Q_{3D}$. (We note that small slopes are hard to distinguish from zero slope since the difference only becomes apparent after a long simulation time.)

If $Q$ is increased further, a point $Q_{2D}$ is reached beyond which the slope becomes independent of $Q$. Above this second critical point, the flow becomes exactly 2D \citep{benavides_alexakis_2017}. The value of $Q_{2D}$ increases with $\Rey$ as $Q_{2D}\propto \Rey^{1/2}$. 
This scaling is verified in this work as well and shown in the right panel of figure \ref{fig:slopes}. The two critical points $Q_{3D}$ and $Q_{2D}$ at this early stage of development of the inverse cascade have been studied in detail in the past \citep{celani2010morethantwo, benavides_alexakis_2017}.

\begin{figure}                                                                          %
\centering                                                                              %
\begin{subfigure}[b]{0.485\linewidth}                                                   %
\includegraphics[width=\textwidth]{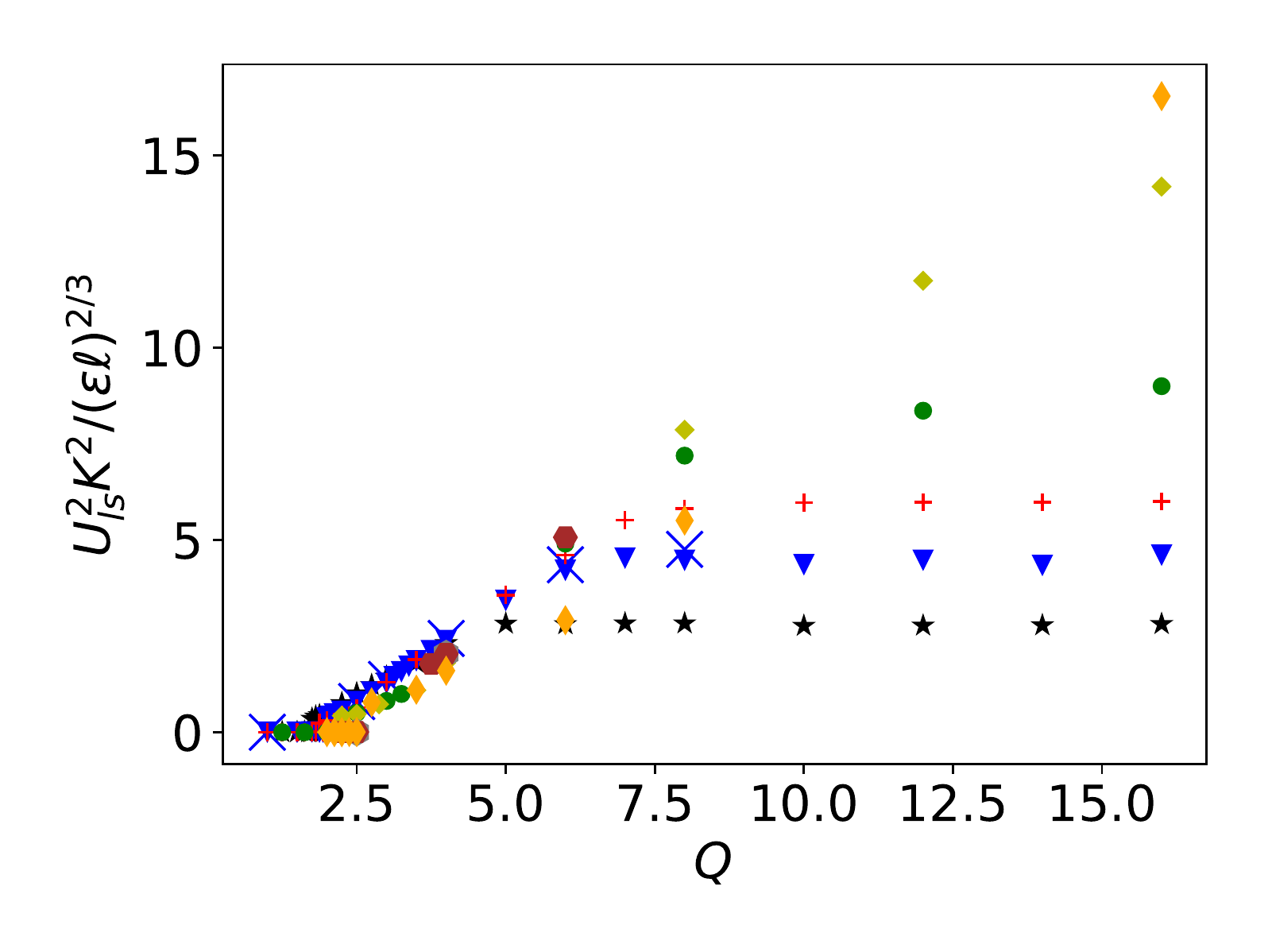}  \caption{\label{fig:U2_nonres}}  %
\end{subfigure}                                                                         %
\begin{subfigure}[b]{0.485\linewidth}                                                   %
 \includegraphics[width=\textwidth]{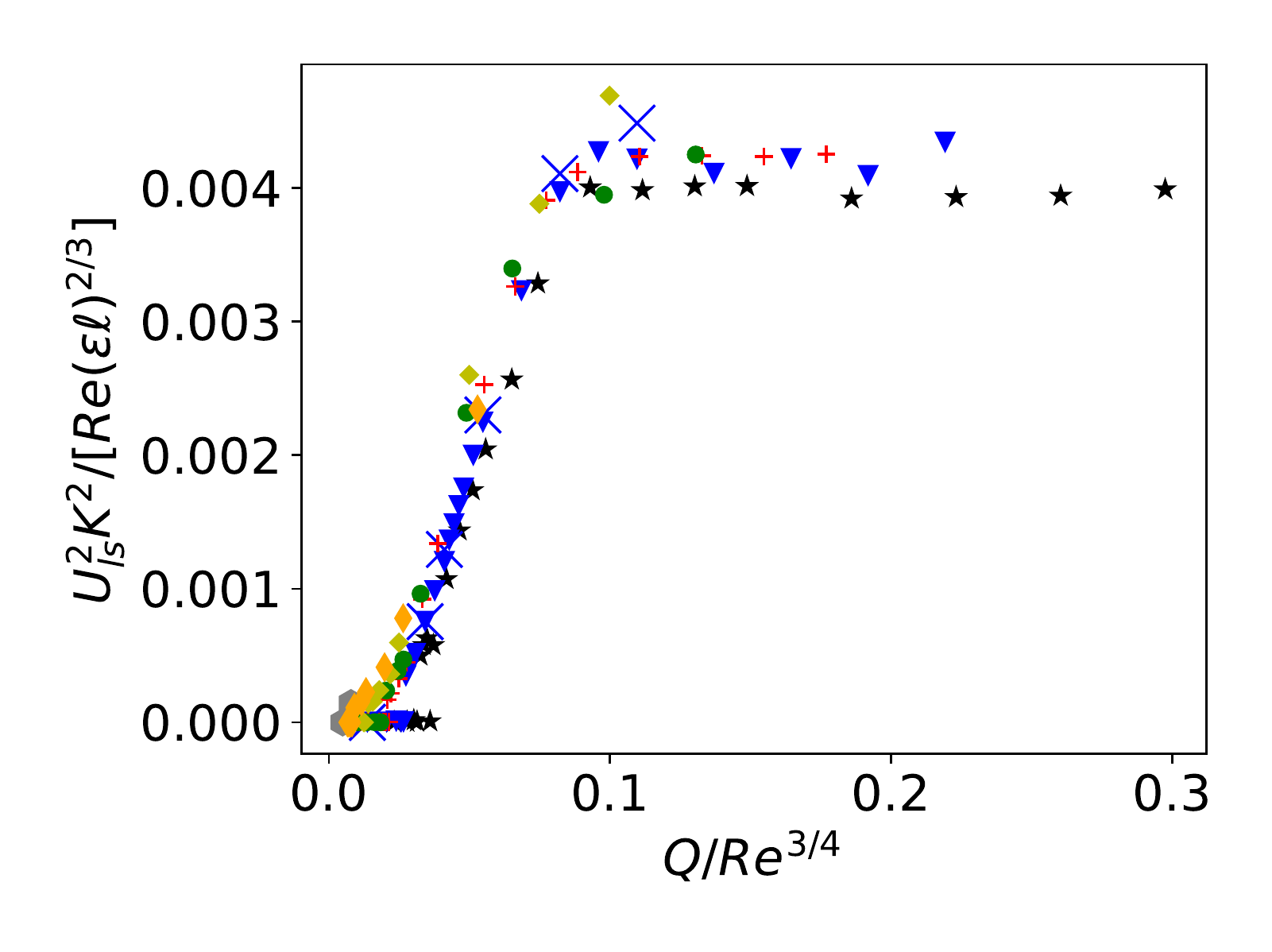} \caption{\label{fig:U2_res}}  %
\end{subfigure}                                                                         %
\caption{                                                                               %
Left panel: $U_{ls}^2$ as defined in eq. (\ref{eq:horiz_ls_energy}),                    %
nondimensionalised by $(\epsilon \ell)^{2/3}/K^2$ as a function of $Q$.                 %
Right panel: the same data (excluding hyper-viscous run),
with large-scale energy rescaled by $1/Re$ and plotted vs. $Q/\Rey^{3/4}$ showing    %
a satisfactory data collapse. The value $Q_{3D}/\Rey^{3/4}$ 
(where $\ULS$ plateaus) coincides for all $\Rey$ at $Q/\Rey^{3/4} \approx 0.09$-$0.1$.}   \label{fig:U2}                                  %
\end{figure}                                                                             %
\begin{figure}                                                                          %
\centering                                                                              %
\includegraphics[width=0.59\textwidth]{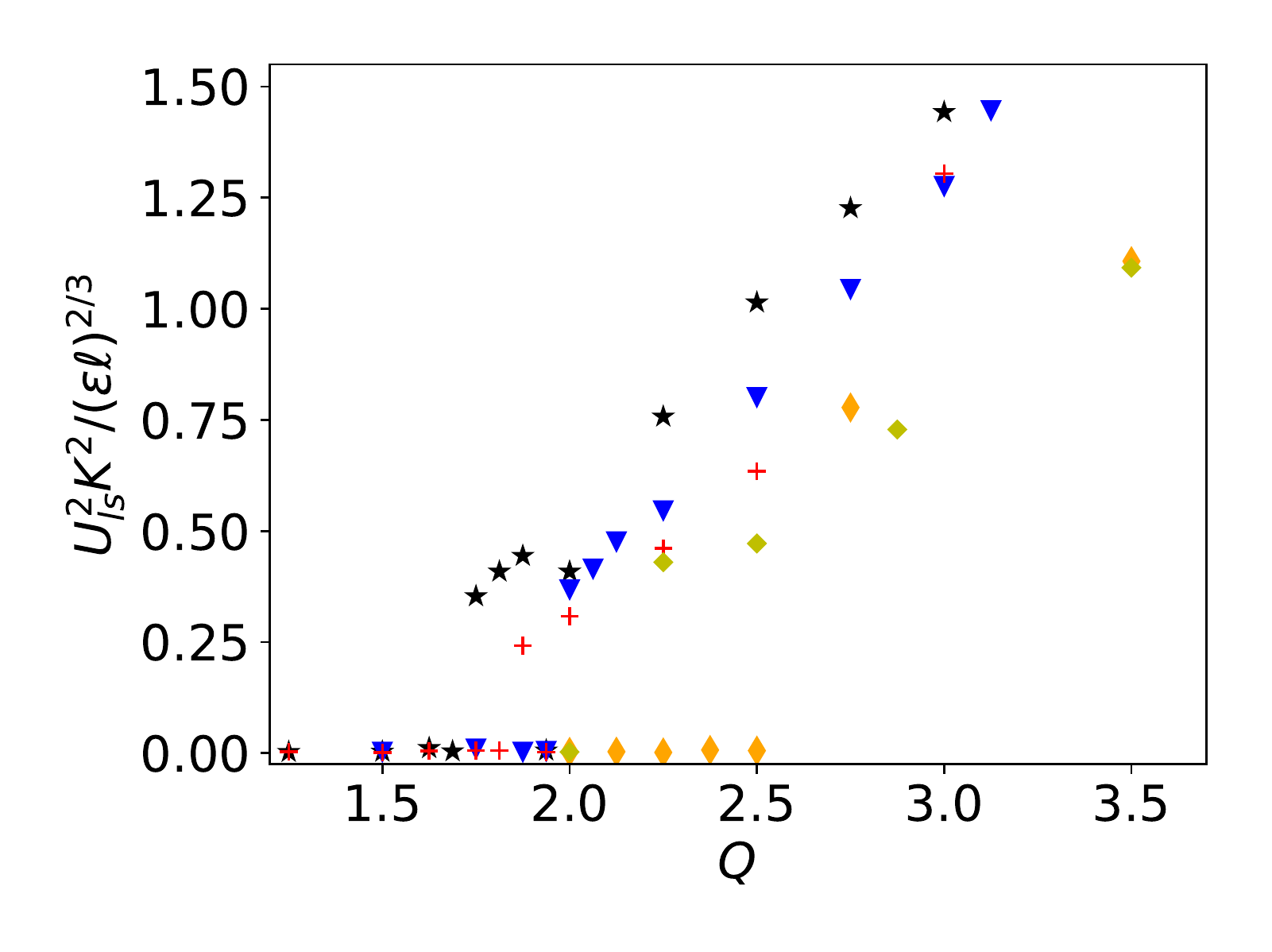}                         %
\caption{Zoomed-in version of figure \ref{fig:U2_nonres} showing that there is a     %
discontinuity in $U_{ls}^2/(\epsilon \ell)^{2/3}$ at $Q_{3D}$ for all Reynolds numbers 
up to the second highest simulated.}
\label{fig:U2_nonres_zoom}                                                              %
\end{figure}                                                                            %
Here we mostly focus on the second stage of evolution: the steady state and the properties of the condensate. Figure \ref{fig:U2} shows the equilibrium  value of $U_{ls}^2$, as defined in equation (\ref{eq:horiz_ls_energy}), non-dimensonalised by the forcing energy scale $(\epsilon \ell)^{2/3}$ and multiplied by $K^2$. In the left panel, it is plotted versus $Q$ (figure \ref{fig:U2_nonres}) and in the right panel it is rescaled by $1/\Rey$ and plotted versus $\eta/H=Q\Rey^{-3/4}$ (figure \ref{fig:U2_res}). 
First consider figure \ref{fig:U2_nonres}. At small $Q$, there is very little energy in the large scales. This corresponds to the values of $Q$ that displayed no inverse cascade at the initial stage. In the absence of an inverse cascade, the large scales only possess a small non-zero energy and are expected to be in a thermal equilibrium state \citep{kraichnan1973helical, dallas2015statistical, cameron2017effect}. 
For $Q>Q_{3D}$ the energy in the large scale takes larger values. In all cases, the energy increases nearly linearly $\ULS \propto (Q-Q_{3D})$ for $Q_{2D}>Q>Q_{3D}$. With the chosen coordinate, the close coincidence of the experiments with $K=1/8$ and $K=1/16$ at $\Rey= 305$ indicates a scaling of $U_{ls}^2 \propto L^2$. If we zoom in on $Q_{3D}$ (see the figure \ref{fig:U2_nonres_zoom}), we observe clear signs of small but discontinuous jumps of $\ULS$ at $Q_{3D}$ that are not visible in the zoomed out figure \ref{fig:U2_nonres}. These cases are examined in more detail in the next section. 

The increase of the large-scale energy stops at the second critical point $Q_{2D}$, where $U_{ls}^2$ becomes independent of $Q$. It is noteworthy that the curves for various values of $\Rey$ all follow the same straight line between their respective $Q_{3D}$ and $Q_{2D}$ with only some deviations at low $Q$. Furthermore, both $Q_{2D}$ and the plateau value of $\ULS$ depend on $\Rey$.
In figure \ref{fig:U2_res}, the same data is plotted, but with rescaled axes. The rescaling collapses the data well, with some deviations at small $Q$ related to the convergence of $Q_{3D}$. This indicates that at large values of $Q$, $\ULS$ scales like $\ULS \propto (\epsilon \ell)^{2/3} \Rey$. This is precisely the scaling of the condensate of 2D turbulence \citep{bofetta2012twodimensionalturbulence}. The critical value where the transition to this maximum value of $\ULS$ occurs is $Q_{2D}\Rey^{-3/4}=\eta/H_{2D} \approx 0.09-0.1$. 

\begin{figure}                                                                       %
\centering                                                                           %
\begin{subfigure}[b]{0.485\linewidth}                                                %
 \includegraphics[width=\textwidth]{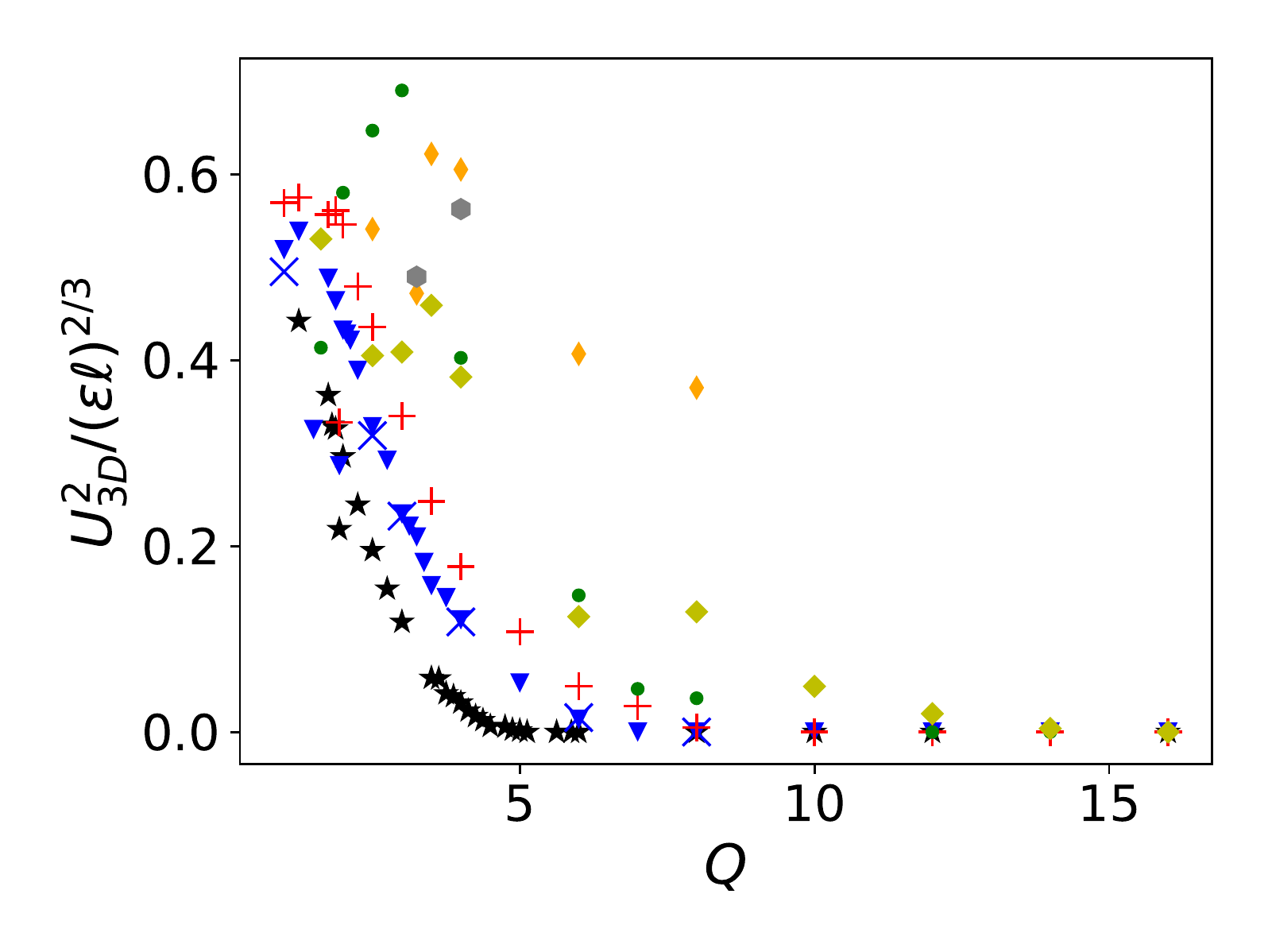}  \caption{\label{fig:U2_3D}}        %
\end{subfigure}                                                                      %
\begin{subfigure}[b]{0.485\linewidth}                                                %
\includegraphics[width=\textwidth]{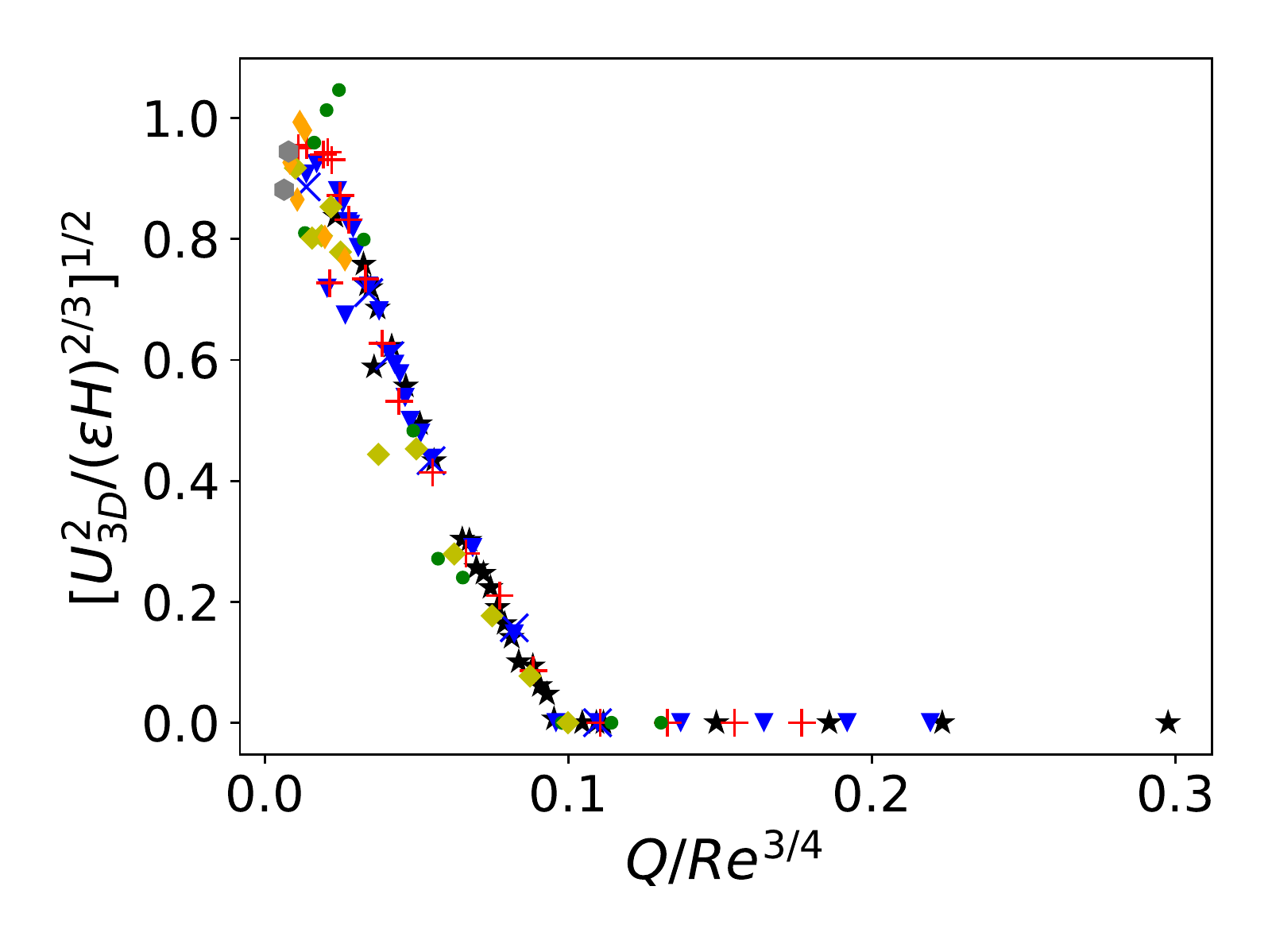}\caption{\label{fig:U3_res}}            %
\end{subfigure} \hspace{0.08cm}                                                      %
\caption{Left panel:  $U_{3D}^2$ as defined in equation          %
(\ref{eq:3D_energy}), non-dimensionalised by $(\epsilon \ell)^{2/3}$ and plotted     %
versus $Q$. 
Right panel: the same information as figure (\ref{fig:U2_3D}), %
but in terms of the square-root of the 3D kinetic energy rescaled by                 %
$(\epsilon H)^{2/3}$, plotted versus $Q/\Rey^{3/4}$. This rescaling indicates 
that $U_{3D}^2 \propto (Q_{2D}-Q)^2$ close to the transition.}
\label{fig:U3D}                                                                      %
\end{figure}

\begin{figure}                                                                       %
\begin{subfigure}[b]{0.485\textwidth}                                                %
\includegraphics[width = \textwidth]{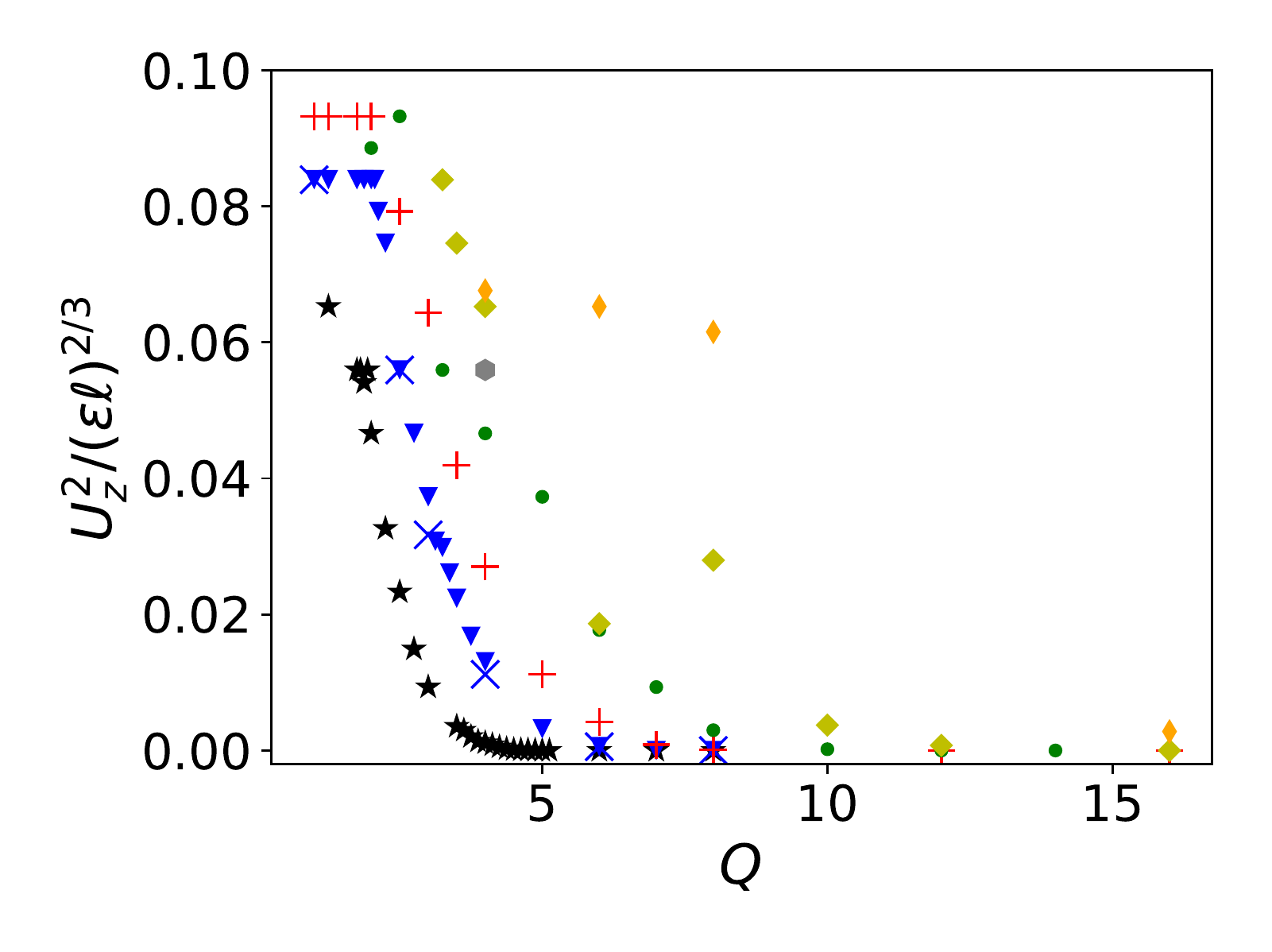} \caption{\label{fig:Uv}}                
\end{subfigure}                                                                      %
\begin{subfigure}[b]{0.485\textwidth}                                                %
\includegraphics[width = \textwidth]{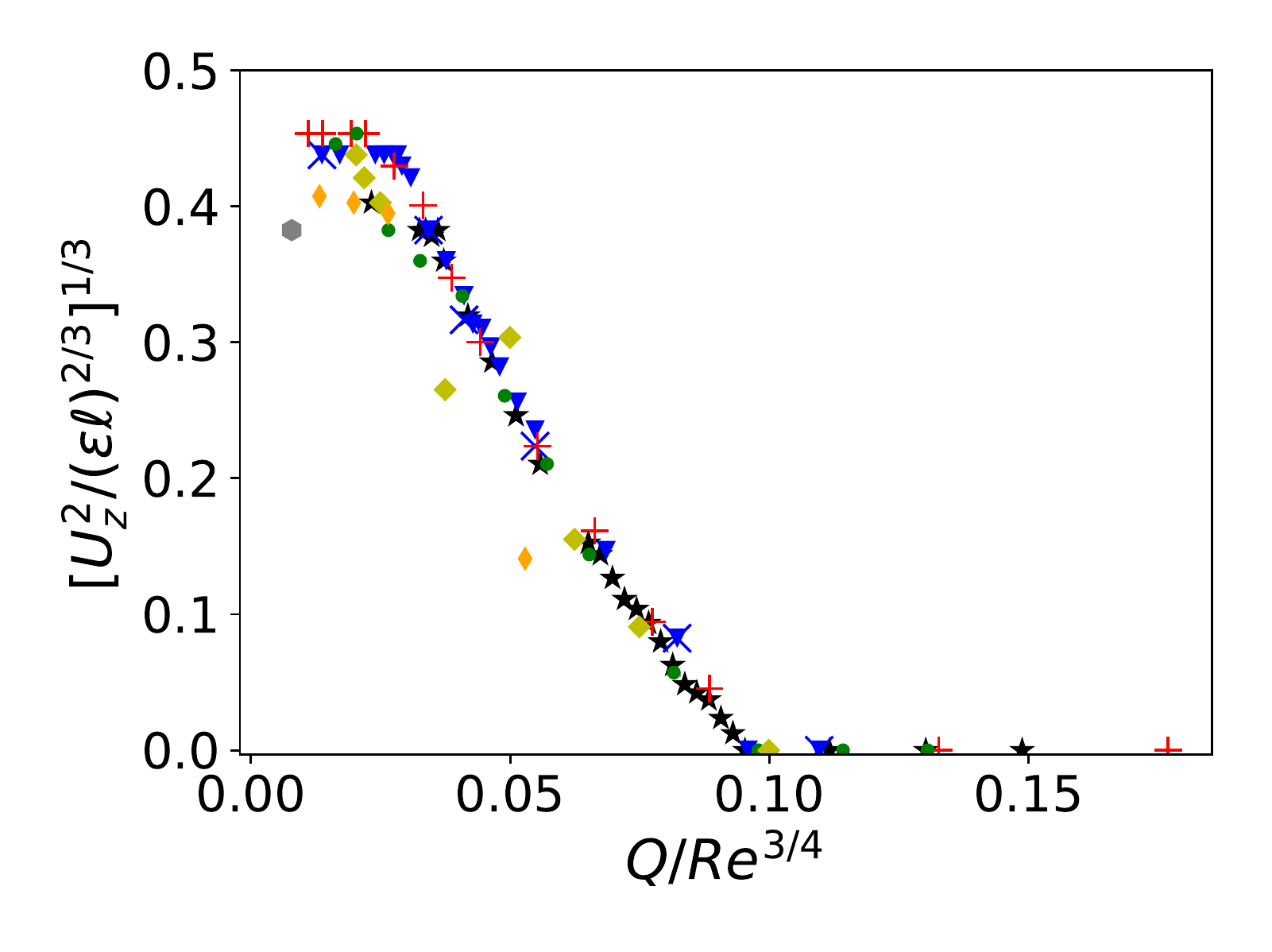}  \caption{\label{fig:Uv_res}}      %
\end{subfigure}                                                                      %
\caption{Left panel: $U_z^2$ as defined in equation      %
(\ref{eq:vert_energy}) as a function of $Q$. 
Right panel: The various curves collapse     %
when the abscissa  $\eta/H$ and the coordinate to be                  %
$U_{3D}^2/(\epsilon H)^{2/3}$. Raising the coordinate to the $1/3$ power, the curve 
becomes linear close to onset. This indicates that close to onset, $U_v^2$
scales as $U_z^2 \approx (Q_c-Q)^{3}$,     %
where $Q_c\approx 0.09-0.1\approx Q_{2D}$. We note that the scaling exponent is 
different from that found for $U_{3D}^2$.}                                            %
\label{fig:Uv_all}                                                                   %
\end{figure}

The scaling allowing to collapse the data in figure \ref{fig:slopes} (transient stage) is different from that in figures \ref{fig:U2},\ref{fig:U3D} and \ref{fig:Uv_all} (condensate state).
This implies that $Q_{2D}\propto \Rey^{1/2}$ estimated during the early stage of the inverse cascade development is different from $Q_{2D}\propto\Rey^{3/4}$  estimated at steady state
where a condensate is fully developed. The reason for this difference is that the transition to exactly 2D motion occurs when the maximum shear in the flow (which produces 3D motion by shear instabilities) is balanced by small-scale viscous dissipation. In the presence of the inverse cascade, an $E(k)\propto \epsilon^{2/3}k^{-5/3}$ spectrum is formed at $k>k_f$, such that the peak of the enstrophy spectrum $k^2E(k)$ is at the forcing scale. Thus the balance between 2D shear and 3D damping is  
$$(\epsilon \ell)^{1/3}/\ell \sim \nu/H^2,$$ 
implying $H_{2D} \sim \ell \Rey^{-1/2}$ \citep{benavides_alexakis_2017}.
In the condensate, however, most of the energy and enstrophy are located in the largest scales and are such that energy injection $\epsilon$ is balanced by large-scale dissipation $\propto \nu \ULS/L^2$. The large-scale shear is thus $U_{ls}/L\propto (\epsilon/\nu)^{1/2}$ which is balanced by the damping rate of 3D perturbations at onset,
$$ (\epsilon/\nu)^{1/2} \sim \nu/H^2,$$
giving the scaling $H_{2D} \propto \epsilon^{1/4} \nu^{3/4} \propto \ell \Rey^{-3/4} $.
We will recover the very same steady state scaling in the section \ref{sec:3mode} from a low-order model. These two scalings imply the interesting possibility that a flow which becomes exactly 2D at the early stages of the inverse cascade for $Q \gtrsim \Rey^{1/2}$ may develop 3D instabilities at the condensate state if $Q \lesssim \Rey^{3/4}$.

Figure \ref{fig:U3D} shows $U_{3D}^2$ as defined in equation (\ref{eq:3D_energy}).
In the left panel it is non-dimensionalised by the forcing energy $(\epsilon \ell)^{2/3}$ and plotted vs. $Q$ (figure \ref{fig:U2_3D}), while in the right panel, it is non-dimensionalised by $(\epsilon H)^{2/3}$, raised to the power $\frac{1}{2}$ and plotted versus $\eta /H=Q/\Rey^{-3/4}$. 
Figure \ref{fig:U2_3D} shows that beyond some non-monotonic behaviour at small $Q$, $U_{3D}^2$ decreases monotonically with $Q$ until it reaches zero at $Q_{2D}$ and remains zero beyond this point. The 3D energy increases with $\Rey$ at a given $Q$. Under the rescaling in figure \ref{fig:U3_res}, the various curves collapse nicely. In particular, the point where $U_{3D}^2$ vanishes is sharp and identical for all $\Rey$, namely $\eta/H\approx 0.1$. Comparing with figure \ref{fig:U2_res}, one sees that this point and $Q_{2D}$ coincide within the range of uncertainties. This means that beyond $Q_{2D}$, not only is $U_{ls}^2$ independent of $Q$, but also $U_{3D}^2$ vanishes. This confirms that $Q_{2D}$ corresponds to the point where the motion becomes invariant along $z$. 

Finally, figure \ref{fig:Uv_all} shows the vertical kinetic energy, non-dimensionalised by $(\epsilon \ell)^{2/3}$, once plotted versus $Q$ and once taken to the $1/3$ power and plotted versus $Q/\Rey^{3/4}$. The general features of figure \ref{fig:Uv} are similar to figure \ref{fig:U2_3D}: like 3D energy, vertical kinetic energy decreases with $Q$ until it reaches zero and it increases with $\Rey$. The curves collapse in figure \ref{fig:Uv_res} and the behaviour close to $Q_{2D}$ becomes linear if the coordinate is raised to the $1/3$ power, indicating an approximate scaling $U_z^2 \approx (Q_c-Q)^{3}$ with $Q_c\approx Q_{2D}$. This indicates that the point beyond which the vertical kinetic energy vanishes is close to $Q_{2D}$, implying that beyond $Q_{3D}$, the motion is not only invariant along $z$ but also restricted to the x-y plane. Hence, for $Q>Q_{2D}$, the flow has two-dimensionalised exactly. 
\section{Behaviour close to the transitions: hysteresis and intermittency}
\label{sec:close_to_transitions}
In this section, we discuss the behaviour close to the two transition points $Q_{2D}$ and $Q_{3D}$. Each transition shows a different non-trivial behaviour. Close to $Q_{3D}$, we observe discontinuous transitions and hysteresis for some range of parameters, while close to $Q_{2D}$, we find both spatial and temporal intermittency with localised bursts of 3D energy.
\subsection{Close to $Q_{3D}$: Discontinuity and Hysteresis}
We begin by discussing the behaviour of the flow for $Q$ close to $Q_{3D}$ where a sharp increase of the large-scale energy was observed.
This sharp increase could indicate the presence of a discontinuity that could further imply the presence of hysteresis. 

\begin{figure}                                                                                                 
\centering                                                                                                     
\includegraphics[width=7cm]{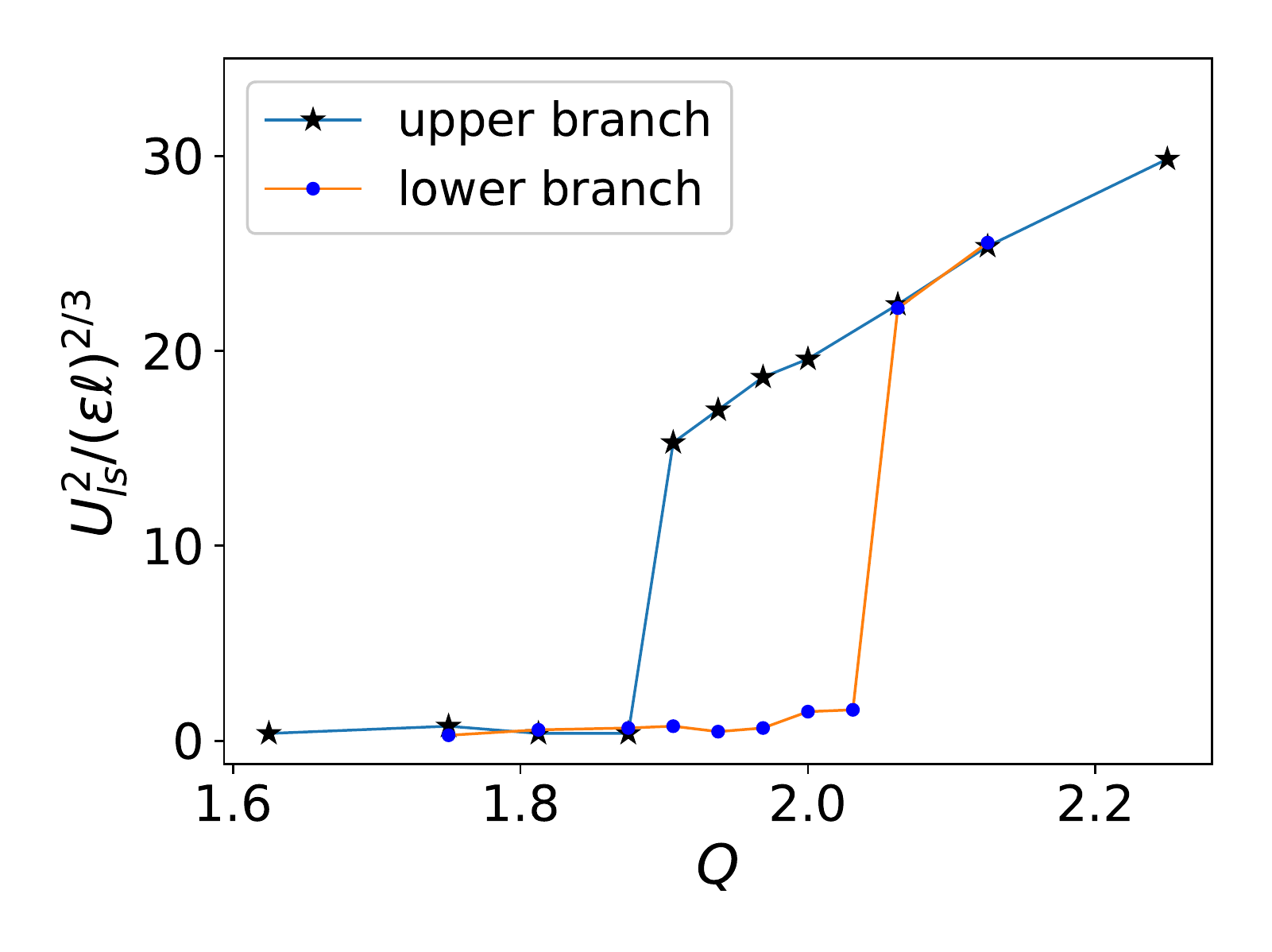}                                                                  
\caption{Hysteresis curve of $\ULS$ non-dimensionalised by the forcing energy. Two experiments are shown, the 'lower branch' starting from small $Q$ (deep layer) and increasing $Q$ and the 'upper branch' starting from large $Q$ (thin layer) and decreasing $Q$. }                                                                                  
\label{fig:hyst}                                                                                               
\end{figure}

\begin{figure}                                                                                                 
\begin{subfigure}[b]{0.48\linewidth}                                                                           
\includegraphics[width=6cm]{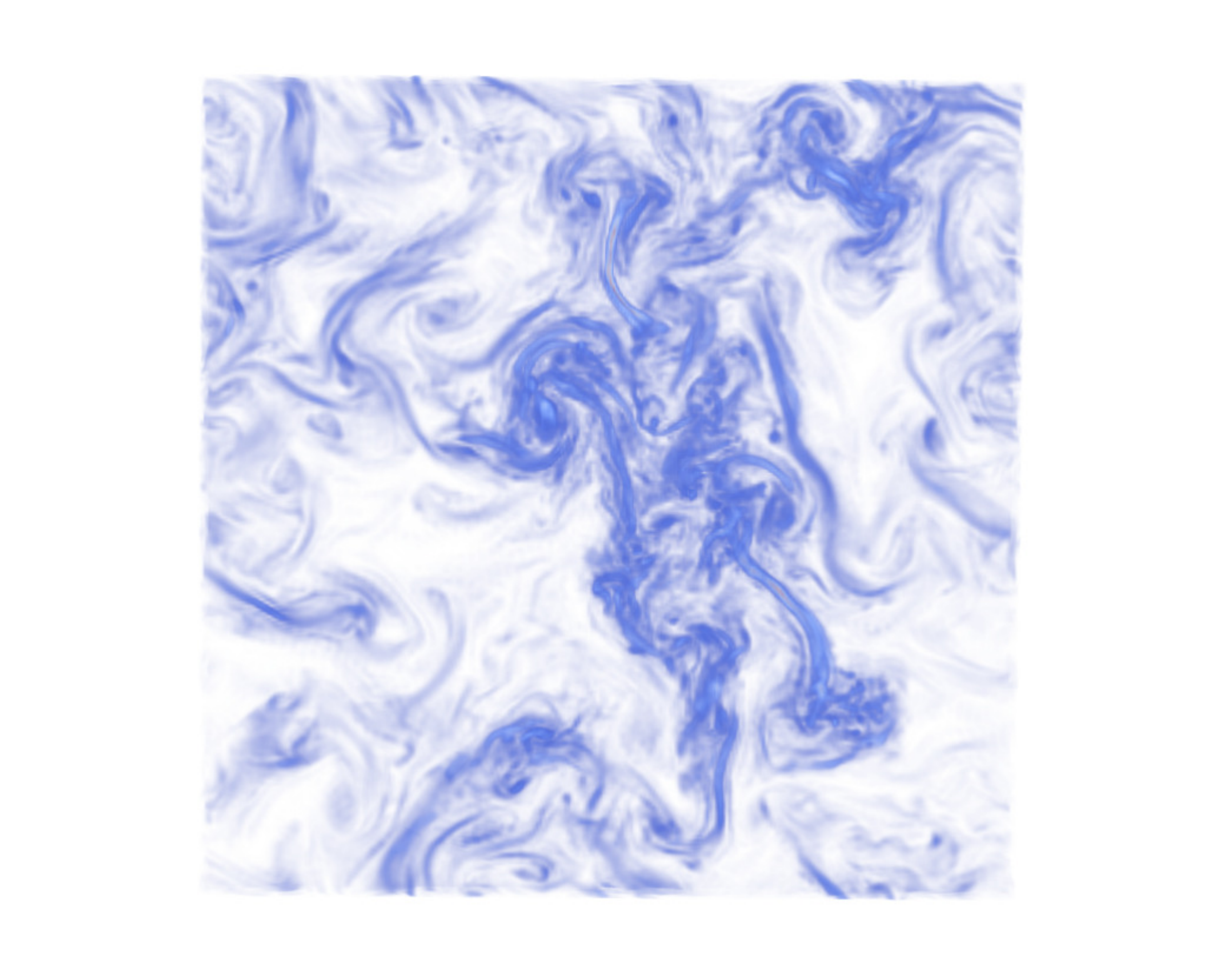} \caption{\label{fig:vis_3D_hyst}}                                      
\end{subfigure}                                                                                                
\begin{subfigure}[b]{0.48\linewidth}                                                                           
\includegraphics[width=6cm]{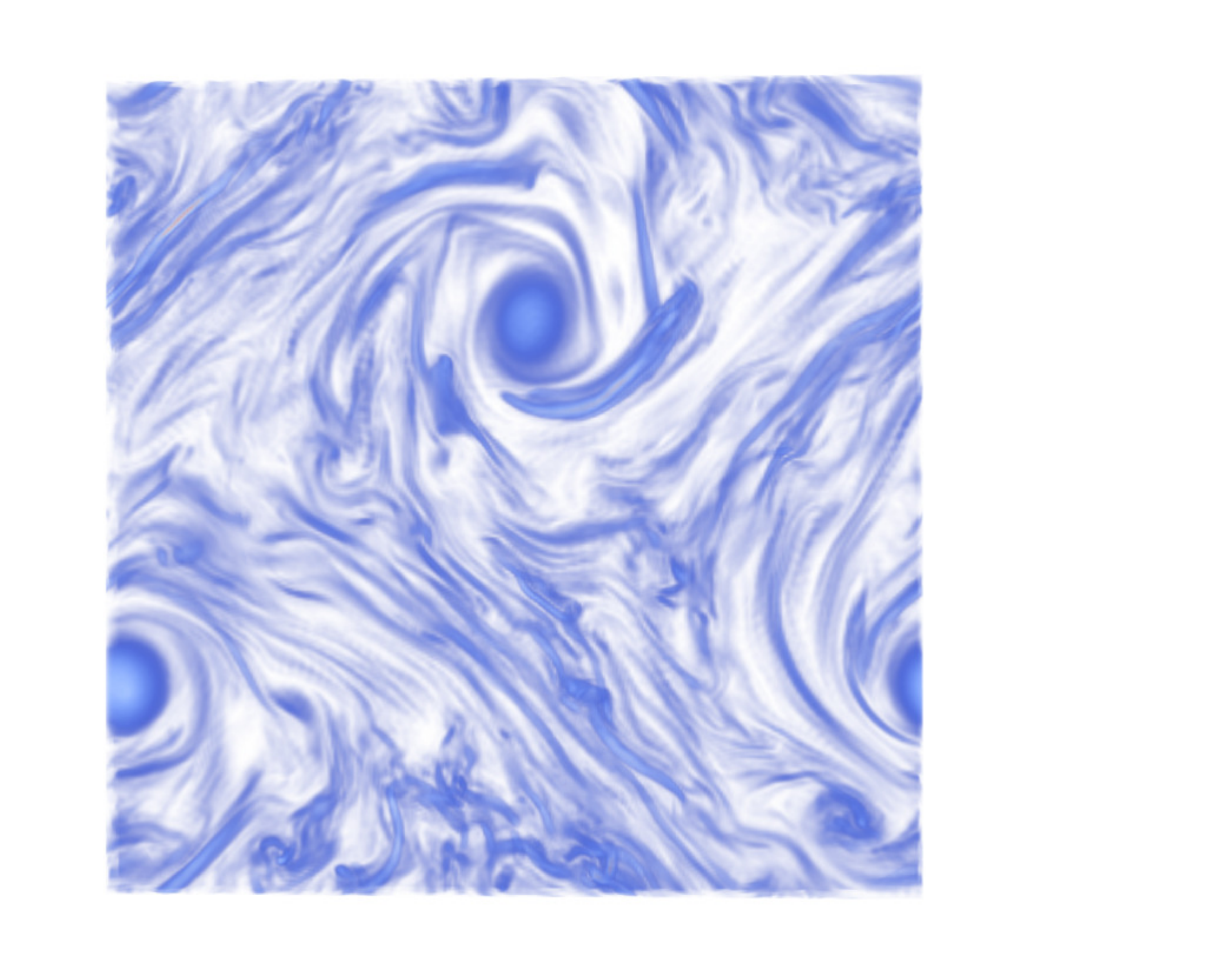}\caption{\label{fig:vis_2D_hyst}}                                   
\end{subfigure}                                                                                                
\caption{Figures \ref{fig:vis_3D_hyst} and \ref{fig:vis_2D_hyst} visualise the typical flow field after long simulation time at $Q\approx1.97$ in the hysteresis experiment on the upper (\ref{fig:vis_2D_hyst}) and lower (\ref{fig:vis_3D_hyst}) branches. The lower branch flow field shows small scale structures and no large-scale organisation, reminiscent of 3D turbulence. By contrast, the upper branch flow field is characterised by two large-scale vortices in addition to smaller-scale structures in between them.  }
\label{fig:hyst2}                                                                                              
\end{figure}

To verify the presence of a discontinuity we need perform many different runs varying $Q$ is small steps as well as veryfing sensitivity to initial conditions.
To do this, a hysteresis experiment has been performed at $\Rey = 406$, consisting of two series of runs, that we refer to as the `{\it upper branch}' and 
the `{\it lower branch}', see figure \ref{fig:hyst}. On the upper branch, we start with random initial conditions and $Q\approx 2.25$ for which the system reaches a condensate equilibrium with an associated non-zero value of large-scale energy. Once the run has equilibrated, we use that equilibrium state to initialise a run at $Q\to Q-\Delta Q$ with $\Delta Q=0.1$. By decreasing $Q$, the physical height of the box is increased. To be able to use the equilibrium state reached at one $Q$ as initial condition for a neighbouring $Q$, the $z$-dependence of the velocity field is scaled and the velocity field is projected onto its diverge-free part, formally $v(x,y,z) \to \mathbb{P} v(x,y,\lambda z)$, where $\mathbb{P}= \mathbb{I} - \nabla^{-2} \nabla (\nabla\cdot\phantom{i}) $. Having changed $Q$ and applied this procedure, we let the system equilibrate to a new condensate state. This is repeated five more times (step size reduced to $\Delta Q= 0.05$ and then $\Delta Q =0.025$) down to $Q\approx 1.9$. When $Q$ is now lowered $0.025$ further, the condensate decays into 3D turbulence and the large-scale energy saturates to close to zero. Reducing $Q$ even more, $\ULS$ remains small, indicating a 3D turbulent state. The lower branch was calculated similarly, with the only difference that the experiment started at low $Q$ and $Q$ was increased in steps of $0.05$ and then of $0.025$. For small $Q$, the two branches coincide, while the lower branch remains at low $\ULS$ (3D turbulence) up to $Q\approx 2.025$. For $Q$ larger than $Q=2.025$, the lower branch merges with the upper branch, closing the hysteresis loop and a condensate is spontaneously formed from 3D turbulence. In other words, for $\Rey = 406$ in the range $1.9 \le  Q \le 2.025 $, there are multiple steady states and to which state the system will saturate depends on the initial conditions. The flow field for two such states starting from different initial conditions for $Q\approx1.97$ is visualized in \ref{fig:hyst2}. \\

The following remarks are in order: although for each $Q$ we ran the simulations until saturation was achieved, since we are dealing with a noisy system, rare transitions can exist between the two branches of the hysteresis loop. To test this, we picked the point $Q\approx 1.97$ on both lower and upper branches and ran them for a long time (thousands of eddy turnover times $\tau = (L^3/\epsilon)^{1/2}$). In neither case did we see a transition between the two branches, indicating that such transitions are rare (if not absent) in the middle of the hysteresis loop. Near the edges of the hysteresis loop at $Q\approx 2.05$ and $Q\approx1.9$, the dependence on simulation time is likely to be stronger, but this has not been investigated. 

Furthermore, we note that the bifurcation diagram of figure \ref{fig:hyst} corresponds to a relatively low Reynolds number $\Rey = 406$. 
Whether this subcritical behaviour persists at larger $\Rey$ and/or larger box sizes (smaller $K$) is still an open question.
Figure \ref{fig:U2_nonres_zoom} suggests that a discontinuity continues to exist at $Q=Q_{3D}$ up to high Reynolds numbers ($\Rey = 2031$ shown there). In addition, we found more points at higher $\Rey$ that showed a dependence on initial conditions but without having enough values of $Q$ to create a hysteresis diagram.
These findings suggest that subcritical behaviour and hysteresis might survive even at high $\Rey$. 
However, due to the high computational cost at higher resolution and the long duration of the runs required to verify that the system stays in a particular state, we could not investigate this possibility in detail. Further simulations at larger $\Rey$ and possibly smaller $K$ (larger boxes) are required to resolve this issue. Similar hysteretic behaviour has recently been reported in rotating turbulence, see \citep{yokoyama2017hysteretic}. More generally, multistability is observed in many turbulent flows, see \citep{weeks1997transitions,Ravelet2004_multistability} as examples.

\begin{figure}                                                                                                  %
\begin{subfigure}[b]{0.48\linewidth}                                                                            %
\includegraphics[width=6cm]{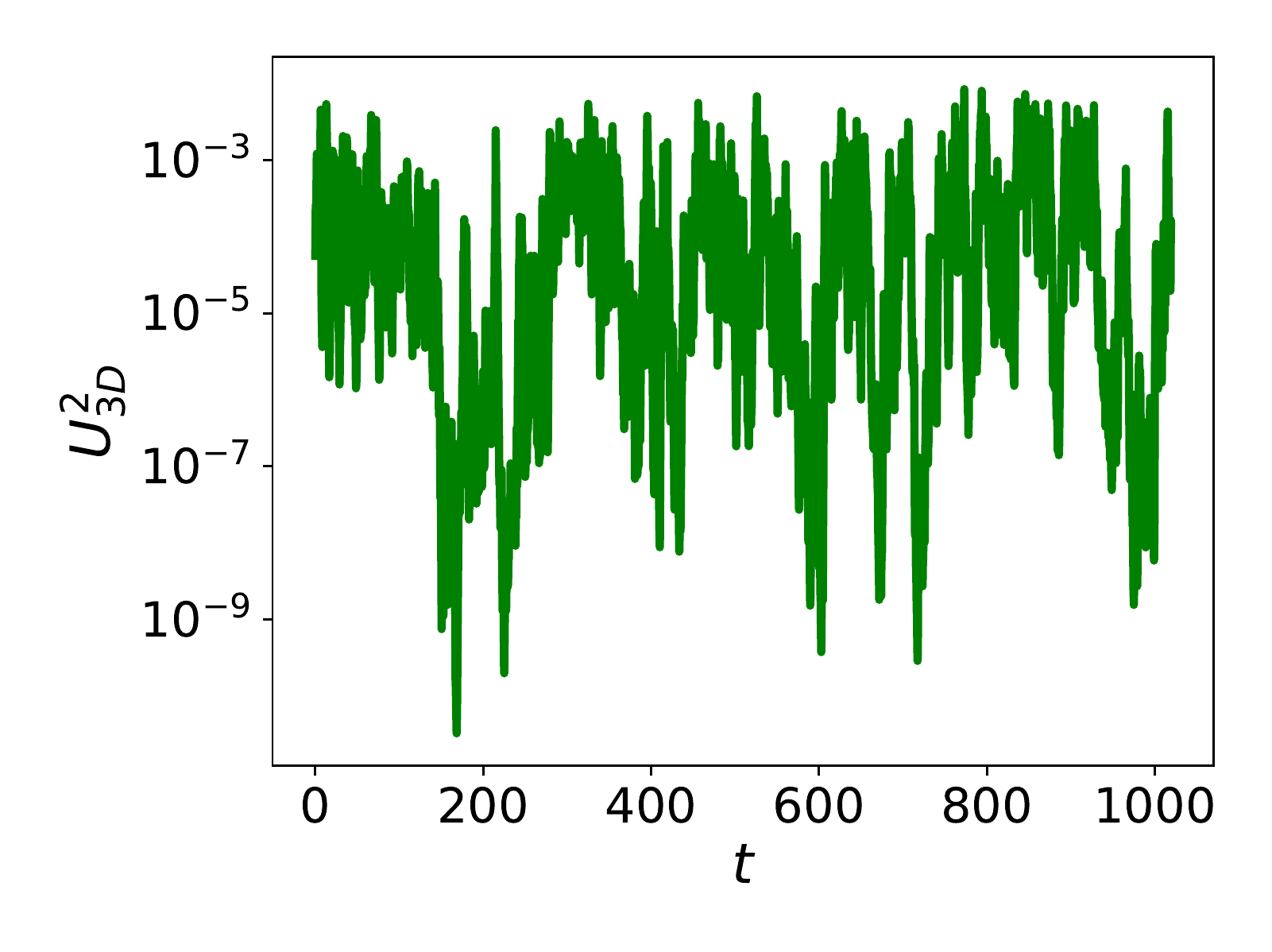} \caption{\label{fig:ts_int}}                              %
\end{subfigure}                                                                                                 %
\begin{subfigure}[b]{0.48\linewidth}                                                                            %
\includegraphics[width=6cm]{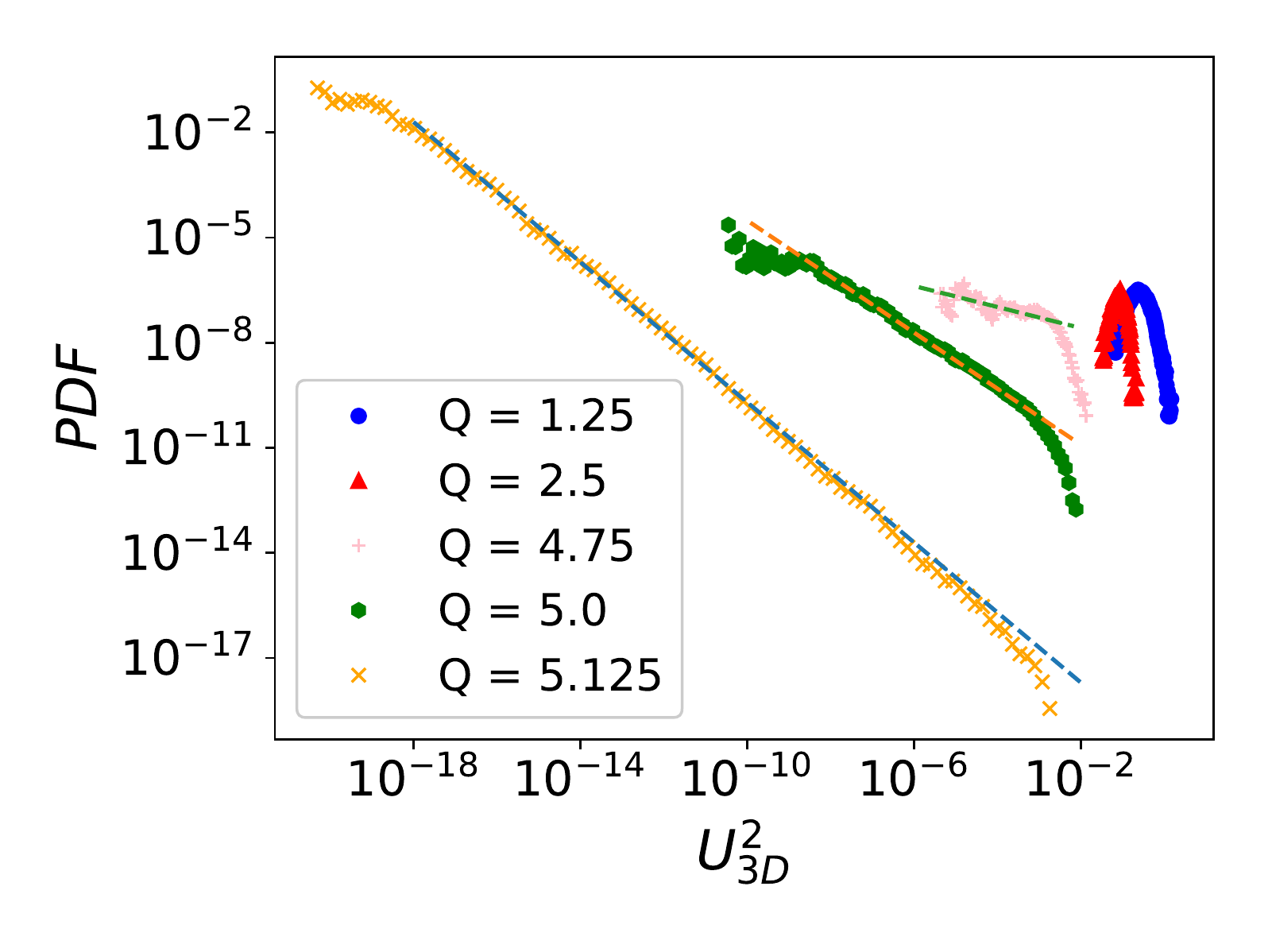}\caption{\label{fig:int_spectra}}                            %
\end{subfigure}                                                                                                 %
\caption{Plots showing temporal intermittency at $\Rey=203$. Figure \ref{fig:ts_int} shows a typical time       %
series (on lin-log axes) of 3D energy close to $Q_{2D}$. Specifically, $Q =5$, while $Q_{2D}\approx 5.13$%
at this value of $\Rey$. Figure \ref{fig:int_spectra} shows PDFs corresponding to this time series as          %
well as for different values of $Q$ (PDFs shifted by a constant factor for better visibility). The ifferent symbols mark different values of $Q$, while the dotted lines correspond to power laws with exponents $-1$ (bottom), $-0.8$ (middle) and $-0.3$ (top) respectively. }            %
\label{fig:temp_int}                                                                                            %
\end{figure}                                                                                                    %

\subsection{Close to $Q_{2D}$: Intermittent bursts}

Next, we discuss the behaviour of the flow close to the second critical point $Q_{2D}$. A typical time series of 3D energy for $Q\lesssim Q_{2D}$ is shown in figure \ref{fig:ts_int}. One observes bursty behaviour and variations over many orders of magnitude, indicating on-off intermittency \citep{fujisaka1985new,platt1993_onoffintermittency}. On-off intermittency refers to the situation where a marginally stable attractor loses or gains stability due to noise fluctuations. When instability is present, a temporary burst is produced before the system returns to the attractor. On-off intermittency predicts that the unstable mode $X$ follows
a power-law distribution $P(X) \propto X^{\delta-1}$ for $X\ll1$ where  $\delta$ measures the deviation from onset (here $\delta\propto (Q_{2D}-Q)/Q_{2D}$) and all moments scale linearly with the deviation  $\langle X^n\rangle \propto \delta$.
\begin{figure}                                                                                            
\centering                                                                                                
\begin{subfigure}[b]{0.28\linewidth}        
\includegraphics[width =1.01\textwidth]{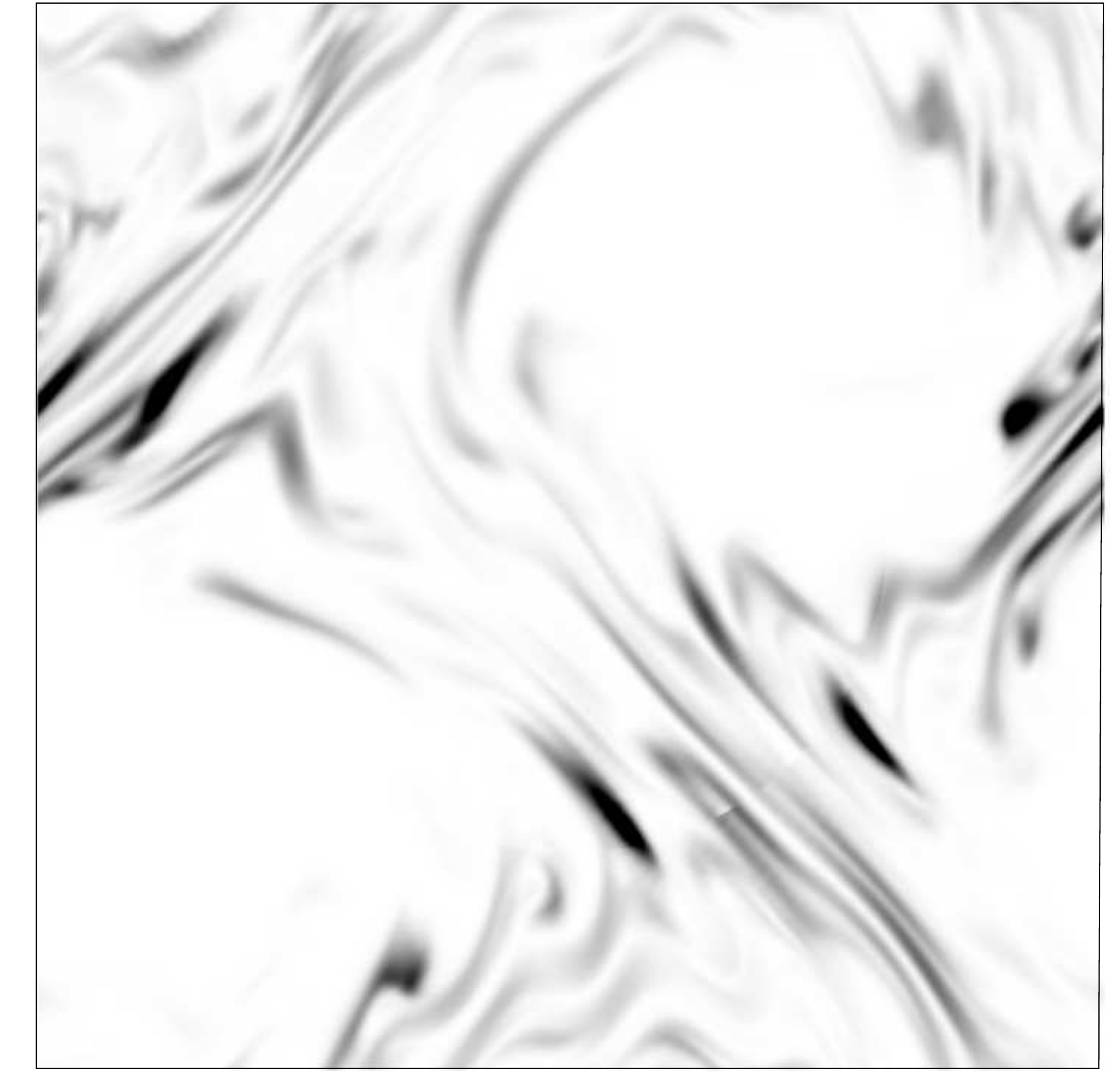} \caption{\label{fig:no_sp_int_U2_3D}} 
 \end{subfigure}    
 \hspace{0.14cm}
\begin{subfigure}[b]{0.28\linewidth}      
\includegraphics[width=1.02\textwidth]{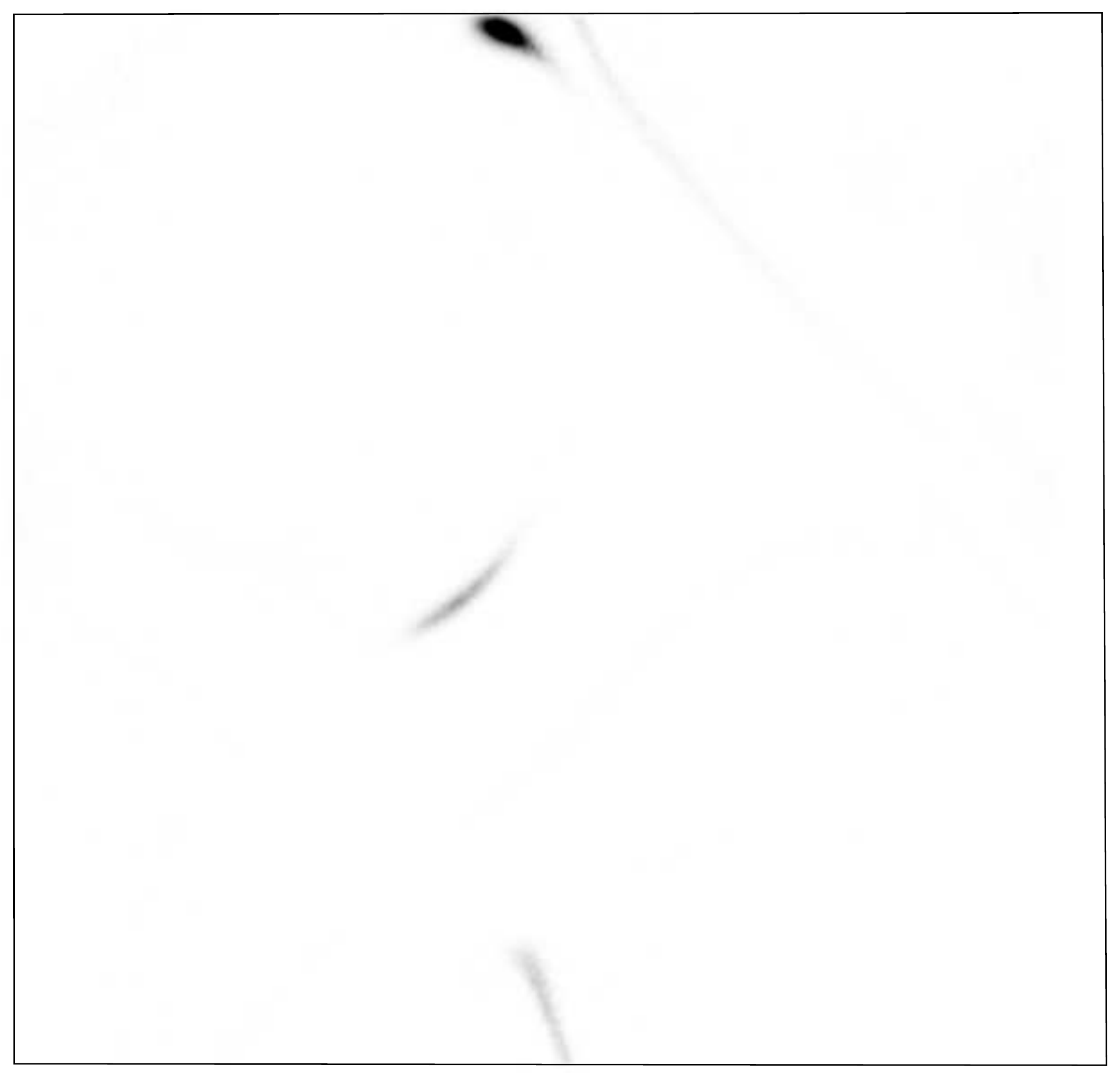} \caption{\label{fig:some_sp_int_U2_3D}}  
\end{subfigure}  
\hspace{0.14cm}
\begin{subfigure}[b]{0.28\linewidth}                                                                      
\includegraphics[width=0.98\textwidth]{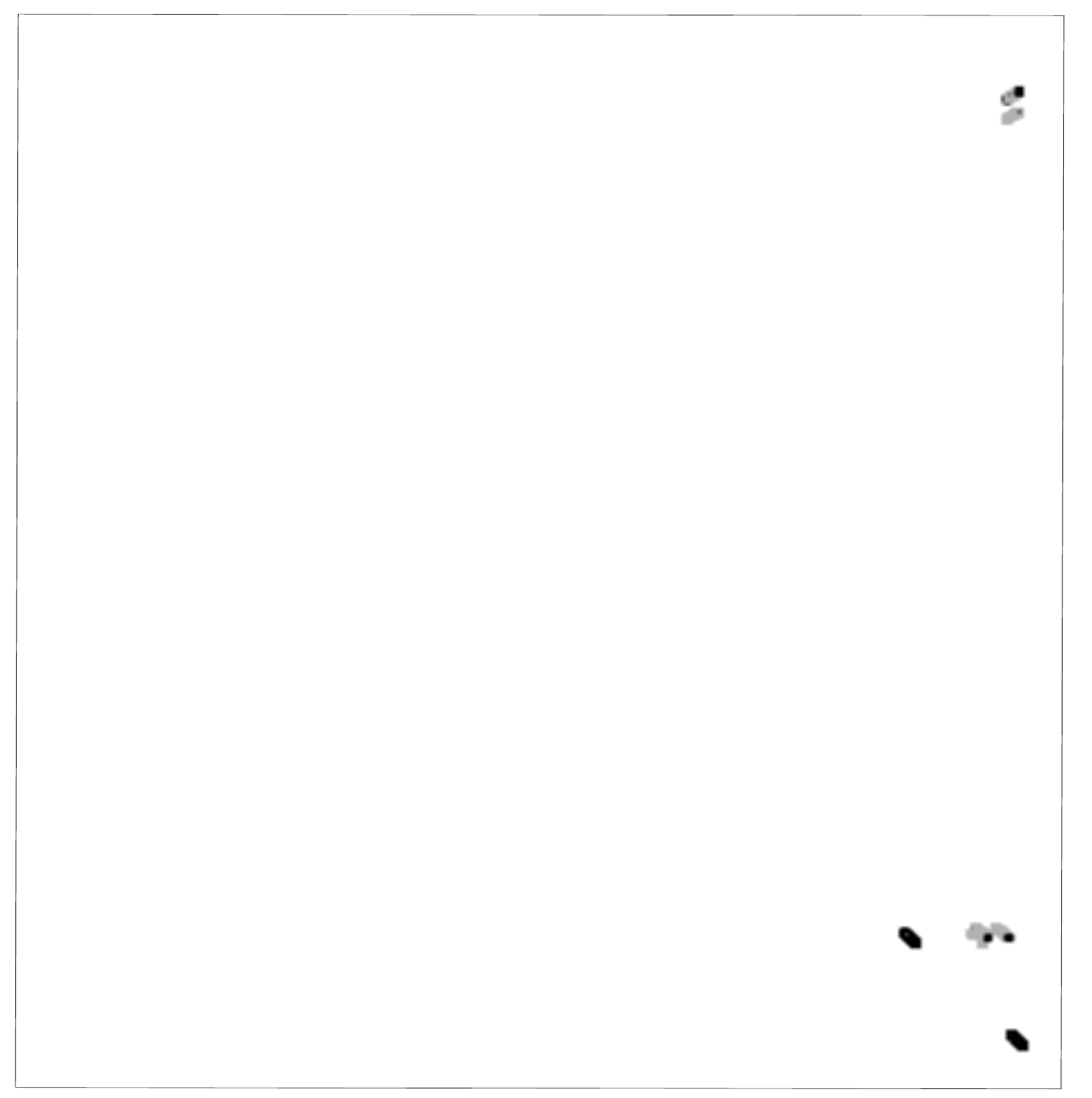} \caption{\label{fig:sp_int_U2_3D}}  
\end{subfigure}  
\begin{subfigure}[b]{0.1\linewidth}  
\includegraphics[height=4.25cm, width=1.3\textwidth]{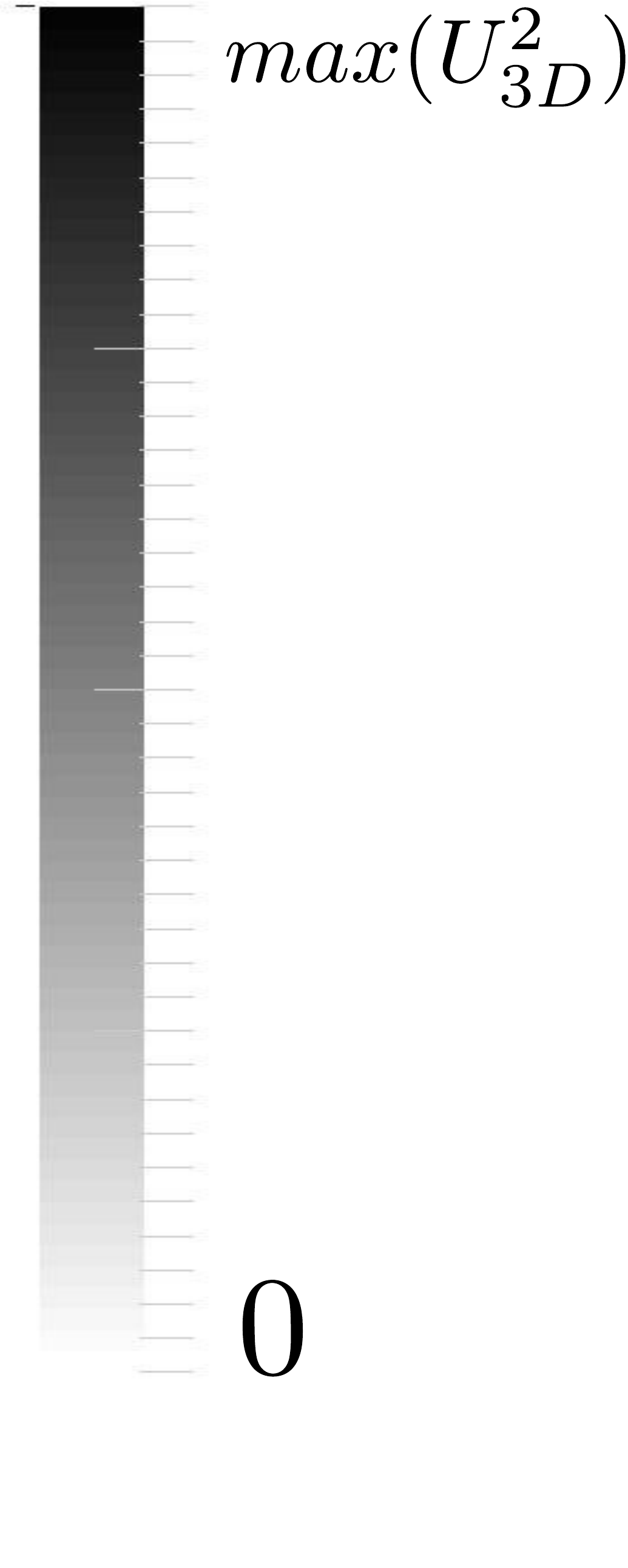}
\end{subfigure}
\caption{Snapshots of $\mathbf{u}_{3D}^2$ for $\Rey = 203$ and $ Q=2.5$ (figure \ref{fig:no_sp_int_U2_3D}), $Q=5.0$ (figure \ref{fig:some_sp_int_U2_3D}) and $Q=5.125$ (figure \ref{fig:sp_int_U2_3D}), 
(corresponding to figure \ref{fig:temp_int}). The colorbar is chosen in each plot such that
the maximum value of $\mathbf{u}_{3D}^2$ is shown in black. As $Q$ increases towards $Q_{2D}\approx 5.13$, $U_{3D}^2$ becomes more and more localised. In figure \ref{fig:sp_int_U2_3D}, $\mathbf{u}_{3D}^2$ is concentrated in small columnar structures (upper and lower right-hand corner) absent in figure \ref{fig:sp_int}.}
\label{fig:int_vis_U3D}                                                                                   
\end{figure}                                                                                          

\begin{figure}                                                                                            
\centering                                                                                                
\begin{subfigure}[b]{0.28\linewidth}                                                                      
\includegraphics[width =1\textwidth]{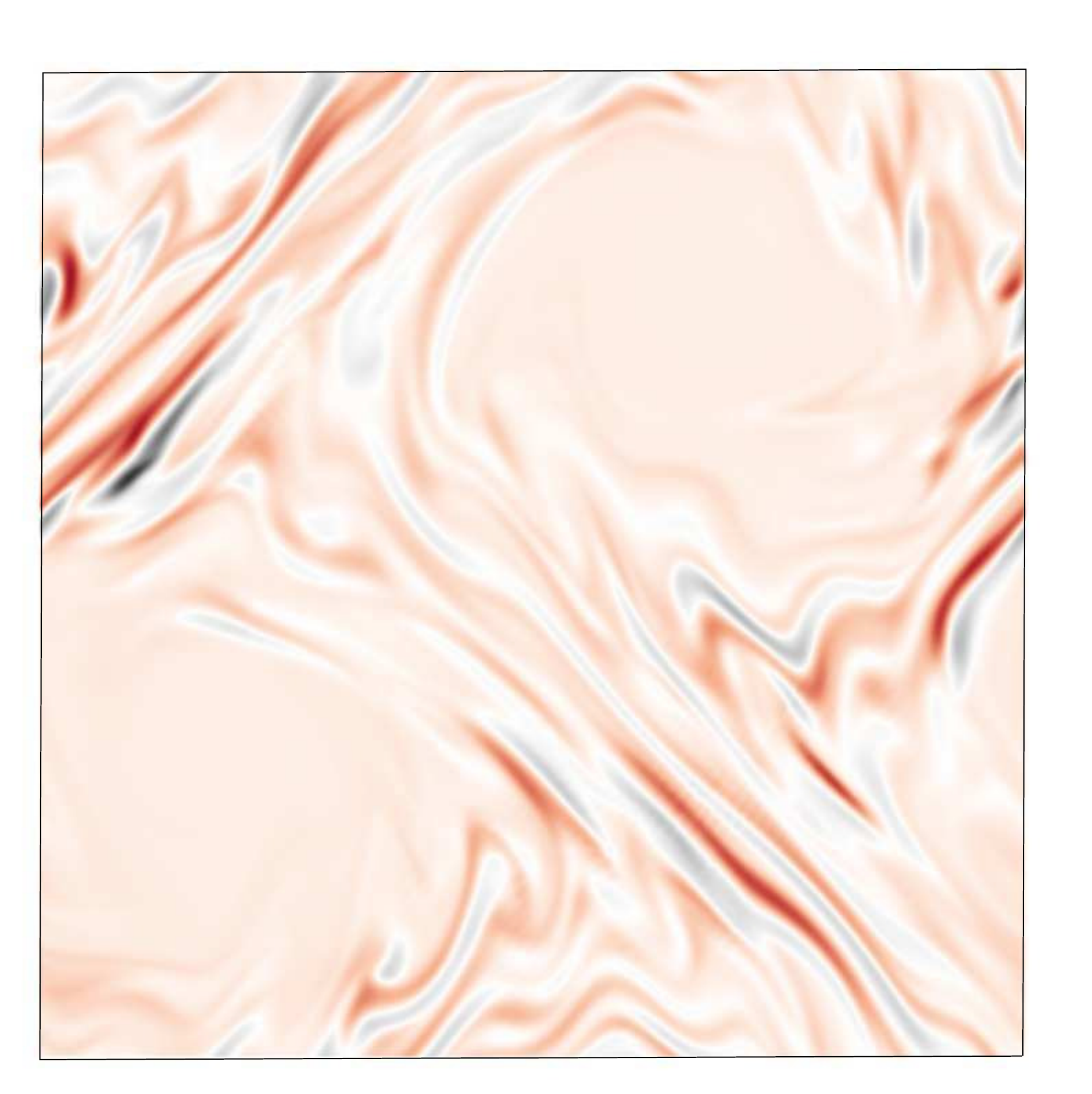} \caption{\label{fig:no_sp_int}}         
\end{subfigure}    
\begin{subfigure}[b]{0.28\linewidth}                                                                     
\includegraphics[width=0.94\textwidth]{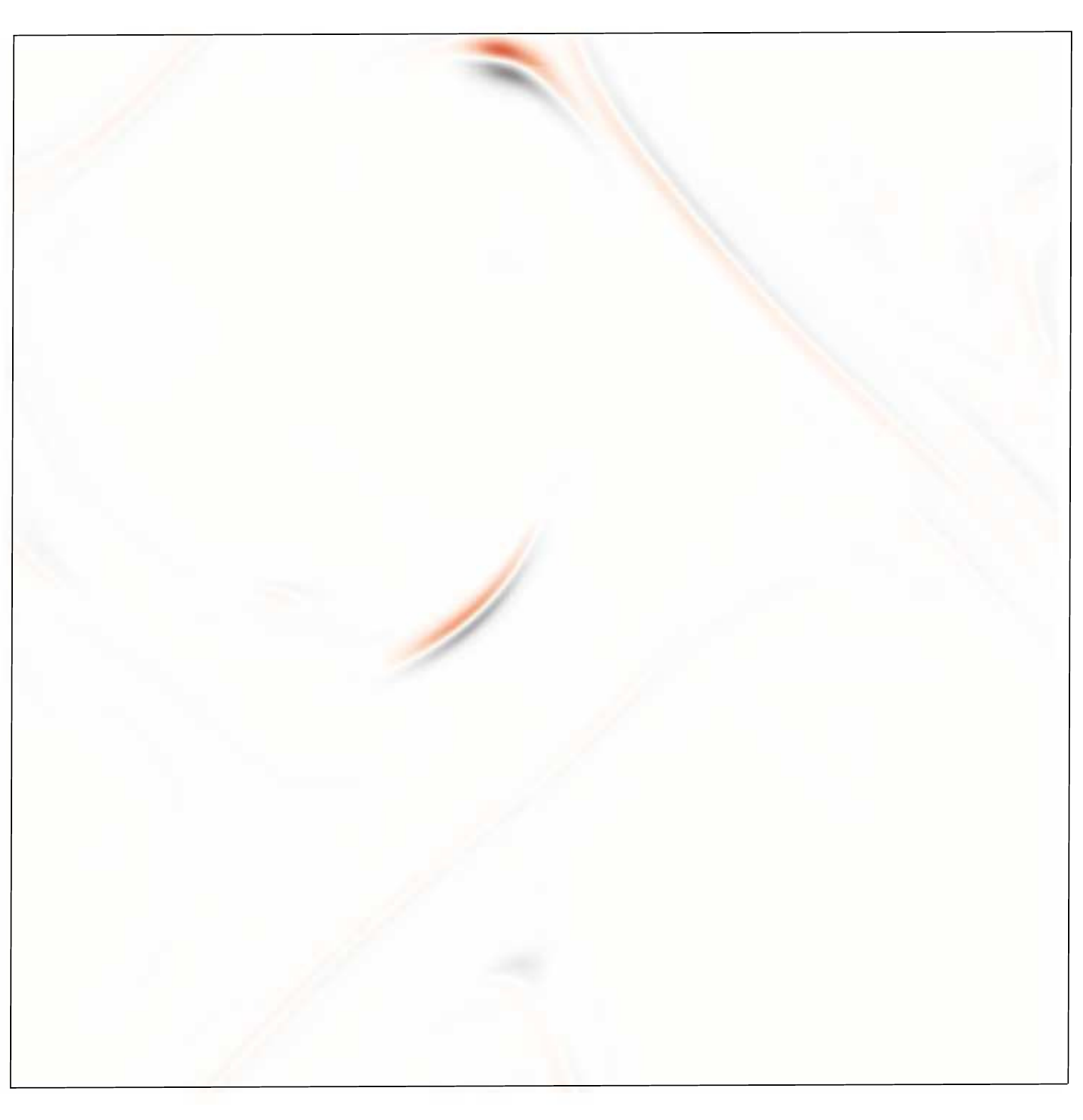} \caption{\label{fig:some_sp_int}}  
\end{subfigure}  
\begin{subfigure}[b]{0.28\linewidth}                                                                      
\includegraphics[width=0.935\textwidth]{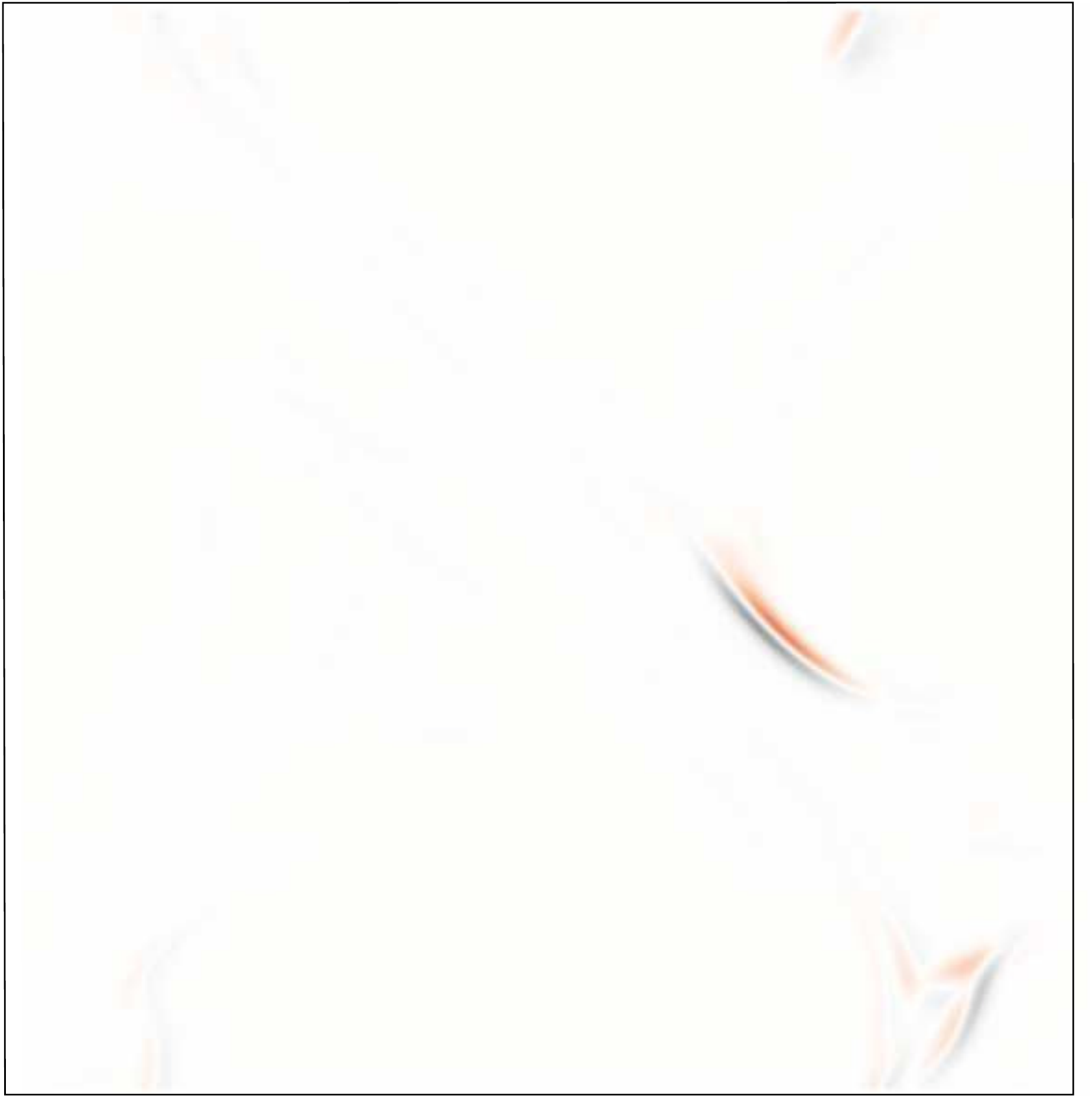} \caption{\label{fig:sp_int}}      
\end{subfigure}  
\begin{subfigure}[b]{0.1\linewidth}  
\includegraphics[height=4.15cm, width=1.1\textwidth]{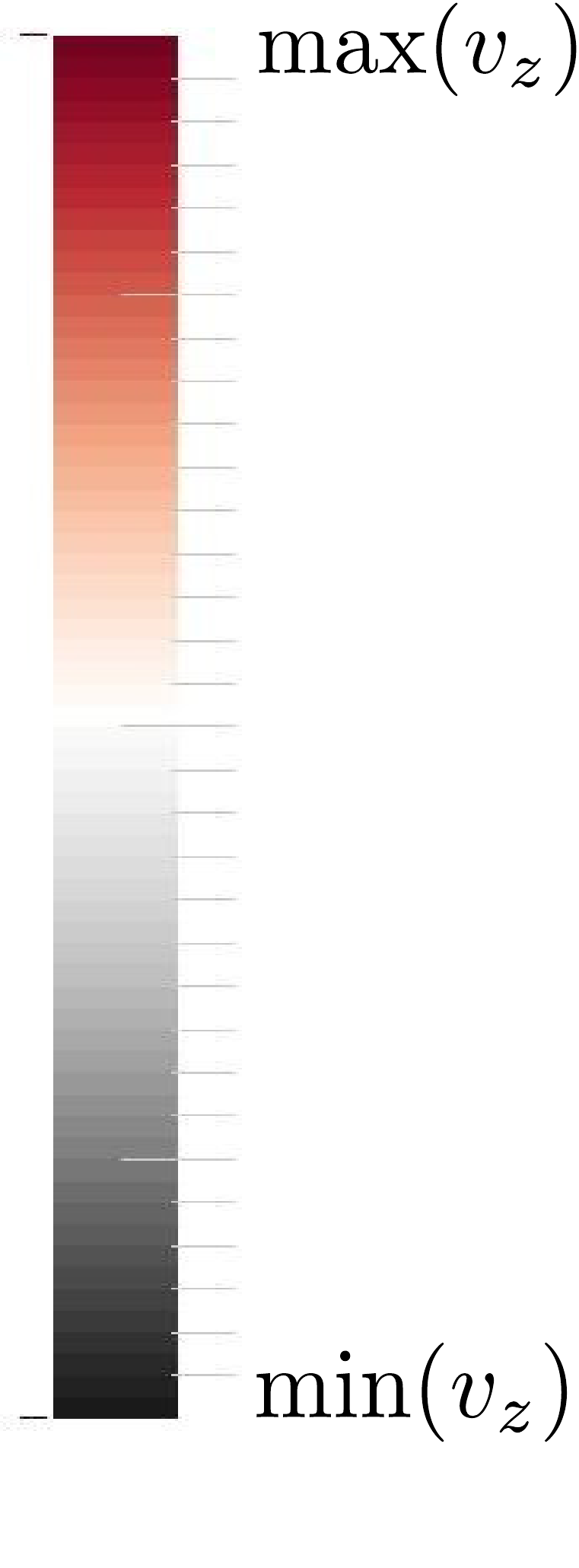}
\end{subfigure}
\caption{Snapshots of $v_z$ for $\Rey = 203$ and $ Q=2.0$ (figure \ref{fig:no_sp_int}),  $Q=5.0$ (figure \ref{fig:some_sp_int}) and $Q=5.125$ (figure \ref{fig:sp_int}). The colour scale on the right is chosen in each plot such that the minimum (negative) is shown in black and the maximum (positive) in red.
(same as in figure \ref{fig:temp_int}). As in figure \ref{fig:int_vis_U3D}, $u_z$ becomes localised in smaller and smaller areas as $Q$ increases, i.e. there is increasing spatial intermittency as $Q_{2D} \approx 5.13$ is approached.}
\label{fig:int_vis_uz}                                                                                
\end{figure}                                                                                          

In our system, the 2D flow forms the marginal stable attractor that loses stability to 3D perturbations depending on the exact realisation of the 2D turbulent flow. To formulate this, we decompose the velocity field into its 2D and 3D parts, $\mathbf{u} = \mathbf{u}_{2D} + \mathbf{u}_{3D} $, where the 2D part is defined as the Fourier sum of $\mathbf{u}$ restricted to modes with $k_z = 0$. Filtering the 3D component of equation (\ref{eq:NS_inc}), dotting with $\mathbf{u}_{3D}$ and integrating over the domain gives
\begin{equation}
\frac{1}{2}\partial_t U_{3D}^2 =-  \left\langle \left\lbrace \mathbf{u}_{3D}\cdot \nabla \mathbf{u}_{2D}  \right\rbrace \cdot \mathbf{u}_{3D} \right\rangle   - \nu \left\langle  | \nabla \mathbf{u}_{3D} |^2 \right\rangle, \label{eq:on_off_int}
\end{equation}
where $\langle \cdot \rangle$ denotes integration over the domain. 
The chaotic 2D motions then act as multiplicative noise while the viscous terms provide a mean decay rate. An important physical mechanism for creating 3D disturbances is 3D elliptic instability of the 2D counter-rotating vortex pair forming the condensate, as described in \citep{le_dizès_laporte_2002}. This may explain the presence of the critical value $Q_{2D}$ itself: instability requires small vertical wavenumbers, but the minimum wavenumber increases with decreasing $H$ and $Q_{2D}$ corresponds to the point where 3D perturbations begin to decay. 

Figure \ref{fig:temp_int} shows that temporal intermittency is present in the thin-layer system. Panel (a) shows a typical time series of 3D energy at $Q\lesssim Q_{2D}$ which fluctuates over six orders of magnitude. In particular, as mentioned before, there are burst-like excursions in 3D energy. In figure \ref{fig:int_spectra}, PDFs constructed from this time series and similar ones for different values of $Q$ are shown along with dotted lines indicating power laws with exponents $-1$, $-0.8$ and $-0.3$. The PDFs are very close to a power law for a significant range of $U_{3D}^2$ and and the exponent converges to minus one as the transition is approached, in agreement with on-off intermittency predictions. However, 
the scaling of 3D energy with deviation from onset shown in figure \ref{fig:U3_res} does not follow the linear prediction of on-off intermittency, but rather $\langle U^2_{3D} \rangle \propto (Q_{2D}-Q)^2 $. For $U_z^2$, figure \ref{fig:Uv_all} seems to suggest yet a different scaling, namely $\langle U^2_{z} \rangle \propto (Q_{2D}-Q)^3$. A similar behaviour was also found in \citep{benavides_alexakis_2017} and was attributed to the spatio-temporal character of the intermittency that not only leads to 3D motions appearing more rarely in time as criticality is approached but also to them occupying a smaller fraction of the available volume. This appears also to be the case in our results, as demonstrated in figures \ref{fig:int_vis_U3D} and \ref{fig:int_vis_uz}, where $\mathbf{u}_{3D}^2$ and the vertical velocity $u_z$ are plotted for three different values of $Q$. As $Q$ approaches the critical value $Q_{2D}$, the structures become smaller for $\mathbf{u}_{3D}^2$ and $u_z$ with the difference that $\mathbf{u}_{3D}^2$ shows spot-like structures in figure \ref{fig:sp_int_U2_3D} which are absent for $u_z$. This difference may be related to the two different scalings observed for $U_{3D}^2$ and $U_z^2$ with $Q_{c}-Q$ small: if the volume fraction of vertical motion depends on $Q_c-Q$ to a different power than that of vertical variations, two different behaviours of $U_z^2$ and $U_{3D}^2$ would follow. A more detailed quantitative investigation of the scaling of volume fraction will be needed to clarify this.

In summary, we have found nontrivial behaviour close to both transitions: we have observed hysteresis near $Q_{3D}$ and spatio-temporal intermittency close to $Q_{2D}$ where the temporal behaviour seems to be described by on-off intermittency. Taking into account these effects will be crucial for understanding the exact nature of the observed transitions. 

\section{Spectra and fluxes}

\label{sec:spec_flux}                                                                            
In this section, we discuss the spectral space properties of the three different regimes described in the previous section. For this purpose, it is necessary to define a few additional quantities. In addition to the total 1D energy spectrum defined in (\ref{eq:spec_tot_ksphr}), is of interest to consider the two-dimensional energy spectrum in the $(k_h,k_z)$ plane. 
\begin{equation}
E(k_h,k_z) = \frac{1}{2}\sum_{\mathbf{k}' \atop {{k_x'^2+k_y'^2=k_h^2}\atop  k_z'=k_z} } \left| \hat{{\bf u}}_\mathbf{k'} \right|^2 .
\label{eq:2D_spectrum}
\end{equation}
Moreover, the total 1D energy spectrum may advantageously be split up into three components: 
the energy spectrum of the (vertically averaged) 2D2C field
\begin{equation}
E_{h}(k_h) = \frac{1}{2}\sum_{\mathbf{k} \atop {{k_x^2+k_y^2=k_h^2}\atop  k_z=0} } \left( \left| \hat{{u}}_\mathbf{k}^{(x)}\right|^2 + \left|\hat{{u}}_\mathbf{k}^{(y)} \right|^2 \right),
\label{eq:spec_tot_kh}
\end{equation}
the energy spectrum of the (vertically averaged) vertical velocity 
\begin{equation}
E_{z}(k_h) = \frac{1}{2}\sum_{\mathbf{k} \atop {{k_x^2+k_y^2=k_h^2}\atop k_z = 0} }\left|\hat{{u}}_\mathbf{k}^{(z)}\right|^2,  
\label{eq:spec_z}
\end{equation}
and the energy spectrum of the 3D flow defined as
\begin{equation}
E_{3D}(k_h) = \frac{1}{2}\sum_{\mathbf{k} \atop {{k_x^2+k_y^2=k_h^2} \atop k_z\neq 0} } \left|\hat{{\bf u}}_\mathbf{k} \right|^2,
\label{eq:spec_3D}
\end{equation} 
satisfying $E_{tot}(k_h)=E_h(k_h)+E_{z}(k_h)+E_{3D}(k_h)$. Furthermore, we introduce three different quantities related to spectral energy flux. First, the total energy flux as a function of horizontal wave number
\begin{equation}
\Pi(k_h) = \langle \mathbf{u}^<_{k_h} \cdot (\mathbf{u}\cdot\nabla) \mathbf{u} \rangle,
\label{eq:Pi1}
\end{equation}
where the low-pass filtered velocity field is $$\mathbf{u}^<_{k_h} = \sum_{\mathbf{k} \atop k_x^2+k_y^2 < k_h^2} \hat{\mathbf{u}}_\mathbf{k} e^{i\mathbf{k}\cdot \mathbf{x}}. $$ 
With this definition, $\Pi(k_h) $ expresses the flux of energy through the cylinder $k_x^2+k_y^2=k_h^2$ 
due to the non-linear interactions.
The 2D energy flux as a function of $k_h$ is defined as
\begin{equation}
\Pi_{2D}(k_h) = \langle \overline{\mathbf{u}}^<_{k_h} \cdot (\overline{\mathbf{u}}\cdot\nabla) \overline{\mathbf{u}} \rangle,
\label{eq:Pi2}
\end{equation}
where the over-bar stands for vertical average
and expresses the flux through the same cylinder due to only 2D2C interactions. 
Finally, we define the 3D energy flux (due to all interactions other than those in (\ref{eq:Pi2})) as a function of horizontal wave number by
\begin{equation}
\Pi_{3D} (k_h) = \Pi(k_h) - \Pi_{2D}(k_h).
\label{eq:Pi3}
\end{equation}
It expresses the flux due to all interactions other than the ones in (\ref{eq:Pi2}).

\begin{figure}
\begin{subfigure}[b]{0.325\linewidth}    
\includegraphics[width=1\textwidth]{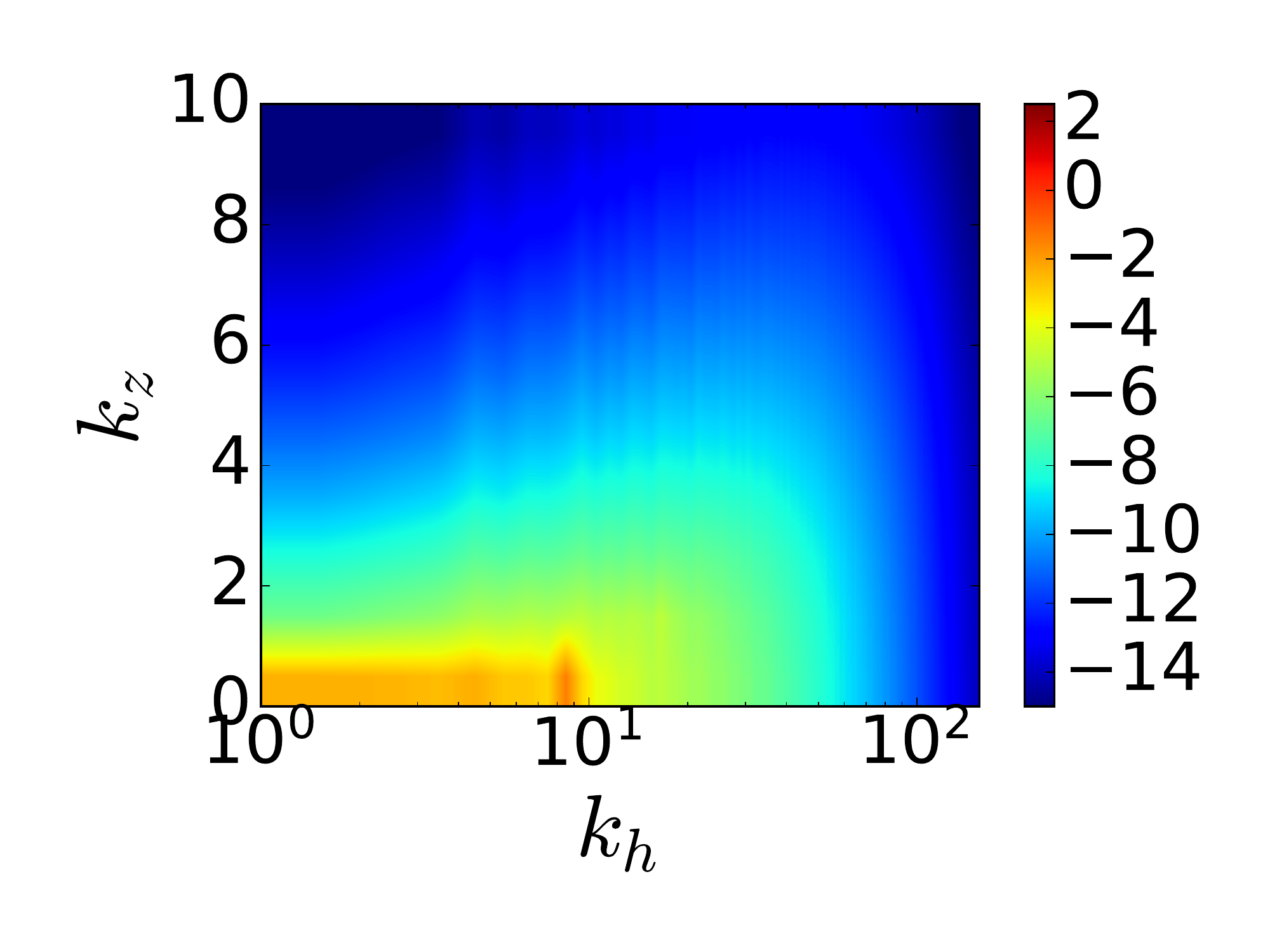} \caption{\label{fig:3Dspec2D}}
\end{subfigure}
\begin{subfigure}[b]{0.325\linewidth}    
\includegraphics[width=1\textwidth]{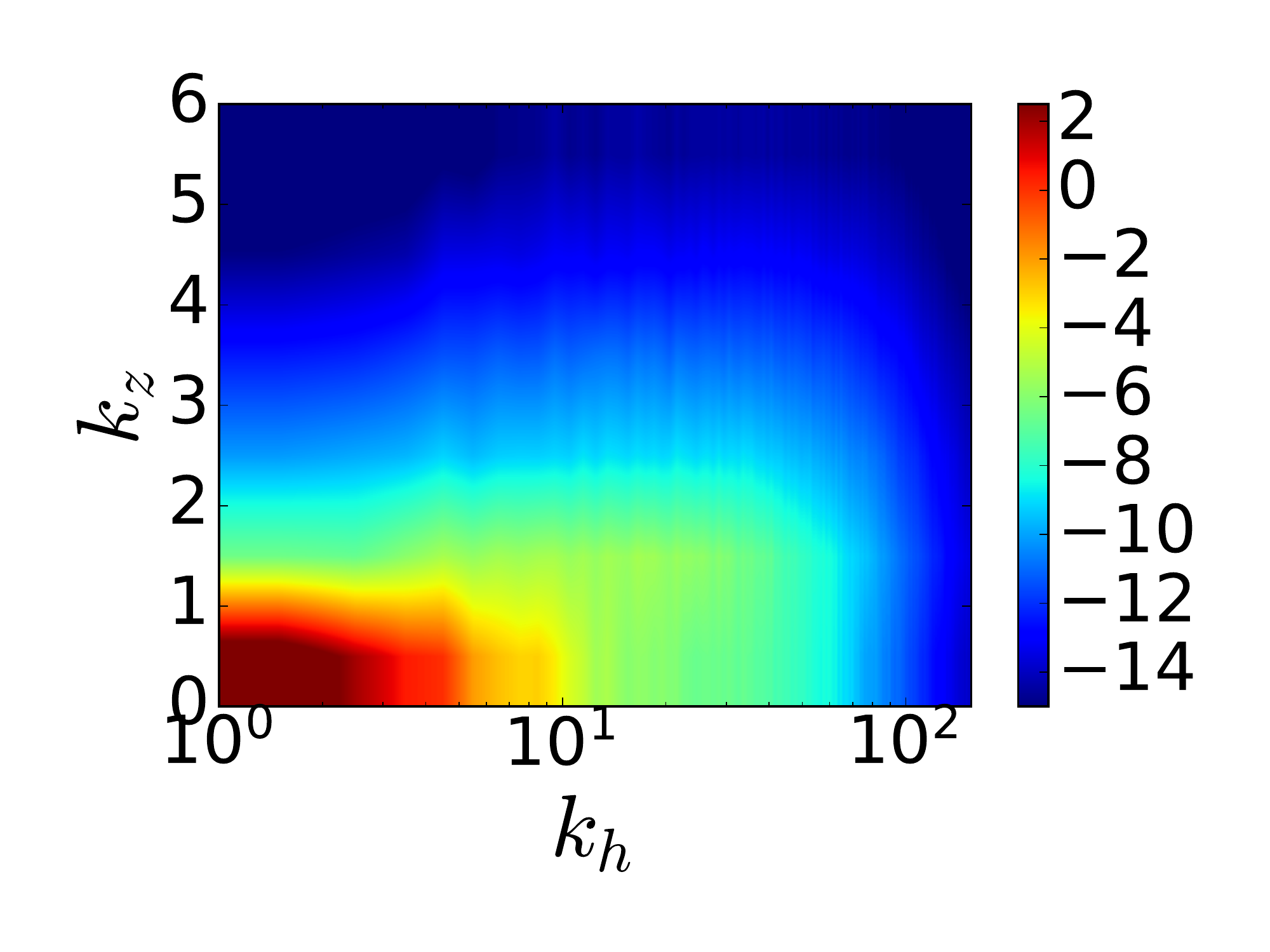} \caption{\label{fig:splitspec2D}}
\end{subfigure}
\begin{subfigure}[b]{0.3
\linewidth}    
\includegraphics[width=1\textwidth]{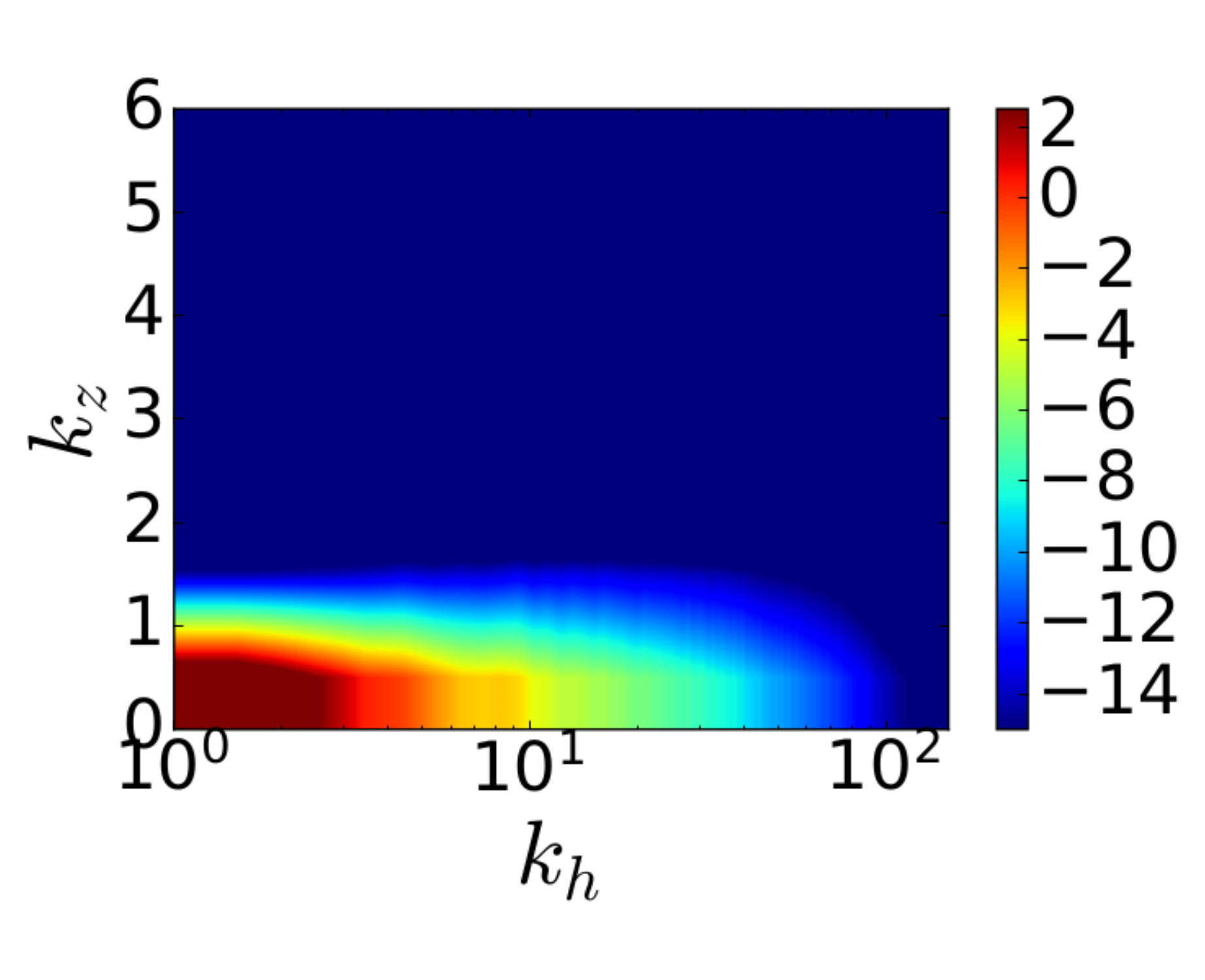} \caption{\label{fig:2Dspec2D}}
\end{subfigure}
\caption{Logarithmic surface plots of $E(k_h,k_z)$ at steady state in the three regimes a) $Q<Q_{3D}$, b) $Q_{3D} < Q<Q_{2D}$ and c) $Q_{2D}<Q$.}
\label{fig:2Dspecs}
\end{figure}
\begin{figure}                                                                            
\begin{subfigure}[b]{0.325\linewidth}                                                       
\includegraphics[width=1\textwidth]{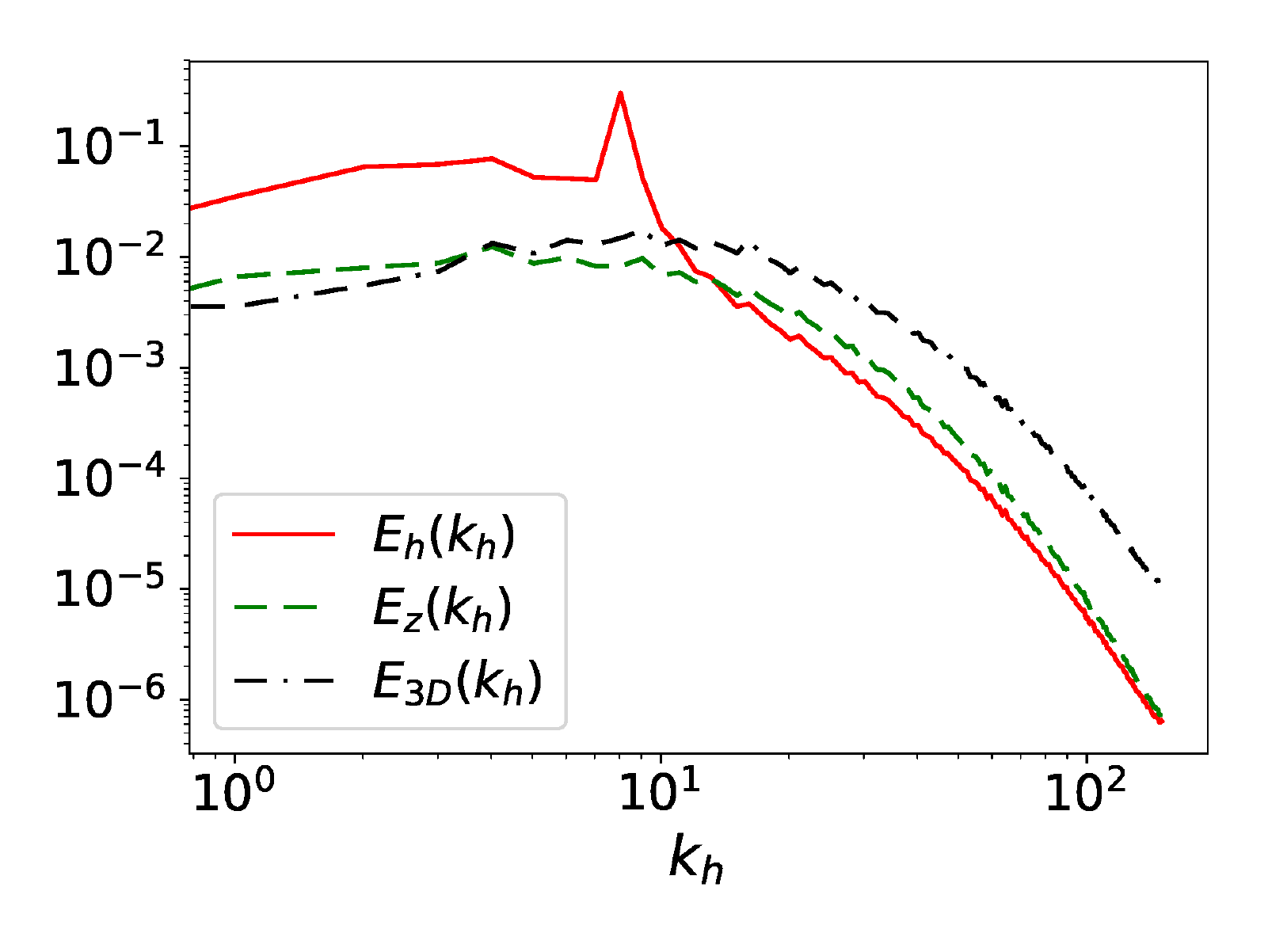} \caption{\label{fig:3Dspec}}
\end{subfigure}
\begin{subfigure}[b]{0.325\linewidth}                                                   
\includegraphics[width=1\textwidth]{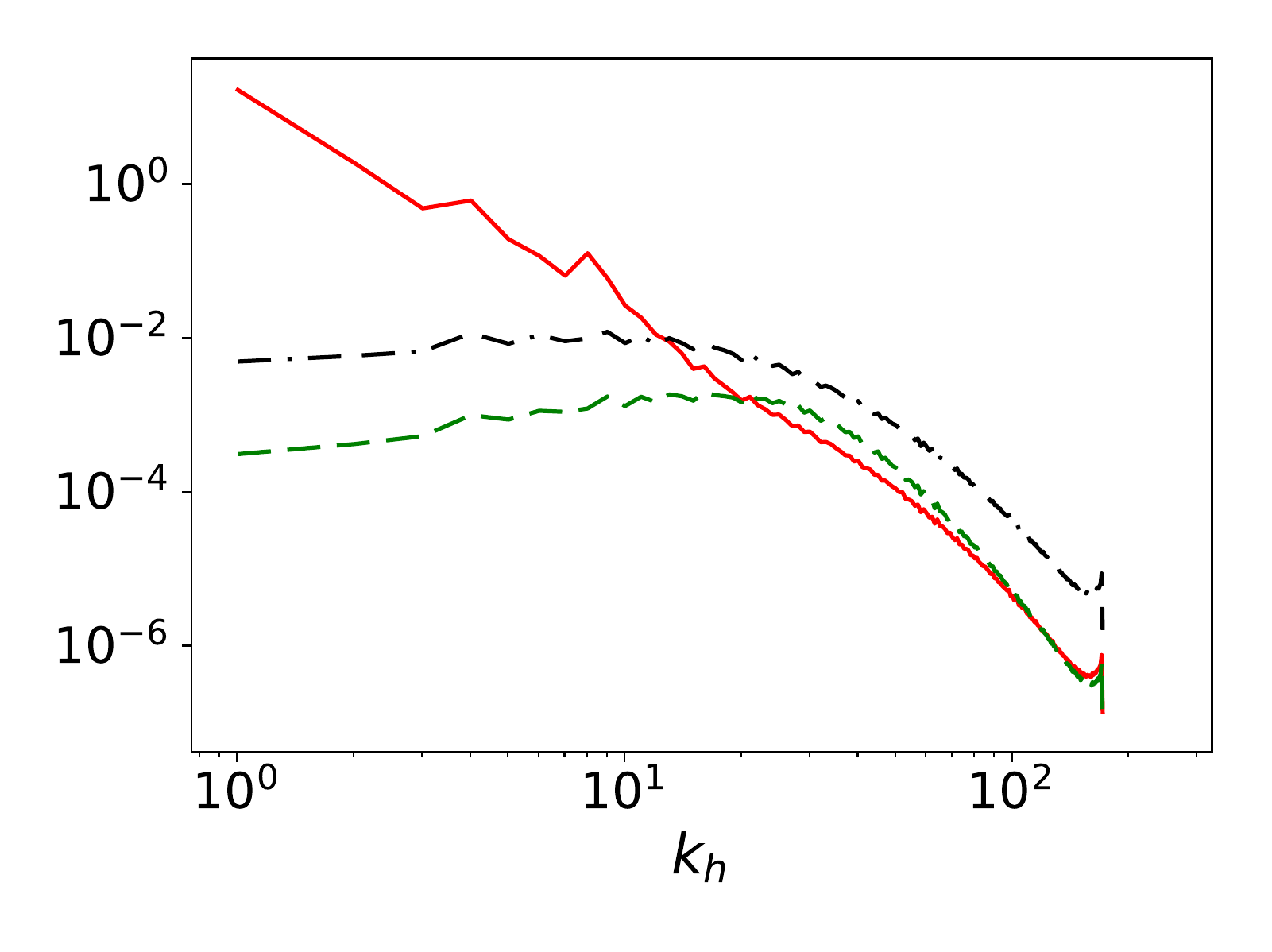} \caption{\label{fig:splitspec}}
\end{subfigure}
\begin{subfigure}[b]{0.325\linewidth}                                                              
\includegraphics[width =1\textwidth]{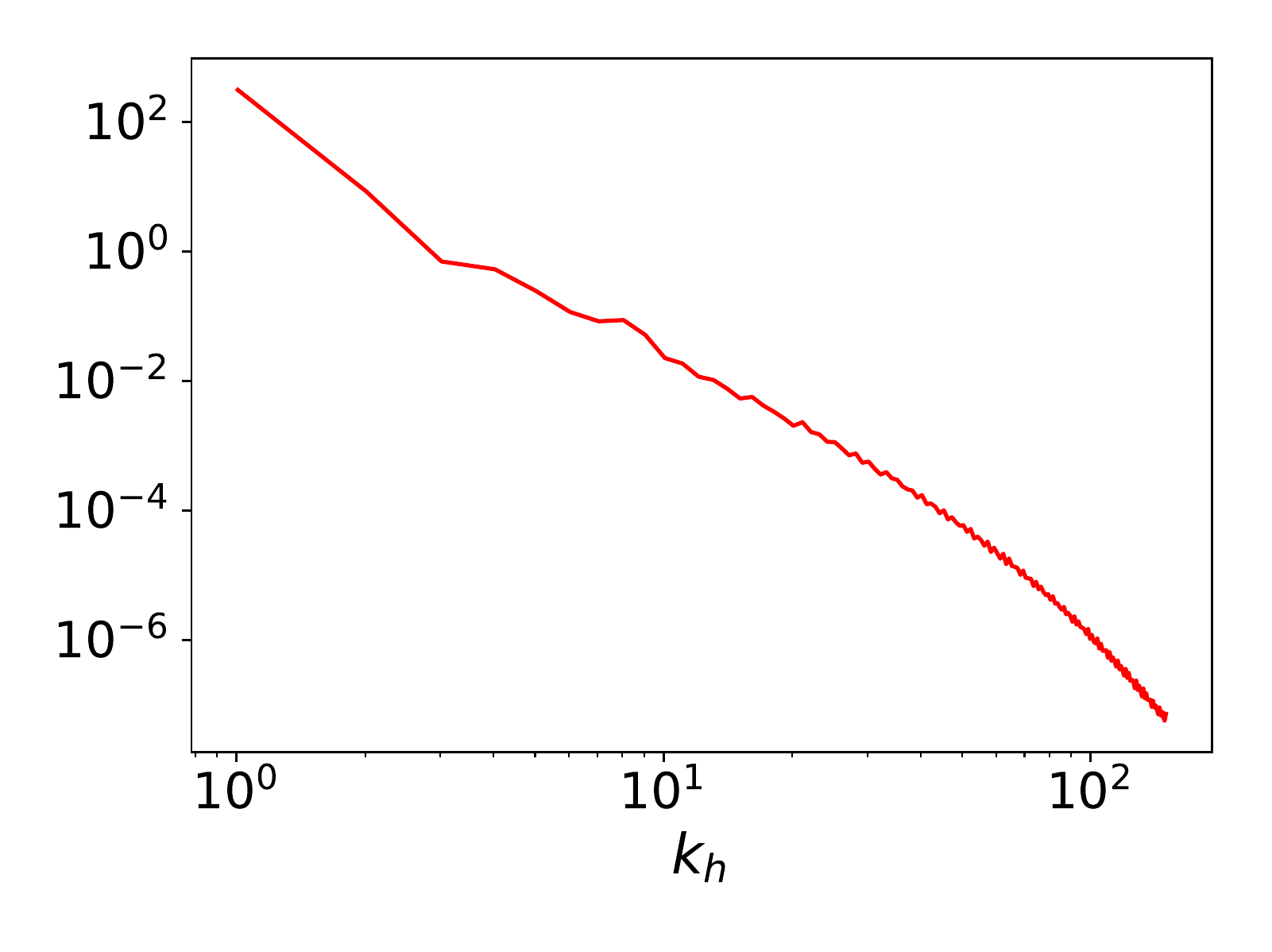} \caption{\label{fig:2Dspec}}     
\end{subfigure}
\caption{Three different energy spectra, $E_{h}(k_h)$, $E_{z}(k_h)$, $E_{3D}(k_h)$ at $\Rey = 609$ for 3D turbulence (figure \ref{fig:3Dflux}), 2D turbulence (figure \ref{fig:2Dspec}) and an intermediate case $Q\in(Q_{3D},Q_{2D})$ (figure \ref{fig:splitspec}) \textit{flux-loop condensate} (cf. main text). For 3D turbulence ($Q=1.25<Q_{3D}$), the 2D energy spectrum peaks at the forcing scale and is an order of magnitude bigger than the other components. In the flux-loop condensate ($Q=4$), 2D energy is maximum at $k=1$ and 3D and vertical energy are non-zero. In 2D turbulence ($Q=16>Q_{2D}$), 2D energy is maximum at $k=1$, but 3D and vertical energy vanish.}
\label{fig:spectra}                                                                              
\end{figure}                                                                                     

\begin{figure}                                                                                   
\begin{subfigure}[b]{0.325\linewidth}                                                             
\includegraphics[width=1\textwidth]{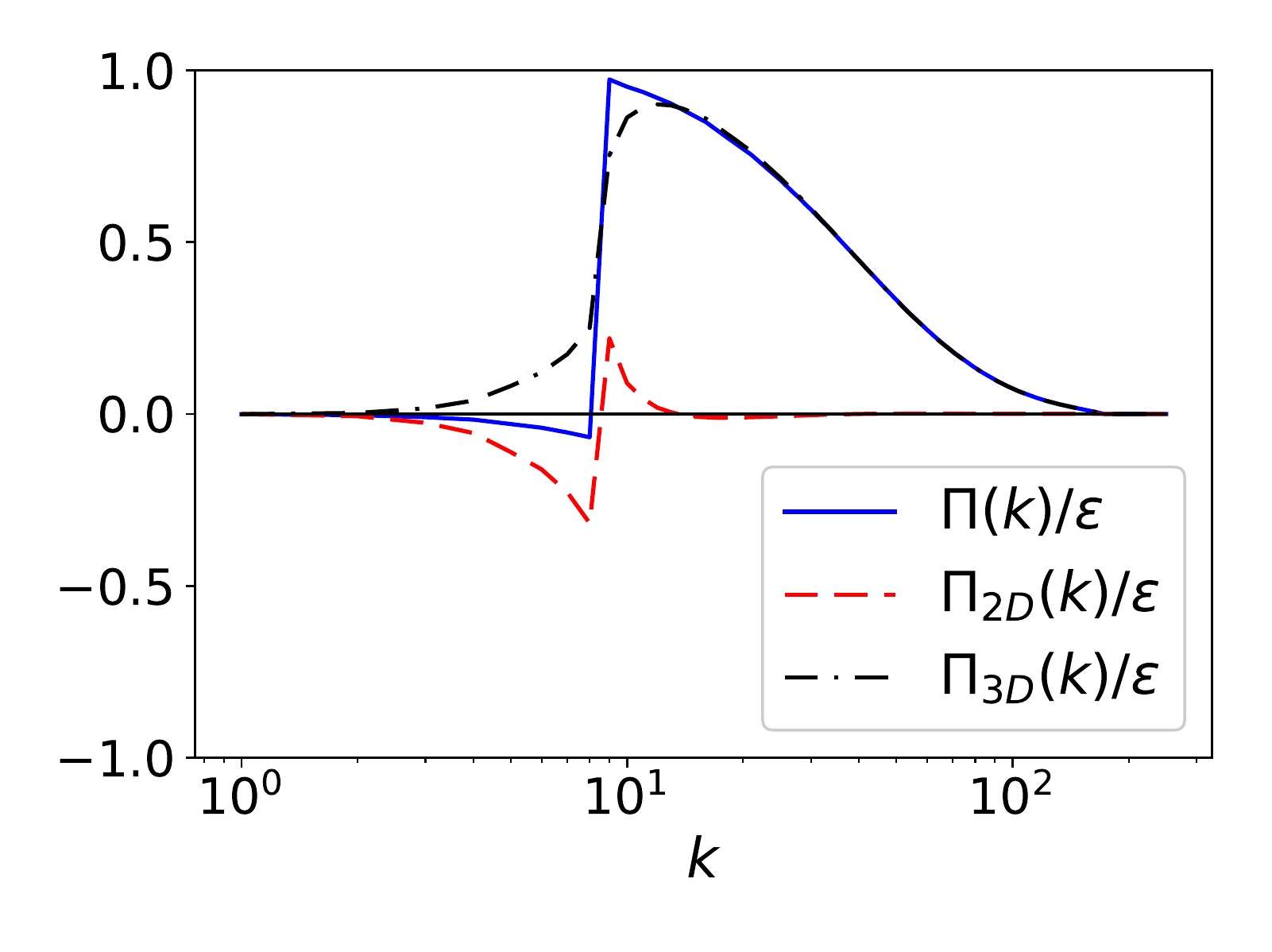} \caption{\label{fig:3Dflux}}
\end{subfigure}
\begin{subfigure}[b]{0.325\linewidth}                                                  
\includegraphics[width=1\textwidth]{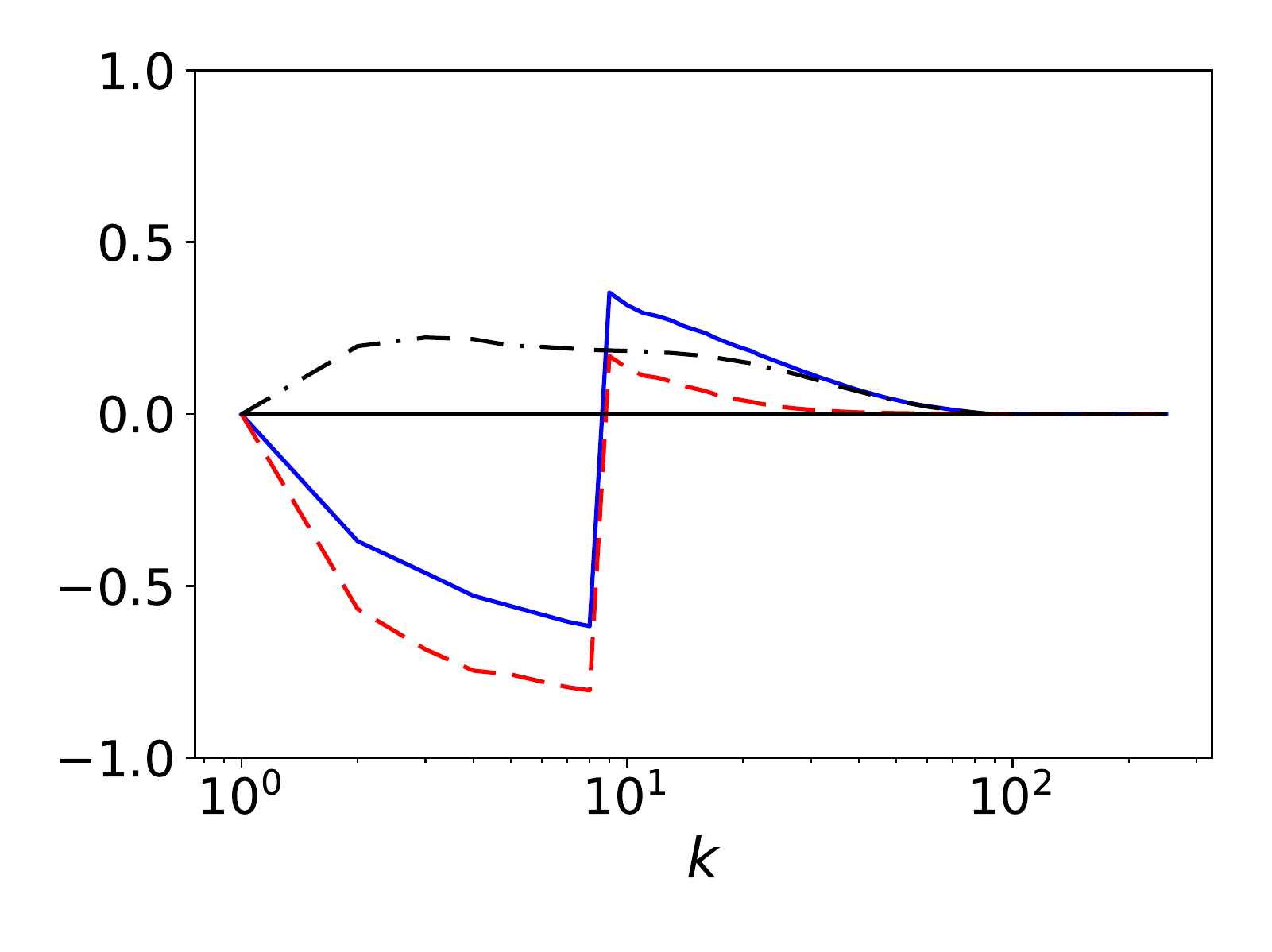} \caption{\label{fig:splitflux}}
\end{subfigure}
\begin{subfigure}[b]{0.31\linewidth}                                                    
\includegraphics[width =1\textwidth]{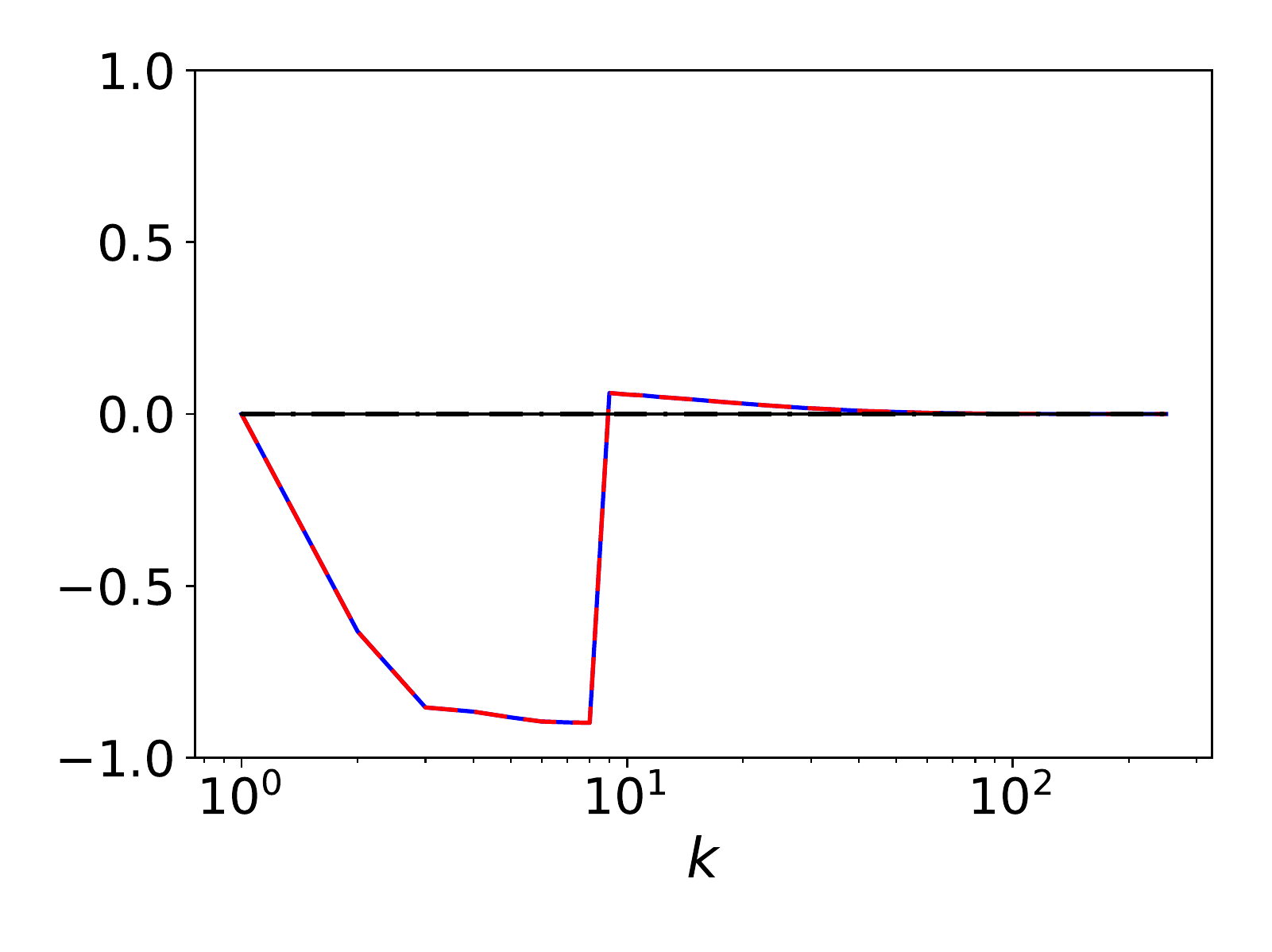} \caption{\label{fig:2Dflux}}     
\end{subfigure}
\caption{Three different components of spectral energy flux, $\Pi(k_h)$, $\Pi_{2D}(k_h)$ and $\Pi_{3D}(k_h)$, are shown for the same three cases and in the same order as in figure \ref{fig:spectra}. 
}
\label{fig:fluxes}                                                                               
\end{figure}                                                                                     

\begin{figure}                                                                                   
\centering                                                                                       
\begin{subfigure}[b]{0.45\linewidth}                                                             
\includegraphics[width=1\textwidth]{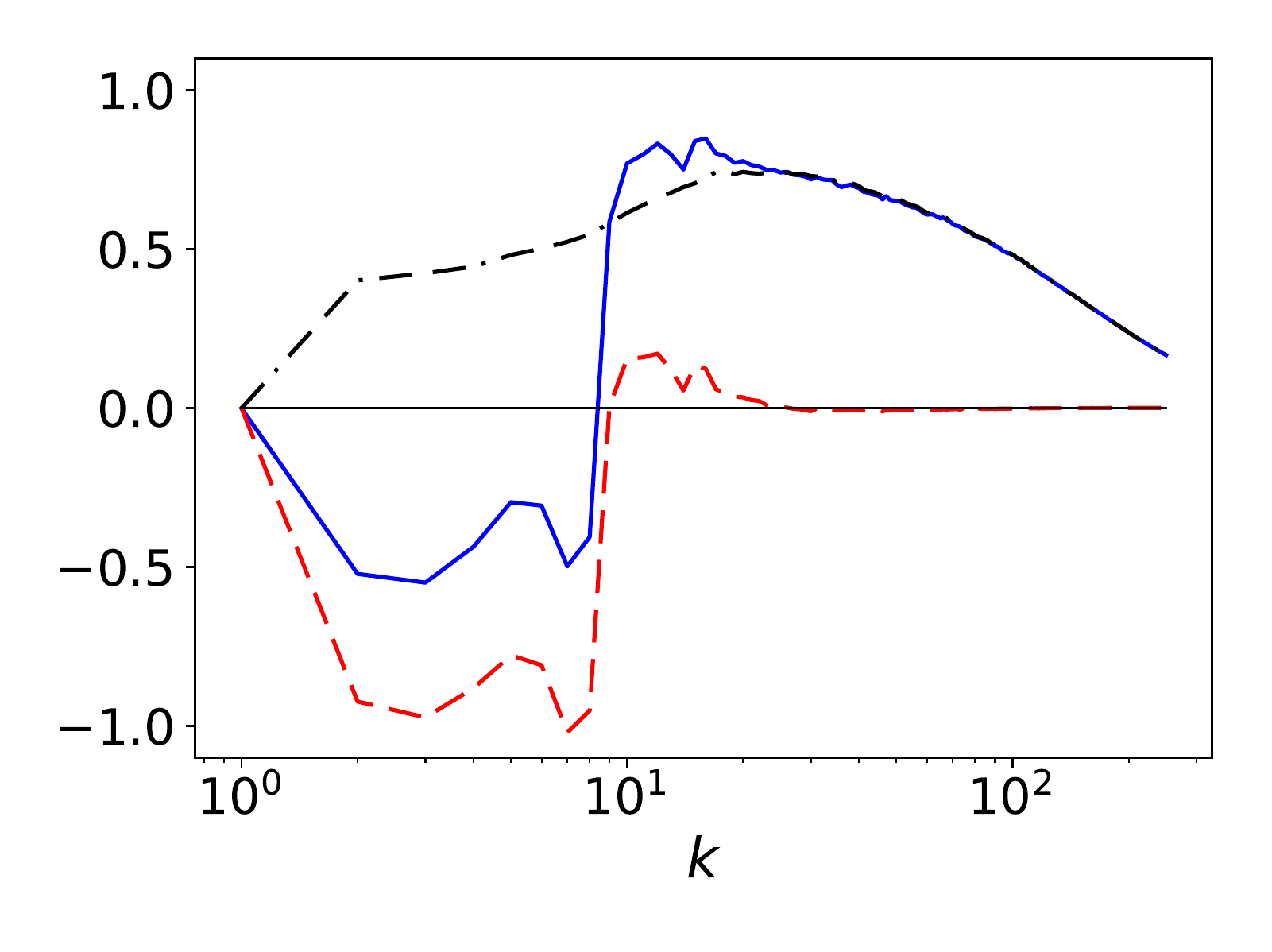} \caption{\label{fig:splitflux_highRe}}
\end{subfigure}
\begin{subfigure}[b]{0.45\linewidth}                                                  
\includegraphics[width=1\textwidth]{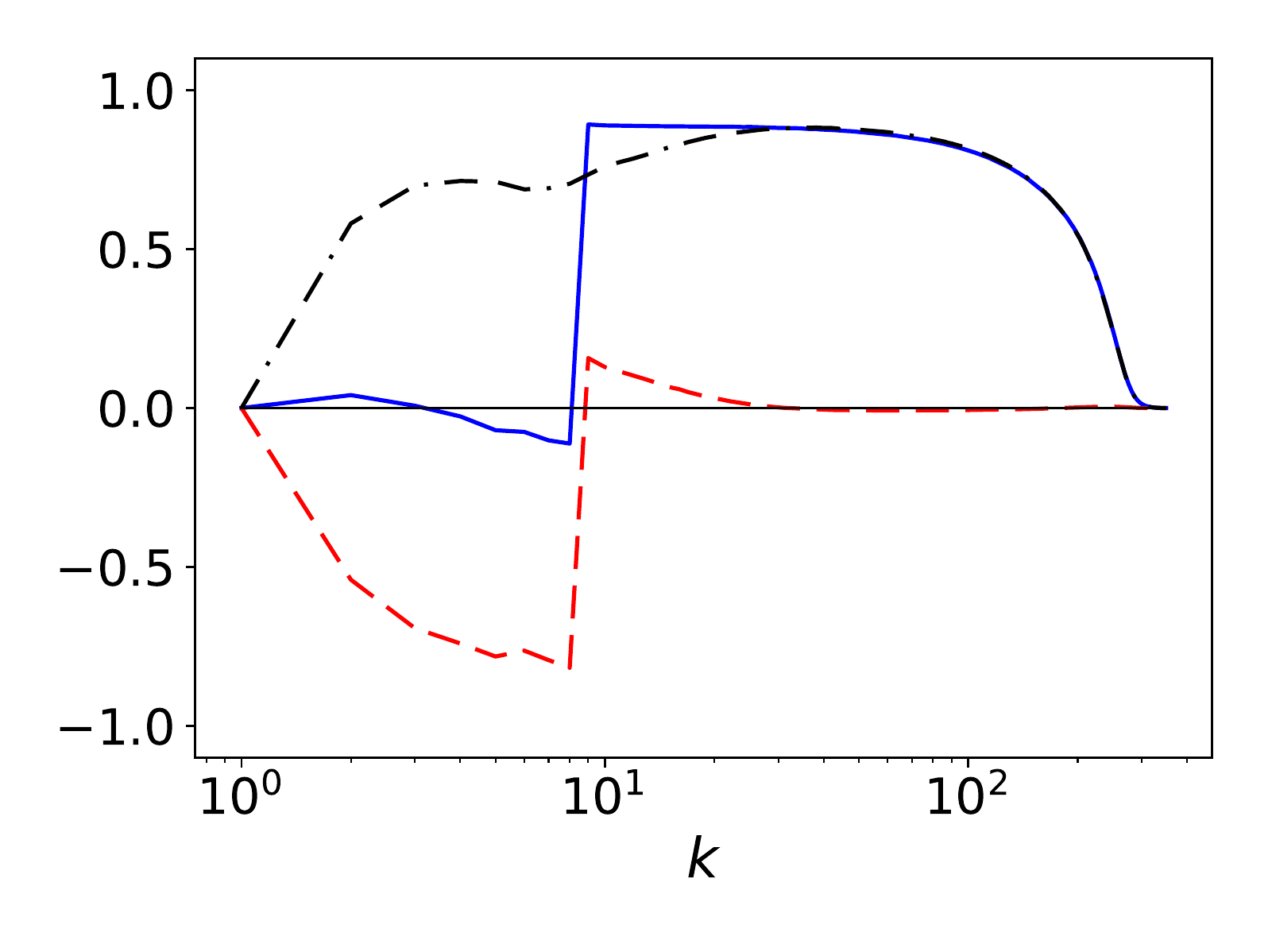} \caption{\label{fig:splitflux_NuH}}
\end{subfigure}
\caption{Flux loop condensate steady state fluxes for $\Rey = 4062$ in panel (a) and the hyperviscous run in panel (b). }
\label{fig:fluxLoopHiRe}                                                                         
\end{figure}                                                                                     

Figure \ref{fig:2Dspecs} shows the steady state 2D energy spectrum in the three different regimes: a) $Q<Q_{3D}$, b) $Q_{3D} < Q<Q_{2D}$ and c)  $Q_{2D}<Q$. In the 3D turbulent case a), the global maximum is at the forcing scale and $k_z=0$, while large $k_z$ modes have a relatively larger fraction of total energy than in cases b) and c). In cases b) and c), a condensate is present with a maximum at the largest wavenumber $k_h=1,k_z=0$. In case b), there is still energy in the $k_z\neq 0$ modes, while in case c), the energy is entirely concentrated in the $k_z=0$ mode. Figure \ref{fig:spectra} shows the energy spectra $E_h(k_h), E_z(k_h)$ and $E_{3D}(k_h)$ for the same three cases a)-c).
In case a) (3D turbulence), all three spectra are of the same order, with a small excess of $E_h(k_h)$ in the large scales and an excess of $E_{3D}(k_h)$ in the small scales. The small scale separation between the forcing and the dissipation scale does not allow us to observe a $k^{-5/3}$ power-law regime.
In case b), $E_h(k_h)$ clearly dominates in the large scales, forming a steep spectrum (close to $E_h(k_h)\propto k_h^{-4}$). However, at wavenumbers larger than the forcing wavenumber $k_f=8$, $E_z(k_h)$ and $E_{3D}(k_h)$ become of the same order as $E_h(k_h)$. 
In case c) (2D turbulence), where $Q>Q_{2D}$, the spectra  $E_z(k_h)$ and $E_{3D}(k_h)$ have reduced to values close to the round-off error and are not plotted. The 2D spectrum $E_h(k_h)$ displays again a steep power-law behaviour close to $E_h(k_h)\propto k_h^{-4}$.

Figure \ref{fig:fluxes} shows the energy fluxes as defined in eqs. (\ref{eq:Pi1}-\ref{eq:Pi3}) for the same three cases examined in figure \ref{fig:spectra}. In panel (a), where the case $Q<Q_{3D}$ is examined, there is almost no inverse flux of energy and $\Pi(k_h<k_f)$ is practically zero.
The small inverse flux that is observed for $\Pi_{2D}(k_h)$ at $k<k_f$ does not reach the largest scale of the system and is nearly completely balanced by $\Pi_{3D}(k_h)$, which is forward. At wavenumbers larger than $k_f$, the total flux is positive and is completely dominated by $\Pi_{3D}$. 
This is to be contrasted with the rightmost panel (c) with $Q>Q_{2D}$, where at small wavenumbers, the flux is negative and is dominated by the 2D flow, while at large wavenumbers there is a very small forward flux.
For the intermediate case $Q_{3D}<Q<Q_{2D}$ in panel (b), there is an inverse energy flux.
This flux can be decomposed into a negative 2D part $\Pi_{2D}(k_h)$ and a positive 3D part $\Pi_{3D}(k_h)$. In other words, the 2D components of the flow bring energy to the largest scales of the system, which is then brought back to the small scales by the 3D components of the flow associated with a forward energy flux, thus forming a loop for the energy transfer. For this reason, we refer to this case as {\it flux-loop condensate}.

Due to finite viscosity, part of the energy that arrives at the largest scale (shown in figure \ref{fig:splitflux}) is dissipated. Therefore, the two fluxes are not completely in balance. As $\Rey$ is increased, however, the fraction of the energy that is dissipated in the large scales is decreased and the two opposite fluxes come closer to balancing each other. This is shown in figure \ref{fig:fluxLoopHiRe}, where the energy fluxes for the highest $\Rey$ simulation and for the simulation with hyper viscosity are plotted. The two opposite directed fluxes are closer in amplitude. At $\Rey\to \infty$ it is thus expected that 
the inverse and forward fluxes at large scales will be in perfect balance and all the energy is dissipated in the small scales. It is worth noting, however, that  the inverse cascade (negative flux) due to the 2D components has much stronger fluctuations than the forward cascading flux that has lead to the non-monotonic 
behaviour  of the flux observed in figure \ref{fig:fluxLoopHiRe} at small $k$ due to insufficient time averaging.

\section{A three-mode model} 
\label{sec:3mode}

In this section, we formulate and analyse a simple three-scale ODE model which reproduces certain features of the DNS results described in section \ref{sec:DNS}. 

\begin{figure}                                          
\centering                                              
\includegraphics[width=0.55\textwidth]{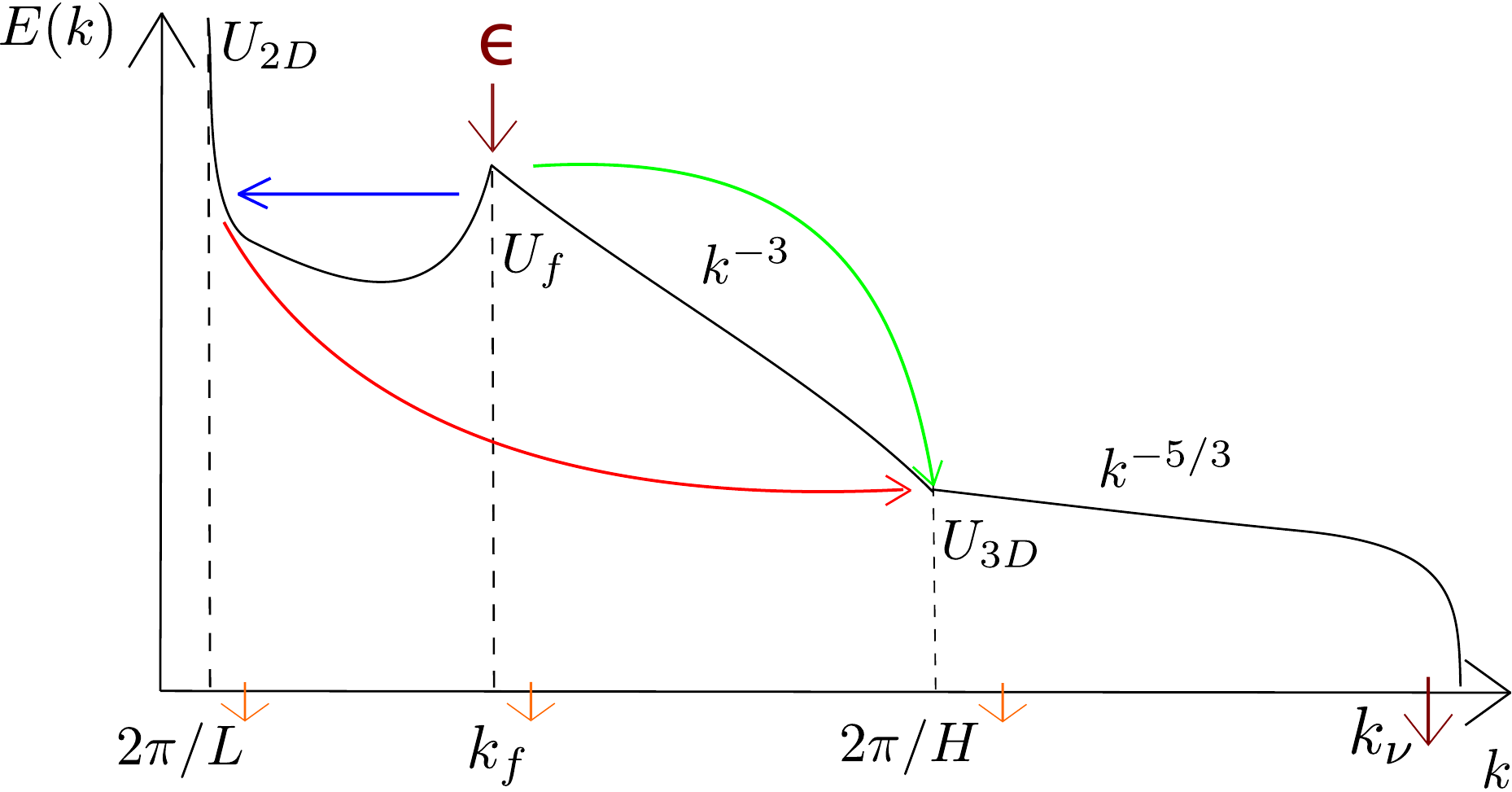}
\caption{Sketch of three-mode model. Solid curve: energy spectrum $E(k)$ of the condensate state. The energy injected at $k_f$ at a rate $\epsilon$ is distributed between large and small scales. Moreover, energy is transferred from large to small scales. Finally viscous dissipation occurs at all scales since $\Rey<\infty$ (short arrows on abscissa) and energy is transferred to the dissipation range (arrow at $k_\nu=2\pi/\eta$). The spectrum $E(k)$ shares certain features with figure 1e) of \citep{xia2011upscale} and figure 3 of \citep{celani2010morethantwo}.}  
\label{fig:3mode_illustr}                                
\end{figure}

As illustrated in figure \ref{fig:3mode_illustr}, our model comprises a 2D mode $U_{2D}$ at the scale $L$ of the domain, a mode $U_f$ at the forcing scale $\ell$ and a 3D mode $U_{3D}$ at the scale of the layer height $H$, whose interactions are spectrally non-local, thus taking into account a major result from section 5. 
The model describes the system at steady state where these scales are well separated, but is not
expected to capture the transient phase where all intermediate scales 
between $L$ and $\ell$ participate due to the inverse cascade. As before, let $Q=\ell/H$, $K=\ell/L$ and $\Rey= (\epsilon \ell^4)^{1/3}/\nu$. Interactions between modes are modelled using eddy viscosity, which amounts to modifying the molecular viscosity $\nu$ by terms involving the small-scale velocities, modelling the effect of small-scale on large-scale motions as diffusive. The conceptual foundations of eddy viscosity were laid by de Saint Venant in his \textit{effective viscosity}, \citep{desaintvenant1843notea} (see \cite{darrigol2017joseph} for a historical review). Eddy viscosity was quantified for the first time by \citep{boussinesq1877essai} and later widely popularised through the works of Taylor \citep{gi1915eddy,taylor1922diffusion}, see also \citep[][]{kraichnan1976eddy}. It has been estimated in various limits both in 2D and 3D flows, \citep{Yakhot1987negative3D, hefer1989inverse, gama1994negative, Dubrulle91, meshalkin1962investigation, sivashinsky1985negative,  bayly1986positive, cameron2016large, alexakis20183d}.

There are two notable cases where the dependence of eddy viscosity $\nu_{E}$ on the flow amplitude $U_s$ and length scale $l_s$ is known.
For $\Rey \to \infty$, one expects that $\nu_{E}$ becomes independent of $\nu$ and the only dimensionally consistent possibility 
for $\nu_E$ is given by
\begin{equation}
\nu_{E} = c_1 U_s l_s. \label{eq:highRe_eddyvisc}
\end{equation}
where $c_1$ is a non-dimensional number.
In the low-$\Rey$ limit, on the other hand, an exact asymptotic expansion can be carried out \citep[see][]{Dubrulle91} which reveals that 
\begin{equation}
\nu_E =  c_2 \frac{ U_s^2  l_s^2 }{\nu } + O(U_s^4  l_s^4 /\nu^2),
\label{eq:lowRe_eddyvisc}
\end{equation}
where the non-dimensional number $c_2$ can be evaluated by the expansion. It may seem counter-intuitive that the low-$\Rey$ limit could have any relevance for the turbulent problem, but since we have established in the DNS that the presence of $Q_{2D}$ is a finite-$\Rey$ phenomenon ($Q_{2D}\propto \Rey^{3/4}$), we clearly need to include a finite $\Rey$ ingredient to describe it and the exact result (\ref{eq:lowRe_eddyvisc}) is selected for this purpose. The sign of the prefactors $c_1,c_2$ depends on the exact form of the small-scale flow and in particular its dimensionality. While 2D flows tend to have negative eddy viscosities and transfer energy upscale, 3D flows are expected to have positive eddy viscosities and transfer energy downscale.
For our model, we are going to consider that interactions among the three different scales $L>\ell>H$ are such that 
the flow at the smaller scale acts as an eddy viscosity on the flow at the larger scale.
These interactions are illustrated in figure \ref{fig:3mode_illustr}. 
In particular, the energy injected at the forcing scale $k_f$ at a rate $\epsilon$ is transferred both to the large 
scale $L$ (by a negative eddy viscosity $-\mu$) and to the small scales (by a positive eddy viscosity $\sigma$). 
The large scales lose energy directly to the small scales (via a positive eddy viscosity term $\eta$), while
the small scales dissipate energy by transfer to the dissipation range, modelled by a non-linear energy sink. 
In addition, viscosity is finite, s.t. all scales dissipate locally. 
The set of equations below formalises these ideas:

\begin{subeqnarray}
\frac{\mathrm{d}}{\mathrm{d}t} U_{2D}^2 =& - \left(\nu - \mu + \eta   \right)\frac{U_{2D}^2}{L^2}, \\
\frac{\mathrm{d}}{\mathrm{d}t} U_{f}^2\,\,\,\, = & \epsilon - \left(\nu + \sigma  \right) \frac{U_f^2}{\ell^2}  - \mu \frac{U_{2D}^2}{L^2}, \\
\frac{\mathrm{d}}{\mathrm{d}t} U_{3D}^2 =&  \eta \frac{U_{2D}^2}{L^2} + \sigma \frac{U_f^2}{\ell^2} - \frac{U_{3D}^3}{H} - \nu \frac{U_{3D}^2}{H^2}.
\label{eq:3scale_schematic}
\end{subeqnarray}
Note in particular that eddy viscosities do not dissipate energy, but merely redistributes it between different scales. Adding the three model equations leads to 
\[ \frac{\mathrm{d}}{\mathrm{d}t} ( U_{2D}^2 + U_{f}^2 + U_{3D}^2 ) = 
\epsilon - \nu \left(\frac{U_{2D}^2}{L^2} +  \frac{U_f^2}{\ell^2} + \frac{U_{3D}^2}{H^2} \right) - \frac{U_{3D}^3}{H} , \]
showing that the total kinetic energy only changes due to molecular viscosity $\nu$, energy injection $\epsilon$ and the sink term representing the 3D energy cascade to the dissipation range, $U_{3D}^3/H$.  Depending on $\Rey$, either of the two expressions for eddy viscosity (eqs. \ref{eq:highRe_eddyvisc},\ref{eq:lowRe_eddyvisc})  may be expected to yield an adequate description of the multi-scale interactions in the problem. A model that interpolates smoothly between the large and small $\nu$ limits, thus taking into account the finite-$\Rey$ information necessary for describing $Q_{2D}$, is given by
\begin{equation}
\mu =  \alpha \frac{ U_f^2 \ell^2 }{\nu + U_f \ell},\quad
\eta = \beta\frac{ U_{3D}^2 H^2 }{\nu+ U_{3D}H},\quad
\sigma =  \gamma \frac{ U_{3D}^2 H^2}{\nu + U_{3D} H},
\label{eq:param_eddy_visc}
\end{equation}
with $\alpha,\beta,\gamma>0$ non-dimensional coupling constants. In the limits $\nu\to 0$ and $\nu\to\infty$, the above expressions converge to the formulae for eddy viscosities described before.
The nonlinear dynamical system thus defined possesses a varying number of fixed points depending on parameters. To classify them, first note that $\epsilon\neq 0 \Rightarrow U_f\neq 0$  at any fixed point by (\ref{eq:3scale_schematic}\textit{b}) and the definition of $\mu$ in (\ref{eq:param_eddy_visc}). Hence there are four possibilities:
\begin{enumerate}
\item[(a) ] {\it laminar} state: $U_{2D}=U_{3D}=0$ (all energy in forcing scale),
\item[(b) ] {\it 3D turbulence} state: $U_{2D}=0$ and $U_{3D}\neq 0$,
\item[(c) ] {\it 2D condensate} state: $U_{2D}\neq 0$ and $U_{3D}=0$ and
\item[(d) ] {\it flux-loop condensate} state: $U_{2D} \neq 0$ and $U_{3D}\neq 0$.
\end{enumerate}
 As shown in appendix \ref{sec:appA}, in the zero viscosity limit, there is neither a laminar state nor a 2D condensate fixed point in the model. This emphasises the importance of including finite-$\Rey$ information into the model for describing both $Q_{3D}$ and $Q_{3D}$ in a single model. 
The laminar state appears for values of $\Rey\equiv(\epsilon\ell^{4})^{1/3}/\nu$ 
below a critical value $\Rey_c$ for which there is no 
transfer, neither to large nor to small scales. Above this critical value, one of the three other states is stable, depending on the value of $Q=\ell/H$. For small values of $Q$ (large $H$), the system is in the 3D turbulence state, where energy is only exchanged between the forcing scale $\ell$ and the small scale $H$. Above the critical value $Q_{3D}$, 
the system transitions to the flux-loop condensate state where part of the injected energy
is transferred to the large scales and then back to the small scales, thus forming a loop. Finally, at sufficiently large $Q$ above a second critical point $Q_{3D}$, the system transitions to 
the 2D condensate where it follows 2D dynamics and there is only a transfer of the injected energy to 
the large scales. \\
\begin{figure}
\centering
\includegraphics[width=0.4\textwidth]{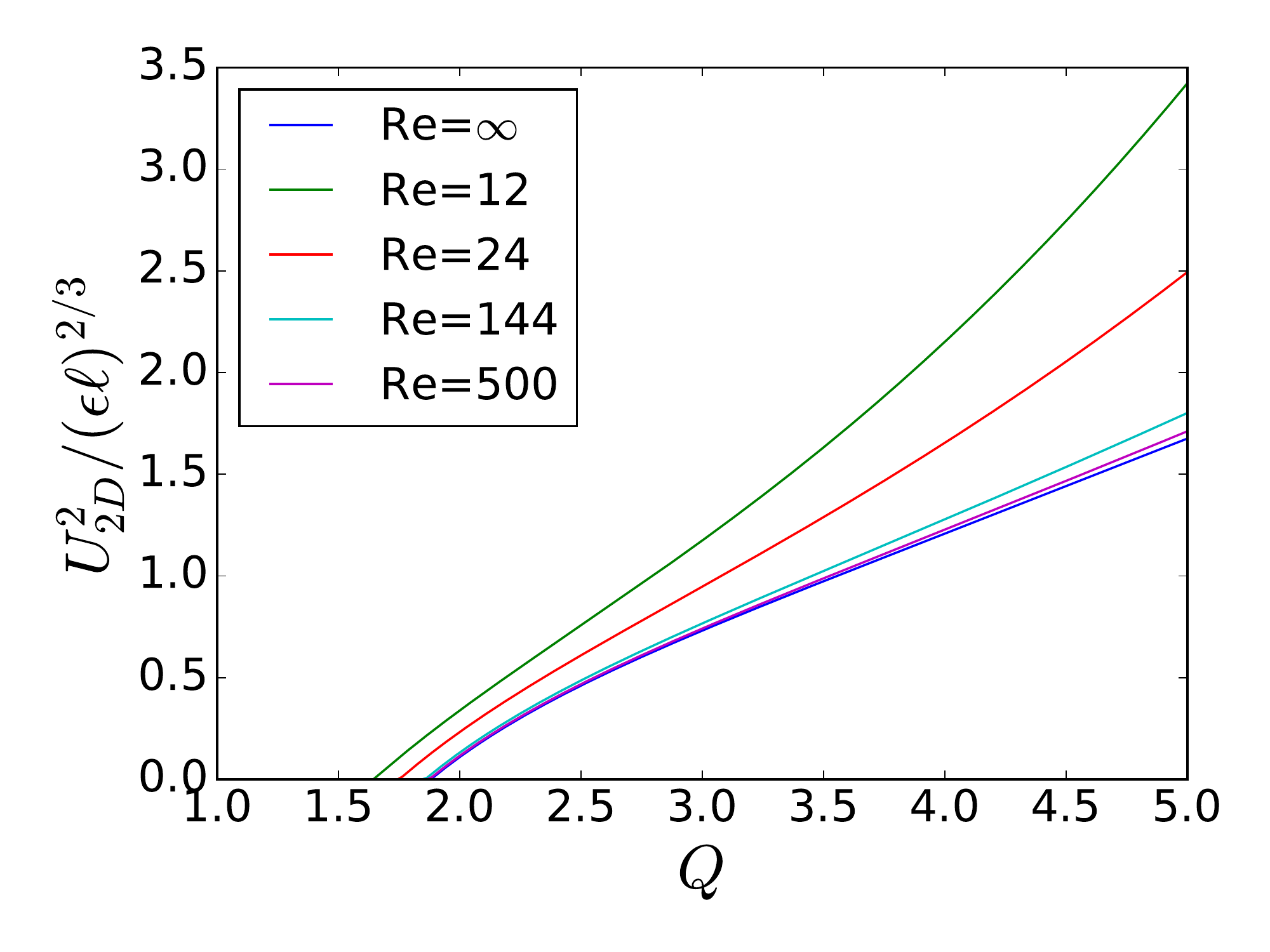}
\caption{Steady state $U_{3D}^2$ from the full model for four different $\Rey$ as well as for the $\Rey=\infty$ limit. The $\Rey=500$ and the $\Rey = \infty$ cases are almost indistinguishable. The parameters used are $L = \ell = 1$, $\alpha = 1$, $\beta = 5$, $\gamma = 0.5$.}
\label{fig:convergence_Q3}
\end{figure}
\begin{figure}                                                        
 \centering                                                           
 \begin{subfigure}[b]{0.4\linewidth}
 \includegraphics[width=\textwidth]{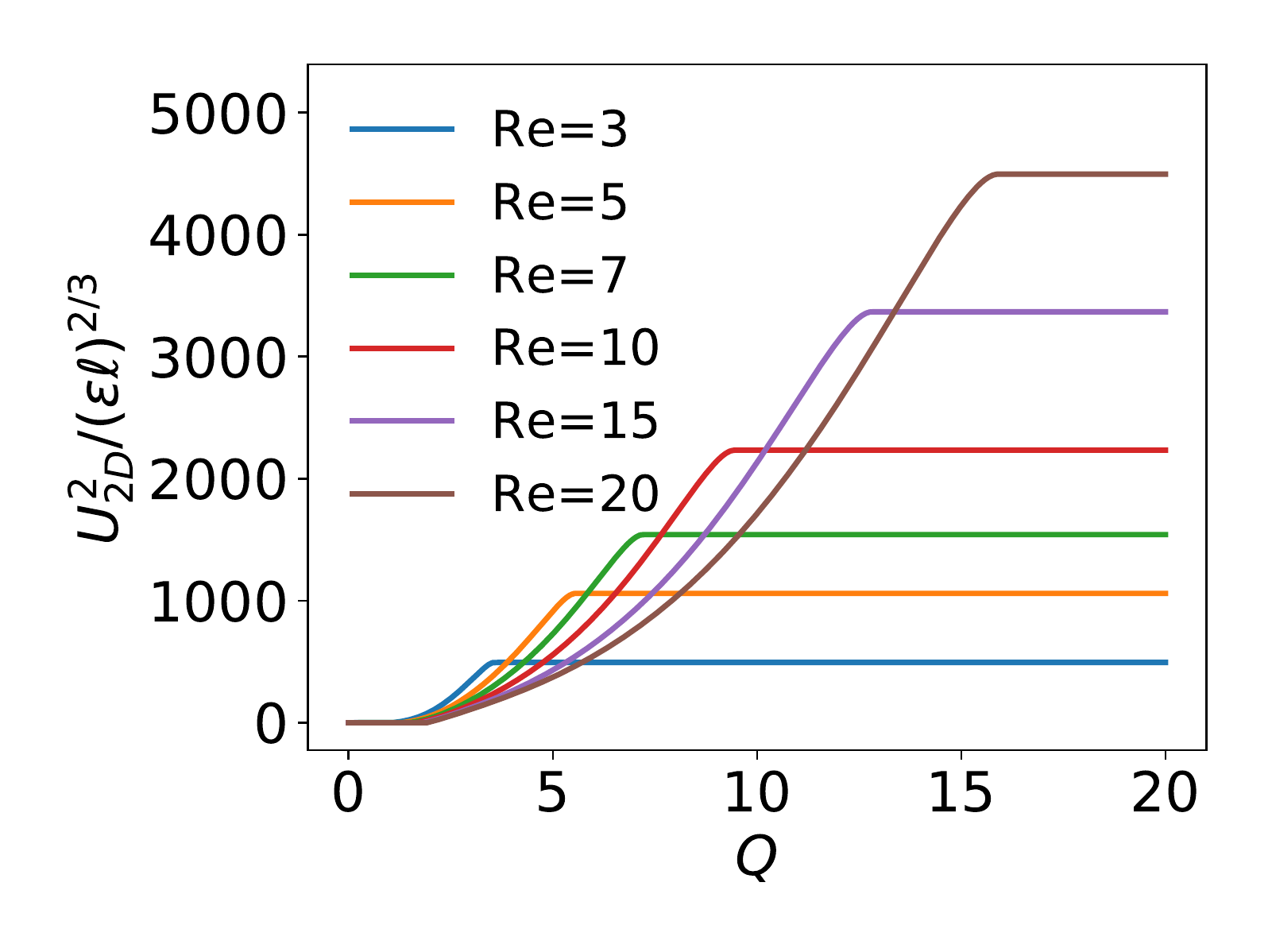} \caption{\label{fig:bif_diag_intRe}}
 \end{subfigure}
 \begin{subfigure}[b]{0.4\linewidth}
 \includegraphics[width=\textwidth]{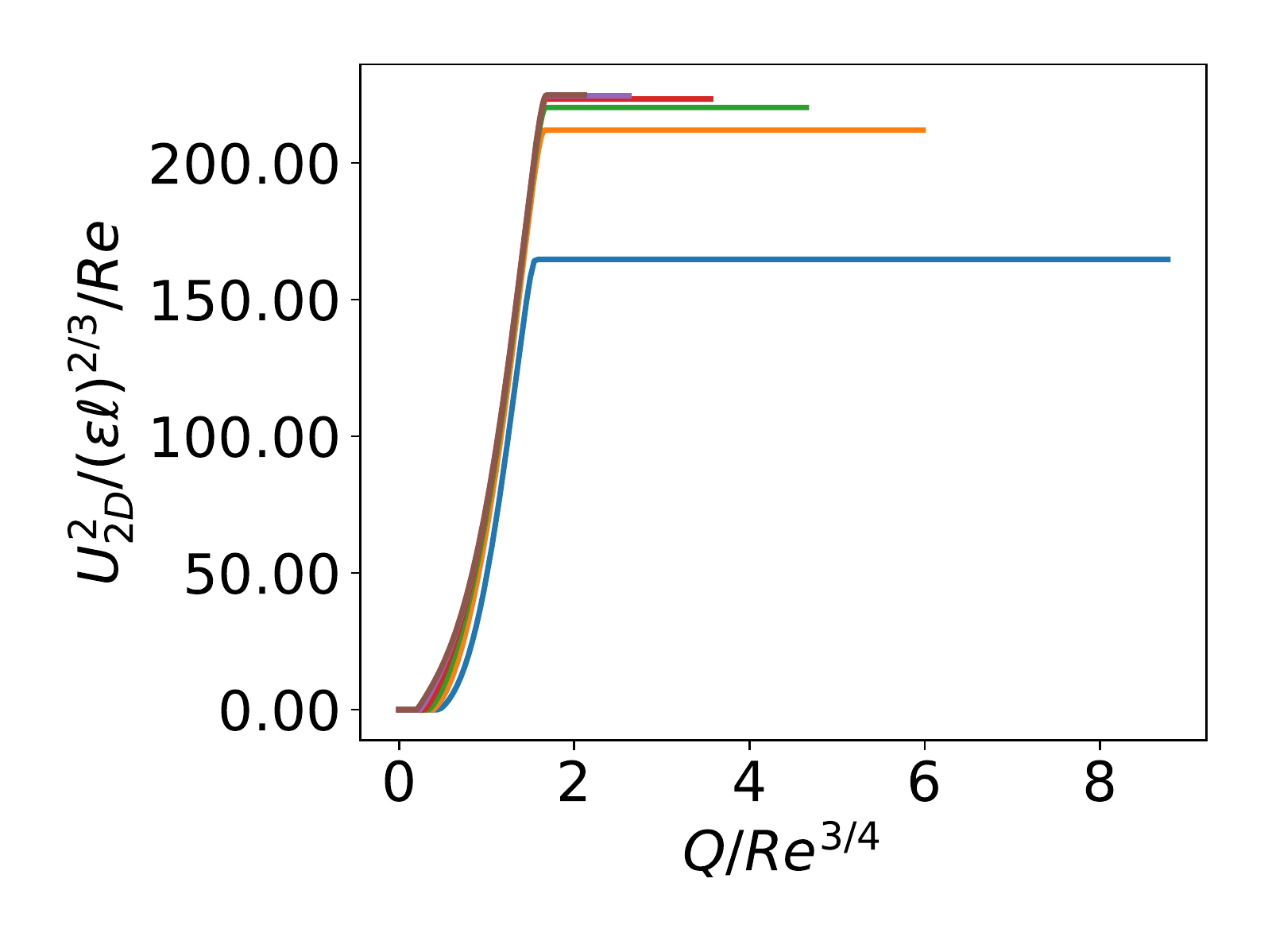} \caption{\label{fig:bif_diag_intRe_resc}}
 \end{subfigure}
 \caption{ Model steady state $U_{2D}^2$ with and without rescaling. The overall structure of the plots is strongly reminiscent of corresponding DNS results in figures \ref{fig:U2_nonres} and \ref{fig:U2_res}. In figure \ref{fig:bif_diag_intRe}, $U_{2D}$ vanishes at small $Q$, increases monotonically between $Q_{3D}$ and $Q_{2D}$ and remains constant for $Q>Q_{2D}$. Figure \ref{fig:bif_diag_intRe_resc} shows the same data as \ref{fig:bif_diag_intRe} with the abscissa rescaled by $\Rey^{3/4}$ and the coordinate rescaled by $\Rey$ as in figure \ref{fig:U2_res}. The collapse improves with increasing $\Rey$. Parameters: $\alpha=0.51$, $\beta=8$, $\gamma =0.1$, $\ell=L/15=1$.} \label{fig:3scale_bifdiag}      
 \end{figure}                                                         

 From this simple model, three major predictions may be derived: 
\begin{itemize} 
\item Firstly, the critical point $Q_{3D}$ is predicted to converge to a $\Rey$-independent value at large $\Rey$ as is shown in figure \ref{fig:convergence_Q3}. In fact, in the infinite $\Rey$ limit of the  model, there remains only one bifurcation, namely that at $Q_{3D}$ between two-dimensional turbulence and the split cascade state. 

\item  Secondly, the critical point $Q_{2D}$ is predicted to obey
\begin{equation}
Q_{2D} \propto \Rey^{3/4},
\end{equation}
\item and thirdly, for $Q>Q_{2D}$, i.e. in the 2D turbulent state, the steady state energy is predicted to be 
\begin{equation}
U_{2D}^2 = (\epsilon\ell)^{2/3 }\left(\frac{L}{\ell}\right)^2\Rey .
\end{equation}\text{ }
\end{itemize}
The detailed derivations of these results are given in Appendix \ref{sec:appA}. All these three main features are in agreement with the DNS and therefore the diagram 
that displays the different phases of the model, shown in figure \ref{fig:bif_diag_intRe}, resembles the corresponding figure \ref{fig:U2_nonres} from the DNS. Indeed, the same rescaling collapses the curves in both cases, see figures \ref{fig:U2_res} and \ref{fig:bif_diag_intRe_resc}. We also note that for $0<Q_{2D}-Q\ll 1$, it is predicted that $U_{3D}^2 \sim (Q_{2D}-Q)^{2}$ (see appendix \ref{sec:appA2}), again in agreement with the DNS.

We understand the present ODE model as a mean field description which captures the global system behaviour and averaged quantities, but does not take fluctuations into account. Due to the importance of fluctuations near criticality, the ODE model does not reproduce the detailed behaviour there. However, when a fluctuating energy input is taken into account by replacing $\epsilon \to \epsilon + \sigma \zeta $ in (\ref{eq:3scale_schematic} b) ($\zeta$ being white Gaussian noise), on-off intermittency is found close to $Q_{2D}$ where the PDF of $U_{3D}^2$ follows a power law with an exponent tending to $-1$ as $Q\to Q_{2D}$ from below (see appendix \ref{sec:appA3}), just as in the DNS. This is a consequence of the structure of the model equations.

To conclude this section, we reiterate that the model presented above successfully captures the the location of the critical points (up to a scaling factor) as well as the amplitude of the condensate $U_{2D}^2$, while not producing a hysteresis. The intermittency close to $Q_{2D}$ found in DNS is reproduced by the model when additive noise is included. 

\section{ Conclusions}

We present the first detailed numerical study of the steady state of thin-layer turbulence as a function of the system parameters using an extensive set of high-resolution simulations. 

It is shown that the split cascade observed at early times of the flow evolution \citep{celani2010morethantwo,benavides_alexakis_2017, musacchio2017split} leads to the formation of condensate states in the long-time limit. 
Three different states were found for large $\Rey$. 
   (a) For very thick layers the system saturates in a regular 3D turbulence state with no inverse cascade and negligible dissipation at large scales.
   (b) At intermediate layer thickness, a flux-loop condensate is formed in which part of the energy transferred to the condensate by the 2D motions is transferred back to the small scales by the 3D motions.
   (c) For very thin layers, the system becomes two-dimensional and forms a 2D-turbulence condensate, where the inversely cascading energy is balanced by the dissipation due to viscosity at large scales.
The transition from  3D turbulence to the flux loop condensate occurs at a critical height $H_{3D}$ $(Q_{3D})$
that is a decreasing (increasing) function of $\Rey$, but saturates at a $\Rey$ independent value for large $\Rey$.
For values of $H$ slightly smaller than $H_{2D}$ the amplitude of the large-scale velocity $\ULS$ jumps discontinuously
to a large value and increases linearly after that. Close to the threshold, a hysteresis diagram was constructed where the
system saturates to a different attractor (3D turbulence or flux-loop condensate) depending on the initial conditions.
Whether this hysteresis behaviour persists at larger $\Rey$ and larger box-sizes $1/K$ remains an open question.
The flux loop condensate transitions to a 2D turbulence condensate at a critical height $H_{2D}$ that scales like
$H_{2D} \propto \ell \Rey^{-3/4}$ unlike the early stages of the development where $H_{2D} \propto \ell \Rey^{-1/2}$
\citep{benavides_alexakis_2017}. For the 2D turbulence condensate, the large-scale energy was found to be inversely proportional to $\Rey$ and independent from $H$. The transition from a flux loop condensate to a 2D turbulence condensate showed strong spatio-temporal intermittency leading to a scaling of the average 3D energy as the square of the deviation from onset $U_{3D}^2\propto (H-H_{2D})^2$, similarly as in \citep{benavides_alexakis_2017}. \\

A three-mode model has been proposed which reproduces the DNS scalings of the the critical points $H_{2D}$ and $H_{3D}$ as well as the amplitude of the condensate in the 2D turbulence regime. The model demonstrates the basic mechanisms involved: A 2D flow that moves energy from the forcing scale to the condensate and a 3D flow that takes away energy both form the large scales and the forcing scales. The model does not describe bistability or discontinuity close to $Q_{3D}$. Nonetheless, it captures the occurrence of both transitions observed in the DNS and provides several correct quantitative predictions. 

We stress once more that the present work is the first numerical study of thin-layer turbulent condensation. Previous studies of the thin-layer problem were restricted to the transient inverse cascade regime due the long computation time needed to reach the condensate state. Therefore, the present study is novel and provides an important first step towards a better understanding of thin-layer turbulent condensates (of which Earth's atmosphere and ocean may be viewed as examples, despite the idealised nature of our set-up), many open questions remain. The complexity of the physics involved close to criticality goes beyond the mean field model and requires further targeted studies. We are convinced that both critical points deserve more detailed investigations by means of numerical simulations, experiments and modelling. Another important remaining open problem is the formation of an inverse cascade from a 3D forcing. To study this, the amplitude of the 3D components of the forcing should be varied compared to the 2D components in a future study. 3D forcing will make a connection with more natural forcing mechanisms like convection that also display condensates \citep{favier2014inverse,rubio2014upscale,guervilly2014large}.

Concerning the realisability of the present numerical results in an experiment, it needs to be stressed that this study only considers the triply periodic domain for simplicity. When attempting to transfer the results to non-slip boundary conditions, a word of caution is therefore in order: viscous boundary layers may lead to large-scale drag, which is explicitly left out from the model set-up used here.  Also, 3D turbulence in boundary layers may infect the interior flow, thereby affecting even high wavenumbers and the two-dimensionalisation even in the bulk of the flow. However, the wealth of experimental observations of turbulent condensates in thin layers, as referenced in the introduction and summarised in \citep{xia2017two}, suggests that the condensation phenomenon at finite height is robust between different boundary conditions as well as between the different forcing methods used in experiment and numerical simulations. In particular, it would be very interesting to probe the discontinuity and associated phenomena reported here in an experiment. This has not been done before and experimental studies of thin-layer turbulent condensates have the advantage of allowing higher Reynolds numbers and much better time statistics. 

\label{sec:discussion}
\section*{Acknowledgements}

We thank three referees for their helpful and valuable suggestions which have helped to improve this paper substantially. This work was granted access to the HPC resources of MesoPSL financed by the R\'egion 
Ile de France and the project Equip@Meso (reference ANR-10-EQPX-29-01) of 
the programme Investissements d'Avenir supervised by the Agence Nationale pour 
la Recherche and the HPC resources of GENCI-TGCC-CURIE \& CINES	Occigen 
(Project No. A0010506421, A0030506421 \& A0050506421) where the present numerical simulations have been performed.
This work has also been supported by the Agence nationale de la recherche
(ANR DYSTURB project No. ANR-17-CE30-0004). AvK was supported by Deutscher Akademischer Austauschdienst and Studienstiftung des deutschen Volkes.


\appendix
\section{Derivation of mean field model predictions} 
\subsection{Scalings of critical points and condensate amplitude}
\label{sec:appA}
 In the low viscosity limit, the eddy viscosities given in equation (\ref{eq:param_eddy_visc}), take the form (\ref{eq:highRe_eddyvisc}) and the resulting system of equations reads
\begin{subeqnarray}
\partial_t U_{2D}^2 =& \frac{(\alpha U_f \ell) U_{2D}^2 }{L^2} - \frac{(\beta U_{3D} H) U_{2D}^2}{L^2}\\
\partial_t U_{f}^2  \phantom{x} =& \epsilon - \frac{(\alpha U_f \ell) U_{2D}^2}{L^2} - \frac{(\gamma U_{3D} H) U_f^2}{\ell^2} \\
\partial_t U_{3D}^2 =& \frac{(\beta U_{3D} H) U_{2D}^2}{L^2} + \frac{( \gamma U_{3D}H) U_f^2}{\ell^2} - \frac{U_{3D}^3}{H}
\label{eq:3scale_highRe}
\end{subeqnarray}
One can easily see that these equations do not permit a fixed point with $U_{3D}=0$ when $\epsilon \neq 0$. To show this, first note that, as in the finite $\Rey$ case, the forcing scale velocity $U_f$ cannot vanish at a fixed point if $\epsilon \neq 0$. Assume there exists a fixed point with $U_{3D}=0$. Then equation (\ref{eq:3scale_highRe}\textit{c}) is trivially satisfied, while (\ref{eq:3scale_highRe}\textit{a}) implies that $U_{2D}=0$ or $U_f=0$. Since $U_f$ must be non-zero, we have $U_{2D}=0$, which leads to a contradiction in equation (\ref{eq:3scale_highRe}\textit{b}) for any $\epsilon \neq 0$. Hence neither a laminar flow state nor a 2D condensate state exists in the system in the infinite $\Rey$ limit. The only two remaining fixed points are 3D turbulence and the flux-loop condensate. The former is given by
\begin{equation}
U_{2D}^2 = 0,\quad 
U_f^2= \frac{\epsilon^{2/3} \ell^2}{\gamma H^{4/3}},\quad
U_{3D}^2 = (\epsilon H)^{2/3}
\end{equation}
Using this result and considering equation (\ref{eq:3scale_highRe}\textit{a}), we can find that the 3D turbulence fixed point becomes unstable to 2D perturbations at
\begin{equation}
H =  \left(\frac{\alpha^2 }{\beta^2 \gamma}\right)^{1/4}\ell
\end{equation}
and thus we obtain that
\begin{equation}
Q_{3D}=  \left(\frac{\beta^2 \gamma}{\alpha^2 }\right)^{1/4}.
\end{equation}
Hence, in the low viscosity limit of our three-scale model, there remains only one bifurcation, namely that at $Q_{3D}$ between two-dimensional turbulence and the split cascade state. The second critical point $Q_{2D}$ vanishes to infinity as $Q_{2D}\propto\Rey^{3/4}$ in this limit. 
Figure \ref{fig:convergence_Q3} demonstrates close to $Q_{3D}$ that the full model converges
to the solution obtained from the asymptotic form of the equations \ref{eq:3scale_highRe} as $\Rey$ increases. This is consistent with the convergence observed in the DNS in figure \ref{fig:U2_res}.

At finite viscosity, one has to solve the full equations,(\ref{eq:param_eddy_visc}) which is difficult analytically for the 2D condensate state. In order to facilitate analytical progress in deriving predictions from the model, one may formally take the high viscosity limit in which the different eddy viscosities take the form of equation (\ref{eq:lowRe_eddyvisc}). The model equations then become
\begin{subeqnarray}
\partial_t U_{2D}^2  = & -\left(\nu - \alpha \frac{U_{f}^2 \ell^2}{\nu} + \beta \frac{U_{3D}^2 H^2}{\nu} \right) \frac{U_{2D}^2}{L^2}, \\
\partial_t U_{f}^2\phantom{x} = &  \epsilon - \left( \nu + \gamma\frac{ U_{3D}^2 H^2 }{\nu} \right) \frac{U_f^2}{\ell^2} - \alpha\frac{ U_f^2 \ell^2}{\nu} \frac{U_{2D}^2}{L^2},\\
\partial_t U_{3D}^2 =   &  \beta \frac{ U_{3D}^2 H^2}{\nu} \frac{U_{2D}^2}{L^2} + \gamma \frac{ U_{3D}^2 H^2}{ \nu} \frac{U_{f}^2}{\ell^2} - \frac{U_{3D}^3}{H} - \nu \frac{U_{3D}^2}{H^2} . \label{eq:3scale_lowRe}
\end{subeqnarray}
To obtain this limiting form of the equations, it is assumed that $\nu\gg U_f \ell ,U_{3D} H$, while no restriction is imposed on $U_{2D}$; in particular, the case of a large scale-based Reynolds number in the large scales $U_{2D}L/\nu$, which is most relevant in the condensate state, is included. The laminar flow is unstable to 3D perturbations when $Q<\gamma^{1/4} \Rey^{3/4}$ and unstable to 2D perturbations when 
\begin{equation}
\Rey> 1/\alpha^{1/3}.
\label{eq:remod}
\end{equation}
When the latter condition is satisfied and $H$ is sufficiently small ($Q$ sufficiently large), the system is attracted to the 2D condensate state, given by 
\[ 
U_{2D}^2=\frac{L^2}{\nu} \left( \epsilon - \frac{\nu^3}{\ell^4 \alpha}  \right),\quad 
U_f^2 = \frac{\nu^2}{\alpha \ell^2},  \quad U_{3D}= 0 .
\]
Note that $U_{2D}^2$ is inversely proportional to the viscosity and proportional to $L^2$ in agreement with 
the scaling of the data in figure \ref{fig:U2_res}. The 2D condensate state ceases to be an attractor of the system when $H$ is sufficiently large such that $U_{3D}$ becomes unstable. This occurs when  
\begin{equation}
H^4 > \left(\beta \frac{\epsilon}{\nu^3} + \frac{\gamma-\beta}{\alpha \ell^4}  \right)^{-1}. 
\end{equation}
Hence, we conclude that 
\begin{equation}
Q_{2D} = \left( \beta\frac{ \epsilon\ell^4  }{ \nu^3} + \frac{\gamma-\beta}{\alpha}\right)^ {1/4} = \left( \beta \Rey^3 + \frac{\gamma-\beta}{\alpha} \right)^{1/4}. \label{eq:H2_lowRey}
\end{equation}
Thus, for moderate values of $\Rey$, there is an approximate scaling $Q_{2D} \propto \Rey^{3/4}\propto \eta$,  the dissipation length (note that $ Re^3 > 1/\alpha$ due to eq. (\ref{eq:remod})), in agreement with the results obtained in section \ref{sec:DNS}, where we showed that the $U_{2D}^2$ data points collapse under rescaling such that $Q \Rey^{3/4}=\eta/H$ is on the abscissa and $U_{2D}^2 K^2/[\Rey (\epsilon \ell)^{2/3}]$ on the coordinate. 
Results from the full model equations (\ref{eq:3scale_schematic}) and (\ref{eq:param_eddy_visc}) are shown in figure  \ref{fig:3scale_bifdiag} where the same scaling is applied. The corresponding plots for equations (\ref{eq:3scale_lowRe}) are very similar. Furthermore, an asymptotic analysis close to $Q_{2D}$ described in section \ref{sec:appA2} of this appendix reveals that the scaling for $U_{3D}^2\propto (Q_{2D}-Q)^2$ which is the same as in the DNS results shown in figure \ref{fig:U3_res} although no intermittency is present. The other critical point $Q_{3D}$, where the $3D$ turbulence solution changes stability, can be evaluated numerically and is found to increase with $\Rey$ indefinitely. This, however, is an artefact of the high viscosity asymptotic form of the eddy viscosities used in this subsection.

\subsection{Behaviour of $U_{3D}$ near $Q_{2D}$}\label{sec:appA2}
Here, we derive the behaviour close to $H_{2D}$ in the three-scale model. First consider $H=H _{2D}(1+\delta)$ and let $\mathbf{x} = (x,y,z)^T = (U_{2D}^2,U_{f}^2,U_{3D}^2)^T$, $   \tilde{\mathbf{x}}= (\tilde{x},\tilde{y},\tilde{z})^T  = \mathbf{x} - (x_{2D},y_{2D},0)^T $, where $x_{2D}$ and $y_{2D}$ are the values of $U_{2D}^2$ and $U_{f}^2$ respectively at $H=H_{2D}$. Then equations (\ref{eq:3scale_lowRe}) can be rewritten exactly in the form
\begin{equation}
 \frac{\mathrm{d}}{\mathrm{d}t}\tilde{\mathbf{x}} = \begin{pmatrix} -\frac{\nu}{L^2} & \frac{\alpha \ell^2 x_2}{\nu L^2} & - \frac{\beta x_2 H^2}{\nu L^2} \\ -\frac{\alpha y_2}{\nu L^2} & -\frac{\nu}{\ell^2} & - \frac{\gamma H^2 y_2}{\nu \ell^2} \\ 0 & 0 & C \end{pmatrix} \tilde{\mathbf{x}} + \begin{pmatrix} 0 \\ 0   \\ -1/H_{2D}\end{pmatrix} \tilde{z}^{3/2} +\tilde{\mathbf{x}}^T \textbf{B} \tilde{\mathbf{x}},
\end{equation}
where $C= -\frac{\nu}{H^2} + \frac{\beta H^2 x_2}{ \nu L^2} +\frac{\gamma H^2 y_2}{\nu \ell^2}$ and the specific coefficients of the quadratic term are irrelevant here. By definition of $H_{2D}$, $C(\delta=0) = -\frac{\nu}{H_{2D}^2} + \frac{\beta H_{2D}^2 x_2}{\nu L^2} +\frac{\gamma H_{2D}^2 y_2}{\nu \ell^2}=0$. Hence, for small $\delta$, $C\propto \delta$. Specifically,
\begin{equation}
C \stackrel{\delta \ll 1}{\sim} \left( \frac{2\nu}{H_{2D}^2} + \frac{2\beta x_2}{\nu L^2} + \frac{2\gamma y_2}{\nu \ell^2}  \right) \delta.
\end{equation}
Hence, considering the $\tilde{z}$ component and balancing the linear term with the $\tilde{z}^{3/2}$ term, we deduce that 
\begin{equation}
\tilde{z} \stackrel{\delta \ll 1}{\sim} \left( \frac{2\nu}{H_{2D}^2} + \frac{2\beta x_2 H_{2D}^2}{\nu L^2} + \frac{2\gamma y_2H_{2D}^2}{\nu \ell^2}  \right)^2 H_{2D}^2 \delta^2.
\label{eq:asymp_U3D_delta}
\end{equation}
This means that $U_{3D}^2 \propto \delta ^2$, which is precisely the scaling observed in figure \ref{fig:U3_res}. It is important to note however that the asymptotic result (\ref{eq:asymp_U3D_delta}) is only valid for very small $\delta$ and cannot be extended to $\delta \sim O(1)$ where the quadratic terms are dominant.

\subsection{On-off intermittency in the three-scale model}
\label{sec:appA3}
\begin{figure}                           
 \centering                              
 \begin{subfigure}[b]{0.4\linewidth}
\includegraphics[width=\textwidth]{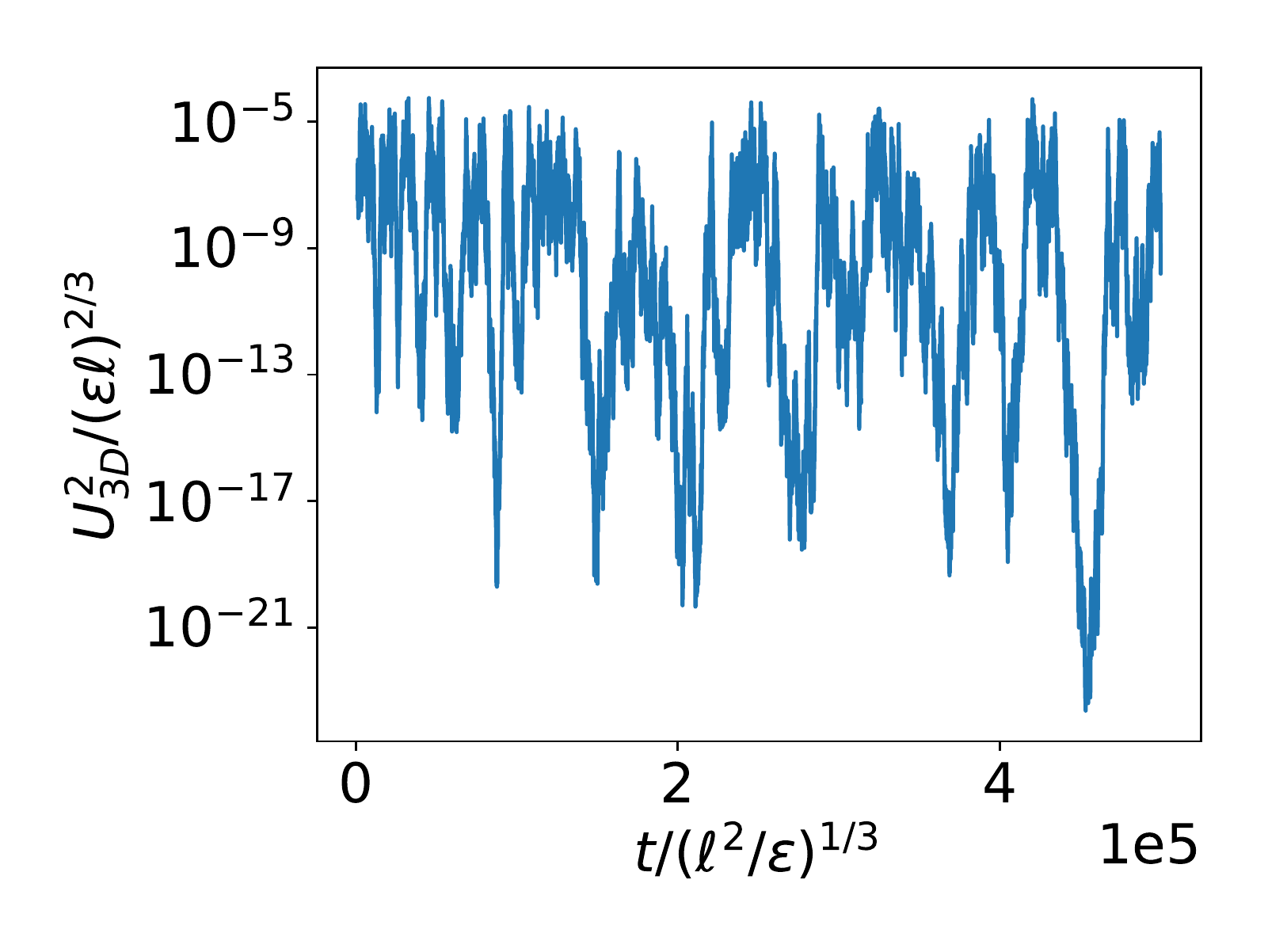} \caption{\label{fig:ts_int_3mode}}
 \end{subfigure}
 \begin{subfigure}[b]{0.4\linewidth}
\includegraphics[width= \textwidth]{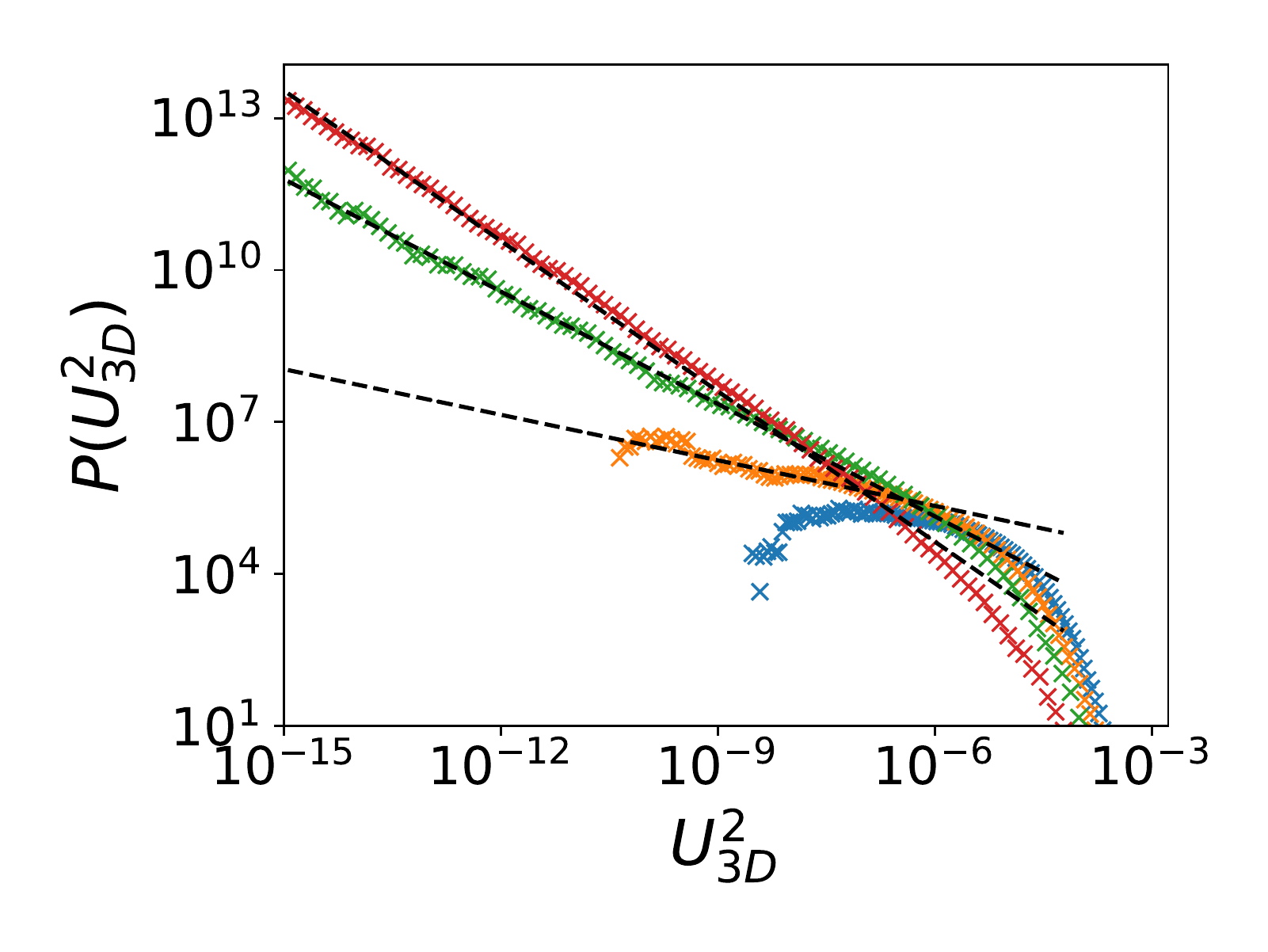} \caption{\label{fig:pdf_int_3mode}}
 \end{subfigure}
 \caption{Time series and PDF showing on-off intermittency close to $Q_{2D}\approx 1.638$ in the 3-scale model with a fluctuating energy injection rate. Parameters: panel (a) $Q=1.63775$, $L=10\ell$, $\Rey = 2$, $\alpha = 0.001$, $\beta = 10$, $\gamma =0.9$, $\sigma =0.1$, panel (b) $q=1.635,\text{ }1.636,\text{ }1.637,\text{ }1.63775$ (bottom to top), other parameters identical. Dashed lines in panel (b) are power laws with exponents $-0.3$ $-0.74$, $-0.99$ (bottom to top). Cf. figures \ref{fig:temp_int} \& \ref{fig:U3_res}.} \label{fig:int_3scale} 
 \end{figure}                             
When a fluctuating energy injection rate is taken into account in the model by replacing $\epsilon\to \epsilon +\sigma \zeta$, where $\zeta\sim \mathcal{N}(0,1)$ is Gaussian white noise, on-off intermittency in $U_{3D}^2$ can be observed in the three-scale model. This is illustrated in figure \ref{fig:int_3scale} in terms of the time series of $U_{3D}^2$ and the corresponding PDF, which approaches a power law with exponent $-1$ as $Q \to Q_{2D}$.

\bibliographystyle{jfm}
\bibliography{LayerTurbulence}

\end{document}